\documentclass[
  aps,prb,
  reprint,
  amsmath,amssymb,
  floatfix,
  nofootinbib,
  superscriptaddress
]{revtex4-2}

\usepackage{graphicx}
\usepackage{bm}
\usepackage{mathtools}
\usepackage{booktabs}
\usepackage{float}
\usepackage{placeins}
\usepackage{xcolor}
\usepackage{xr-hyper}
\usepackage[colorlinks=true,allcolors=blue]{hyperref}
\hypersetup{
  pdftitle={Channel concentration of critical quantum geometry},
  pdfauthor={Qian-Rui Lee and Daw-Wei Wang}
}

\graphicspath{{figures/}}

\newcommand{\Var}{\operatorname{Var}}
\newcommand{\Real}{\operatorname{Re}}
\newcommand{\Imag}{\operatorname{Im}}

\renewcommand{\d}{\mathrm{d}}

\newcommand{\ket}[1]{\lvert #1 \rangle}
\newcommand{\bra}[1]{\langle #1 \rvert}
\newcommand{\braket}[2]{\langle #1 \vert #2 \rangle}
\newcommand{\ketbra}[2]{\lvert #1 \rangle\!\langle #2 \rvert}

\newcommand{\abs}[1]{\left\lvert #1 \right\rvert}
\newcommand{\norm}[1]{\left\lVert #1 \right\rVert}

\newcommand{\Qgt}{\mathcal{Q}}              % quantum geometric tensor
\newcommand{\Berry}{\mathcal{F}}            % Berry curvature (imag part)
\newcommand{\chann}{c}                      % channel index
\newcommand{\Nchan}{\mathcal{N}}            % number of channels

\newcommand{\KF}{K_{F}}
\newcommand{\Ptwo}{P_{2}}
\newcommand{\Pfour}{P_{4}}
\newcommand{\Pn}[1]{P_{#1}}
\newcommand{\Neff}{N_{\mathrm{eff}}}
\newcommand{\xk}{x_{k}}                     % unnormalized channel weight (k)

\newcommand{\hc}{h_{c}}
\newcommand{\Lk}{\Lambda_{k}}               % single-particle gap
\newcommand{\thetak}{\theta_{k}}            % Bogoliubov angle
\newcommand{\gam}{\gamma}

\newcommand{\Phisc}{\Phi}                   % universal scaling function (TFIM)
\newcommand{\Psisc}{\Psi}                   % Lifshitz longitudinal function
\newcommand{\kp}{\kappa}                    % rescaled momentum index

\newif\ifdraftnotes
\draftnotesfalse

\begin{document}

% WPM-BLOCK:B0001
\title{Channel concentration of critical quantum geometry}

% WPM-BLOCK:B0002
\author{Qian-Rui Lee}
\affiliation{Department of Physics, National Tsing Hua University, Hsinchu 30013, Taiwan}

% WPM-BLOCK:B0003
\author{Daw-Wei Wang}
\affiliation{Department of Physics, National Tsing Hua University, Hsinchu 30013, Taiwan}
\affiliation{Center for Theory and Computation, National Tsing Hua University, Hsinchu 30013, Taiwan}
\affiliation{Center for Quantum Technology, National Tsing Hua University, Hsinchu 30013, Taiwan}

% WPM-BLOCK:B0004
\date{September 11, 2026}

% WPM-BLOCK:B0005
% ----------------------------------------------------------------------

% WPM-BLOCK:B0006
\begin{abstract}
The quantum metric quantifies the total ground-state response along a
parameter direction, but does not resolve how excitations share that response.
We define channel concentration (CC) as the sum of squared normalized response
weights over specified excitation channels, such as momentum blocks. An exact
finite-size theorem yields thermodynamic CCs of 2/3 for the field response of
the critical transverse-field Ising model and 1/3 for the half-filled XX
pairing response, despite the same leading metric scaling. In finite-size
approaches to the XY Lifshitz point, field and anisotropy perturbations yield
different concentrations despite a common limiting Hamiltonian with quadratic
dispersion. In a unitary 1+1-dimensional conformal field theory (CFT) on a
circle, we consider a nondegenerate vacuum in a fixed sector perturbed by one
spatially integrated scalar primary. We derive complete zero-momentum
energy-level response weights, including descendants. For scaling dimension
\(0<\Delta<3/2\), these weights determine the normalized response distribution
and an exact universal concentration function. The expression
reproduces the exact Ising lattice limit 2/3 and gives approximately 0.8515
for the three-state Potts thermal field, compared with approximately 0.800
from an exponent-only approximation. Finite-size interacting calculations
compare concentrations and ranked response weights over many-body energy
levels. These exact benchmarks show which response distinctions total metric
scaling misses and guide comparisons with finite-size interacting spectra.
\end{abstract}

% WPM-BLOCK:B0007
\maketitle

% WPM-BLOCK:B0008
% ======================================================================

% WPM-BLOCK:B0009
\section{Introduction}
% WPM-BLOCK:B0010
\label{sec:intro}
% WPM-BLOCK:B0011
% ======================================================================

% WPM-BLOCK:B0012
The quantum geometric tensor (QGT) describes the change of a quantum state
under parameter displacements~\cite{Provost1980,Berry1984,AnandanAharonov1990}.
Its real part is the Fubini--Study metric and, for one tuning parameter, the
fidelity susceptibility.  Fidelity susceptibility is an established
order-parameter-free probe of quantum criticality
\cite{ZanardiPaunkovic2006,YouLiGu2007,VenutiZanardi2007,ZanardiGiordaCozzini2007,Gu2010,KolodrubetzEtAl2017,CarolloEtAl2020}.
In the spectral representation of fidelity susceptibility, each excitation contributes a positive
weight set jointly by an energy denominator and a perturbation form factor.
Their sum measures how strongly the ground state changes. It does not reveal
whether that change is carried by one soft branch, several symmetry-related
branches, or a broad continuum of resolved excitations.

Critical exponents organize the singular scaling of the total response,
but that scaling does not directly display its composition. We use channel
concentration (CC) as the central quantitative diagnostic of this composition
and the full normalized weight distribution to resolve its structure.
At relativistic fixed points, conformal field theory (CFT) supplies operator,
descendant, and level data that can determine such distributions for
specified perturbations and projectors. We ask how to calculate these
distributions, which distinctions they reveal beyond aggregate scaling,
and how to compare them under declared probe conditions, including
controlled changes of channel resolution.

% WPM-BLOCK:B0575
The critical transverse-field Ising model (TFIM) and XX chain give the first
exact test. We compare the TFIM field tangent at \((h,\gamma)=(1,1)\)
with the pairing tangent of the half-filled XX chain at \((0,0)\).
Both responses use positive-momentum Bogoliubov blocks in the even-parity
Neveu--Schwarz (NS) sector.  For the XX sequence, we take
\(L\equiv0\pmod4\).  Both susceptibilities scale as \(L^2\). Exact folding nevertheless shows
that the TFIM response forms one soft ladder, whereas the XX response
forms two equal soft ladders. Their concentrations consequently tend to
\(2/3\) and \(1/3\), respectively.  Equal leading response scaling therefore does not determine
carrier multiplicity.

Directional dependence at the \(XY\) multicritical point is already
established~\cite{MukherjeePolkovnikovDutta2011}. Here, field and anisotropy
tangents provide a second test through punctured approaches to the
Lifshitz point.  They approach the same limiting Hamiltonian
with the same quadratic, \(z=2\) dispersion. Their numerator form factors,
however, select different normalized envelopes. A shared dispersion therefore
does not determine perturbation selectivity.

% WPM-BLOCK:B0576
To describe these response distributions, we choose an orthogonal family
of excitation projectors. Let \(\bm\lambda\) denote the Hamiltonian
parameter point.
The quantity \(x_{\chann}(\bm\lambda,y)\ge0\) is the Fubini--Study weight of
channel \(\chann\) for tangent direction \(y\).  We use
% WPM-BLOCK:B0577
\begin{align}
  \Ptwo&=\sum_{\chann}x_{\chann},
  &\pi_{\chann}&=\frac{x_{\chann}}{\Ptwo},\notag\\
  \Pfour&=\sum_{\chann}x_{\chann}^2,
  &\KF&=\sum_{\chann}\pi_{\chann}^2=\frac{\Pfour}{\Ptwo^2}.
  \label{eq:KFdef}
\end{align}
% WPM-BLOCK:B0578
Thus \(\Ptwo\) measures the total response, while the order-two concentration \(\KF\)
measures how strongly it is concentrated among the declared channels. The normalized
weights \(\pi_{\chann}\) provide the full resolved composition underlying this
indicator. Distinct distributions can have the same CC, so the scalar
concentration does not identify the full composition. Here
\(\Pfour=\sum_c x_c^2\) is the second
raw moment of the positive response weights. It is a quartic form in the
tangent direction, not a fourth-order fidelity coefficient or connected
fourth cumulant. The inverse concentration, \(\Neff=1/\KF\), is the
order-two Hill or R\'enyi effective number
\cite{Renyi1961,Leinster2021}. It is a participation number, not a literal
count of nonzero modes.

% WPM-BLOCK:B0579
The resolution in Eq.~\eqref{eq:KFdef} specifies the excitation projectors
used to define the response channels. It is part of the probe definition,
rather than an implicit basis choice. For the free chains, translation
invariance makes \((k,-k)\) Bogoliubov blocks labeled by \(k>0\) the
natural channels.  On a periodic spin ring,
the even-parity ground-state sequence maps to antiperiodic NS fermions and
avoids the critical zero mode
\cite{Damski2013,DamskiRams2014,Oshikawa2020}.  A different spin structure or
projector partition asks a different resolved question.  The interacting
calculations therefore declare a distinct many-body level-projector
resolution rather than importing the free-chain values.

% WPM-BLOCK:B0580
Existing exact results give the total fidelity susceptibilities of the
TFIM and \(XY\) chains~\cite{Damski2013,DamskiRams2014,LuoZhaoWang2018}.  The mapping of
an XY chain to two commuting Ising Hamiltonians is also established
\cite{IgloiJuhasz2008}.  Our contribution is the resolved response identity
for specified tangents, NS sectors, and momentum-block projectors.  It gives
the finite-size moments and separates ladder multiplicity from within-ladder
shape.

Other constructions expose complementary frequency, Krylov, spatial, and
tensor information about the response. Spectral
constructions expose excitation-frequency support and recover fidelity
susceptibility by frequency integration
\cite{YouLiGu2007,GuYu2014,DeGrandiPolkovnikov2010,DeGrandiGritsevPolkovnikov2010}.
Adiabatic gauge-potential norms retain total-deformation information
\cite{KolodrubetzEtAl2017,OrlovEtAl2023,PozsgayEtAl2024}.
A complementary Krylov construction normalizes resolvent amplitudes over
Lanczos-generated layers.  It resolves how inverse-energy response explores
a dynamically generated Krylov chain~\cite{AlishahihaVasli2026}.
A related construction resolves quantum Fisher information into
state-metric-orthogonal Krylov components of the symmetric logarithmic
derivative~\cite{AlishahihaTabeshVasli2026QFI}.
Spatial markers and QGT reconstruction retain spatial or tensor
structure~\cite{deSousaEtAl2023,ZhengEtAl2022,ChenEtAl2024}.

For the comparisons here, the channels are a predeclared orthogonal family of
physical excitation projectors. Direct comparisons use compatible specifications of the operator,
sector, spin structure, partition, and limit order. Specific responses with
different operators or parameters can be related by an explicit identity,
as in the TFIM--XX theorem. For the same response, known grouping or
refinement relations permit bounds on the change in concentration.
Unlike an eigenstate inverse participation ratio in a coordinate
basis~\cite{Wegner1980}, CC is formed from fidelity-response weights.
It records how the total response \(\Ptwo\) is divided among excitations
at the declared physical resolution.

% WPM-BLOCK:B0581
Together, these results make response distributions over declared
excitation channels a calculable complement to total metric scaling.
Finite-size TFIM--XX identities distinguish one- from two-ladder composition
for the specified tangents and every R\'enyi index, while the Lifshitz
analysis isolates tangent selectivity. For the response to one spatially
integrated scalar primary in a unitary \(1+1\)-dimensional CFT on a circle
with a nondegenerate vacuum in a fixed sector, complete zero-momentum level
weights determine the full distribution and a universal concentration
function for \(0<\Delta<3/2\). These exact distributions provide fixed
references for finite-size interacting comparisons of scalar concentration
and ranked many-body level weights. Refinement bounds constrain changes
between specified resolutions, and independent Bernoulli counting of
final-Hamiltonian Bogoliubov pairs supplies a conditional zero-amplitude
readout of CC at fixed finite size.

% WPM-BLOCK:B0583
Section~\ref{sec:channel} defines CC and the conditions for comparing probes.
Section~\ref{sec:tfim} gives the exact TFIM--XX theorem, while
Sec.~\ref{sec:lifshitz} analyzes directional Lifshitz scaling and its ordered limits;
Sec.~\ref{sec:cft-response} constructs the complete-level conformal response,
and Sec.~\ref{sec:interacting-benchmarks} compares finite interacting spectra
with fixed references. The Discussion addresses resolution, the limits of
these comparisons, and the conditional counting readout in
Sec.~\ref{sec:operational-access}.

The appendices supply the derivations for these main-text results.
Appendix~\ref{app:spectral-projector} proves the projector and coarse-graining
algebra, and Appendix~\ref{app:cft-cylinder-derivation} derives the conformal
cylinder weights.  Appendix~\ref{app:mode} supplies the two-level metric;
its Sec.~\ref{sec:directional} extends the Lifshitz result to arbitrary
tangent rays.

For the exact chains, Appendix~\ref{app:tfim-scaling} gives the TFIM sums and
scaling derivation.  Appendix~\ref{app:ramond} fixes the boundary-sector
contrast, and Appendix~\ref{app:xx-pairing-theorem} proves the XX folding and
momentum-ladder arithmetic.  Appendix~\ref{app:rg-envelope} derives the RG envelope and the
ultraviolet--infrared correction hierarchy.
Appendix~\ref{app:multicone} states and proves the multiple-ladder
factorization conditions.  The weak-quench feasibility module and numerical
audits are supplied separately in the Supplemental Material
(SM)~\cite{SupplementalMaterial}.

% WPM-BLOCK:B0030
\section{Channel-resolved quantum geometry}
% WPM-BLOCK:B0031
\label{sec:channel}
% WPM-BLOCK:B0032
% ======================================================================

% WPM-BLOCK:B0033
\subsection{Gauge-covariant QGT and channel weights}
% WPM-BLOCK:B0034
\label{sec:channel:qgt}

% WPM-BLOCK:B0035
Let $\ket{\psi(\bm{\lambda})}$ be a normalized representative of a smooth
family of nondegenerate ground-state rays, and define the horizontal
derivative
% WPM-BLOCK:B0036
\begin{equation}
  D_{\mu}\ket{\psi}
  =\bigl(1-\ketbra{\psi}{\psi}\bigr)\partial_{\mu}\ket{\psi},
  \qquad
  \braket{\psi}{D_{\mu}\psi}=0.
  \label{eq:horizontal-derivative}
\end{equation}
% WPM-BLOCK:B0037
The QGT is
% WPM-BLOCK:B0038
\begin{equation}
  \Qgt_{\mu\nu}
  =\braket{D_{\mu}\psi}{D_{\nu}\psi}
  =g_{\mu\nu}-\frac{\mathrm{i}}{2}\,\Berry_{\mu\nu},
  \label{eq:qgt-horizontal}
\end{equation}
% WPM-BLOCK:B0039
so that $g_{\mu\nu}=\Real\!\left(\Qgt_{\mu\nu}\right)$ and
$\Berry_{\mu\nu}=-2\Imag\!\left(\Qgt_{\mu\nu}\right)$.  For a tangent vector
$y=y^{\mu}\partial_{\mu}$, the ordinary QGT metric gives only the total
second-order distinguishability
% WPM-BLOCK:B0040
\begin{equation}
  \Ptwo(\bm{\lambda},y)=g_{\mu\nu}y^{\mu}y^{\nu}
  =\norm{D_y\psi}^{2},
  \qquad D_y=y^{\mu}D_{\mu}.
  \label{eq:P2-total}
\end{equation}

% WPM-BLOCK:B0041
Throughout, we use the fidelity convention
% WPM-BLOCK:B0747
\[
  F(\delta;y)
  \equiv\abs{\braket{\psi(\bm{\lambda})}
                         {\psi(\bm{\lambda}+\delta y)}}
  =1-\frac{\delta^2}{2}\Ptwo(\bm{\lambda},y)+O(\delta^3).
\]
% WPM-BLOCK:B0748
Thus \(\Ptwo\) is the directional fidelity susceptibility in our convention.
If the coefficient of \(\delta^2\) in \(1-F\) is denoted by \(G_{yy}\),
then \(G_{yy}=\Ptwo/2\).

% WPM-BLOCK:B0749
To resolve how this total weight is distributed, choose a physically fixed
orthogonal decomposition of the horizontal space into channels,
% WPM-BLOCK:B0042
\begin{equation}
  \sum_{\chann}\Pi_{\chann}=1-\ketbra{\psi}{\psi},
  \qquad
  \Pi_{\chann}\Pi_{\chann'}=\delta_{\chann\chann'}\Pi_{\chann}.
  \label{eq:projector-resolution}
\end{equation}
% WPM-BLOCK:B0043
A channel is specified together with its resolution convention. Examples
include canonical free-fermion momentum blocks, invariant level projectors,
and excitation classes defined by a protocol or detector.

% WPM-BLOCK:B0045
The channel QGT and its metric part are
% WPM-BLOCK:B0046
\begin{align}
  q^{\chann}_{\mu\nu}
  &=\bra{D_{\mu}\psi}\Pi_{\chann}\ket{D_{\nu}\psi},
  &
  g^{\chann}_{\mu\nu}&=\Real\!\left(q^{\chann}_{\mu\nu}\right),
  \nonumber\\
  \Qgt_{\mu\nu}&=\sum_{\chann}q^{\chann}_{\mu\nu}.
  \label{eq:channel-qgt}
\end{align}
% WPM-BLOCK:B0047
The non-negative channel weight selected by $y$ is
% WPM-BLOCK:B0048
\begin{equation}
  x_{\chann}(\bm{\lambda},y)
  =g^{\chann}_{\mu\nu}y^{\mu}y^{\nu}
  =\bra{D_y\psi}\Pi_{\chann}\ket{D_y\psi}
  \ge0,
  \label{eq:channel-weight-projector}
\end{equation}
% WPM-BLOCK:B0049
Summing \(x_{\chann}\) over channels gives \(\Ptwo\); the normalized weights
\(\pi_{\chann}\), the raw second moment \(\Pfour\), and \(\KF\) are defined in
Eq.~\eqref{eq:KFdef}.  For a nondegenerate many-body spectrum, we can resolve
the channels into individual energy eigenstates by choosing
\(\Pi_n=\ketbra{n}{n}\). Write the directional perturbation as
\(\partial_yH=y^{\mu}\partial_{\mu}H\). For \(n>0\), first-order
perturbation theory gives
% WPM-BLOCK:B0052
\begin{equation}
  x_n(\bm{\lambda},y)
  =\frac{\abs{\bra{n}\partial_yH\ket{0}}^{2}}
        {(E_n-E_0)^{2}}.
  \label{eq:spectral-weight-main}
\end{equation}
% WPM-BLOCK:B0053
The projector derivation is given in Appendix~\ref{app:spectral-projector}.
This spectral form is the standard fidelity-susceptibility/QGT
representation~\cite{VenutiZanardi2007,Gu2010,KolodrubetzEtAl2017}.
Equation~\eqref{eq:spectral-weight-main} also relates these spectral weights
to the adiabatic gauge potential (AGP), with the generator sign fixed by
convention. The horizontal state variation along \(y\) is
% WPM-BLOCK:B0054
\begin{equation}
  D_y\ket{0}
  =\sum_{n>0}\ket{n}\,
    \frac{\bra{n}\partial_yH\ket{0}}{E_0-E_n}.
  \label{eq:agp-state-response}
\end{equation}
% WPM-BLOCK:B0055
The scalar \(\Ptwo\) is the squared norm of the AGP-induced state
deformation~\cite{KolodrubetzEtAl2017}. The CC \(\KF\) is the inverse participation ratio of the normalized
channel weights and quantifies how concentrated the response is across the selected
projector subspaces.

% WPM-BLOCK:B0056
For the translation-invariant free-fermion chains below, we use
positive-momentum \((k,-k)\) Bogoliubov blocks as channels. The Hamiltonian
decomposition and the ideal pair readout in the final Bogoliubov basis fix
this resolution. For the interacting systems, the channels are instead
projectors onto distinct many-body levels. Both resolutions fit the same
formal definition, but they define different resolved observables.
For the free momentum blocks, let \(\thetak\) denote the Bogoliubov angle.
The channel weight then reduces to
% WPM-BLOCK:B0057
\begin{equation}
  \xk(\bm{\lambda},y)
  =g^{(k)}_{\mu\nu}y^{\mu}y^{\nu}
  =\frac14\left(y^{\mu}\partial_{\mu}\thetak\right)^{2},
  \label{eq:xk-angle}
\end{equation}
% WPM-BLOCK:B0058
as obtained from the spin-$\tfrac12$ Fubini--Study form in
Appendix~\ref{app:mode} [Eq.~\eqref{eq:twolevel-qgt-app}].

% WPM-BLOCK:B0059
\subsection{Probe specification and resolution}
% WPM-BLOCK:B0060
\label{sec:channel:scope}

% WPM-BLOCK:B0061
The projector family is part of the resolved observable.  Refining, merging,
or changing those projectors can change \(\KF\) while leaving the total QGT
unchanged.  For fixed projectors, tangent direction, and finite-size spin structure,
\(\KF\) is a well-defined and reproducible shape statistic. Direct comparisons
use the same probe specification.  When parameter points or tangent directions differ, any relation
between the resulting observables requires an explicit identity. The paired
TFIM--XX result below provides such an identity. A change of resolution
defines a separate resolved question.
Merging channels can only increase \(\KF\) and reduce the order-two effective
channel number; Appendix~\ref{app:spectral-projector} gives the complete
coarse-graining algebra.

% WPM-BLOCK:B0584
For the periodic free chains, the NS positive-\(k\) resolution selects a
nondegenerate finite-size ground-state sequence and an odd soft-momentum grid.
Near the positive-field TFIM critical point, the fixed odd-parity Ramond
sector has a different, nondegenerate lowest state. Its concentration
\(\KF\) tends to \(2/5\), as derived in Appendix~\ref{app:ramond}.
For the global ground-state NS sequence used in the theorem, the
concentration tends to \(2/3\).

% WPM-BLOCK:B0070
\subsection{Channel concentration and effective participation}
% WPM-BLOCK:B0071
\label{sec:channel:three}

% WPM-BLOCK:B0072
When \(\Ptwo>0\), Eq.~\eqref{eq:KFdef} defines the normalized channel
distribution \(\pi_{\chann}=x_{\chann}/\Ptwo\) and \(\KF\) as its inverse
participation ratio. Equivalently, \(\KF\) is the probability that two
independent draws from this distribution select the same channel. The same
distribution also defines a family of R\'enyi entropies and effective channel numbers.
% WPM-BLOCK:B0069
For \(\alpha>0\), \(\alpha\ne1\), define
% WPM-BLOCK:B0077
\begin{align}
  S_\alpha^{\mathrm{ch}}[\pi]
  &=\frac{1}{1-\alpha}\ln\sum_{\chann}\pi_{\chann}^{\alpha},\notag\\
  N_{\rm eff}^{(\alpha)}&=e^{S_\alpha^{\mathrm{ch}}[\pi]},\notag\\
  S_2^{\mathrm{ch}}&=-\ln\KF.
  \label{eq:renyi-participation-family}
\end{align}
% WPM-BLOCK:B0078
In the limits \(\alpha\to1\) and \(\alpha\to\infty\), the entropies become
the Shannon entropy
\(S_1^{\mathrm{ch}}=-\sum_c\pi_c\ln\pi_c\) and
\(S_\infty^{\mathrm{ch}}=-\ln\max_c\pi_c\), respectively. For an infinite
ladder, each R\'enyi index is used only when its defining moment is finite.
Equation~\eqref{eq:renyi-participation-family} also makes equivalent-ladder
replication transparent, as shown in Appendix~\ref{sec:xy:multi-ladders}.

% WPM-BLOCK:B0079
At order two, \(\Neff=1/\KF=\Ptwo^2/\Pfour\).  For a finite number
\(\Nchan\) of active channels, \(1/\Nchan\le\KF\le1\). A single channel
saturates the upper bound, while a broader distribution has a larger
\(\Neff\).  We focus operationally on the order-two concentration \(\KF\).
It admits the finite-size power sums below and is returned by the
conditional independent-mode counting protocol developed in
Sec.~\ref{sec:operational-access}.

% WPM-BLOCK:B0096
% ======================================================================
% WPM-BLOCK:B0097
\section{Exact TFIM--XX channel concentration and infrared ladders in the \texorpdfstring{$XY$}{XY} family}
% WPM-BLOCK:B0098
\label{sec:tfim}
% WPM-BLOCK:B0099
% ======================================================================
% WPM-BLOCK:B0100
\subsection{Beyond susceptibility scaling: an exact TFIM--XX composition theorem}
\label{sec:tfim:common-contract}

We use the positive-momentum Bogoliubov--de Gennes (BdG) block
\begin{equation}
  H_k=(h-\cos k)\rho_z+\gam\sin k\,\rho_x,
  \qquad
  \tan\theta_k=\frac{\gam\sin k}{h-\cos k},
  \label{eq:xy-bdg-xx-theorem}
\end{equation}
where \(\rho_a\) acts in the Nambu two-level space and \(\gam\) is the
\(XY\)-chain anisotropy, or pairing, coupling.  For an even periodic spin
ring, the nondegenerate even-parity ground-state sequence maps to
antiperiodic Neveu--Schwarz (NS) fermions with positive momenta
\(k_n=(2n+1)\pi/L\), \(n=0,\ldots,L/2-1\).  The declared channels are the
independent positive-\(k\), \((k,-k)\) Bogoliubov blocks.  For the TFIM
comparison, we use the field tangent \(\partial_h\) on the slice
\(\gam=1\) at \(h=\hc=1\). For the half-filled XX slice
\((h,\gam)=(0,0)\), we use the anisotropy tangent \(\partial_\gam\) and
restrict to \(L\equiv0\pmod4\).

\emph{Theorem.} Under the shared \(XY\)-family BdG convention, even-parity NS
spin structure, and positive-\(k\), \((k,-k)\) block resolution, the specified
distinct points and tangents yield two exact paired sequences.  With
\(L\equiv0\pmod4\) for the XX row,
\begin{equation}
  \begin{aligned}
    \KF^{\mathrm{TFIM}}(\hc,L)
    &=\frac{2(L^2+L-3)}{3L(L-1)}
      \xrightarrow{L\rightarrow\infty}\frac{2}{3},\\
    \KF^{\mathrm{XX},(\gam)}(0,L)
    &=\frac{L^2+2L-12}{3L(L-2)}
      \xrightarrow{L\rightarrow\infty}\frac{1}{3},\\
    \KF^{\mathrm{XX},(\gam)}(0,L)
    &=\frac12\KF^{\mathrm{TFIM}}(\hc,L/2).
  \end{aligned}
  \label{eq:paired-theorem-summary}
\end{equation}

The weight-level proof below explains why the two responses have different
concentrations in Eq.~\eqref{eq:paired-theorem-summary} despite the same
leading \(L^2\) susceptibility scaling. Both responses have the same
inverse-square within-ladder infrared envelope, but the TFIM response
occupies one soft ladder and the XX response two equal copies. The half-size
identity makes this carrier-multiplicity distinction exact at every allowed
finite size. Thus, at the specified points, tangents, and channel resolution,
susceptibility scaling alone does not determine how many soft ladders carry
the response.

We first derive the TFIM field-direction weights and use exact NS half-grid
sums to obtain the theorem's first row and its \(2/3\) limit. We then fold the
half-filled XX anisotropy weights into two equal copies of the half-size TFIM
distribution, obtaining the second row and the exact identity in the third.
The carrier interpretation closes the three-row comparison.

% WPM-BLOCK:B0102
\subsection{TFIM channel weights on the shared NS resolution}
% WPM-BLOCK:B0103
\label{sec:tfim:weights}

% WPM-BLOCK:B0104
\emph{Lattice setup.} To begin the proof of the first row of
Eq.~\eqref{eq:paired-theorem-summary}, we derive the TFIM field-direction
channel weights and fix the momentum ladder used by the exact lattice sum.
We use the same real-space convention as the nonintegrable control
simulations:
% WPM-BLOCK:B0105
\begin{equation}
  H=-\sum_{j}\sigma^{z}_{j}\sigma^{z}_{j+1}-h\sum_{j}\sigma^{x}_{j}.
  \label{eq:tfim}
\end{equation}
% WPM-BLOCK:B0106
The Hamiltonian in Eq.~\eqref{eq:tfim} is related by a global spin rotation
to the common convention with an \(XX\) interaction and a \(Z\) field.
Jordan--Wigner and Bogoliubov
transformations~\cite{LiebSchultzMattis1961,Pfeuty1970,Sachdev2011} reduce it
to the \(\gam=1\) specialization of Eq.~\eqref{eq:xy-bdg-xx-theorem}.  Its dispersion is
\(\Lk^{2}=(h-\cos k)^{2}+\sin^{2}k\).  The Bogoliubov angle obeys
\(\tan\thetak=\sin k/(h-\cos k)\), and hence
\(\partial_{h}\thetak=-\sin k/\Lk^{2}\).  Equation~\eqref{eq:xk-angle}
then gives the field-direction channel weight
% WPM-BLOCK:B0107
\begin{equation}
  \xk(h)=\frac{1}{4}\bigl(\partial_{h}\thetak\bigr)^{2}
        =\frac{\sin^{2}k}{4\,\Lk^{4}}.
  \label{eq:xk-tfim}
\end{equation}
% WPM-BLOCK:B0108
The shared resolution above fixes the even-parity NS sequence used in the exact
lattice sum~\cite{Damski2013,DamskiRams2014,Oshikawa2020}.  This nondegenerate
sequence contains no zero momentum.  A different spin structure changes both
the soft momentum grid and the state being differentiated.
Appendix~\ref{app:ramond} describes the nondegenerate lowest state in the
odd-parity Ramond sector.  It distinguishes this state from both an
unprojected fermionic zero-mode degeneracy and the global NS ground state.

% WPM-BLOCK:B0589
\subsection{Infrared carrier picture: Majorana mass response}
% WPM-BLOCK:B0590
\label{sec:tfim:majorana}

% WPM-BLOCK:B0591
To identify the infrared carriers of the TFIM row in
Eq.~\eqref{eq:paired-theorem-summary}, we expand near \(h=1\) and
\(k=0\).  The lattice block reduces to the massive
Majorana/Bogoliubov form
% WPM-BLOCK:B0592
\begin{equation}
  H_k\simeq m\rho_z+vk\rho_x,
  \qquad m=h-1,
  \label{eq:majorana-low-energy}
\end{equation}
% WPM-BLOCK:B0593
with \(v=1\) in the present units.  Since \(m=h-1\), the field tangent acts
as a mass perturbation.  The resulting change in the Bogoliubov angle is
% WPM-BLOCK:B0594
\begin{equation}
  \partial_m\theta_k\simeq-\frac{vk}{m^2+v^2k^2},
  \label{eq:mass-angle-response}
\end{equation}
% WPM-BLOCK:B0595
and therefore, at criticality,
% WPM-BLOCK:B0596
\begin{equation}
  x_k=\frac14(\partial_m\theta_k)^2\sim\frac{1}{4k^2}.
  \label{eq:mass-response-envelope}
\end{equation}
% WPM-BLOCK:B0597
The inverse-square mass response is sampled on the NS odd-momentum grid.
The concentration therefore depends on both the response envelope and the
discrete ladder.  Linear dispersion alone does not fix it.
With \(k_n=(2n+1)\pi/L\), the normalized infrared
distribution is
% WPM-BLOCK:B0598
\begin{equation}
  \pi_n
  =\frac{(2n+1)^{-2}}{\sum_{m\ge0}(2m+1)^{-2}}
  =\frac{8}{\pi^2}\frac{1}{(2n+1)^2}.
  \label{eq:pi-critical-ising}
\end{equation}
% WPM-BLOCK:B0599
The softest block alone carries about \(81\%\) of the total response
weight, with \(\pi_0=8/\pi^2\). The effective channel number of this
infrared distribution is
% WPM-BLOCK:B0600
\begin{equation}
  \Neff=\frac{1}{\KF}=\frac32.
  \label{eq:neff-three-halves}
\end{equation}
% WPM-BLOCK:B0601
Thus the total susceptibility diverges as \(L^2\), yet its effective channel
number remains finite.  The softest block dominates the infrared distribution.
The exact
finite-size sums in the next subsection convert this carrier picture into the
TFIM row of Eq.~\eqref{eq:paired-theorem-summary}.

% WPM-BLOCK:B0602
\begin{figure*}[t]
  \centering
  \includegraphics[width=\linewidth]{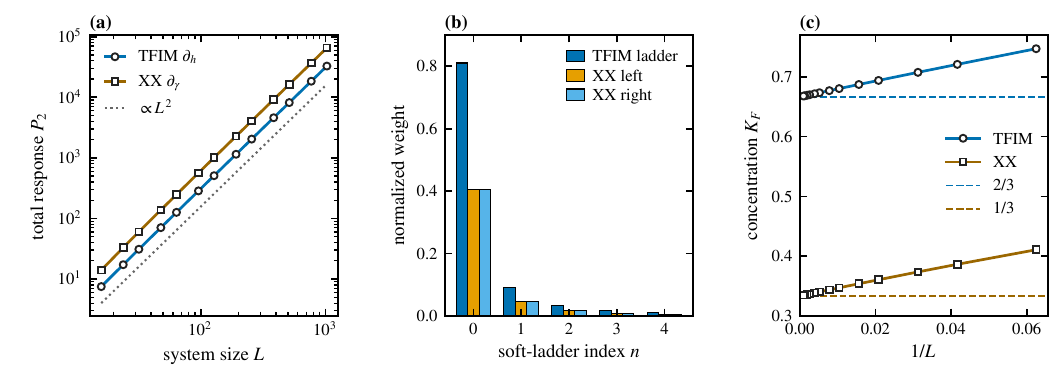}
  \caption{Response scale and resolved composition in the exact TFIM--XX
  comparison.  Here \(\Ptwo\) is the total fidelity susceptibility,
  \(\pi_n\) its normalized block weight, and \(\KF=\sum_n\pi_n^2\)
  the channel concentration.  The TFIM uses the field tangent \(\partial_h\) at
  \((h,\gamma)=(1,1)\); the XX chain uses the pairing tangent
  \(\partial_\gamma\) at \((0,0)\) on the nonsingular
  \(L\equiv0\pmod4\) sequence. All panels use the shared even-parity NS
  positive-momentum-block resolution. (a) Both responses grow as \(L^2\), although
  their amplitudes differ.
  (b) The limiting TFIM distribution occupies one inverse-square soft ladder,
  \(\pi_n=(8/\pi^2)(2n+1)^{-2}\). The XX distribution consists of two
  equal copies with weight \(\pi_n/2\) on each side of its interior soft
  point.  (c) The corresponding exact finite-size concentrations approach
  \(2/3\) and \(1/3\). Because the XX response has two equally weighted
  ladders, exact folding gives its concentration as \(1/2\) of the half-size
  TFIM value, despite the different tangents.
  Thus equal \(L^2\) response scaling does not determine whether one or two
  resolved soft ladders carry the response.}
  \label{fig:twothirds}
\end{figure*}

% WPM-BLOCK:B0109
\subsection{Exact TFIM NS sequence and the \texorpdfstring{$2/3$}{2/3} limit}
% WPM-BLOCK:B0110
\label{sec:tfim:twothirds}

% WPM-BLOCK:B0111
\emph{Exact TFIM sequence.} We now establish the finite-size TFIM row
of Eq.~\eqref{eq:paired-theorem-summary} by summing the critical weights.
At \(\hc=1\), the critical lattice weight is
% WPM-BLOCK:B0112
\begin{equation}
  \xk(\hc)
  =\frac{\sin^{2}k}{4[2(1-\cos k)]^{2}}
  =\frac{1}{16}\cot^{2}\frac{k}{2}.
  \label{eq:xk-crit-exact}
\end{equation}
% WPM-BLOCK:B0113
Its infrared behavior, \(x_k\sim(4k^2)^{-1}\), produces the one-ladder
envelope identified above.  Appendix~\ref{app:tfim-scaling} derives the
required NS half-grid sums analytically from a roots-of-unity identity.
Using these sums in Eq.~\eqref{eq:xk-crit-exact} gives the known exact
critical fidelity susceptibility~\cite{Damski2013,DamskiRams2014} and the
corresponding squared-weight moment:
% WPM-BLOCK:B0120
\begin{align}
  \Ptwo(\hc,L)&=\frac{L(L-1)}{32},
  \nonumber\\
  \Pfour(\hc,L)&=\frac{L(L-1)(L^{2}+L-3)}{1536},
  \label{eq:tfim-critical-moments-exact}
\end{align}
% WPM-BLOCK:B0121
which gives
% WPM-BLOCK:B0122
\begin{equation}
  \KF(\hc,L)
  =\frac{2}{3}\frac{L^{2}+L-3}{L(L-1)}
  =\frac{2}{3}+\frac{4}{3L}+O(L^{-2}).
  \label{eq:twothirds-derivation}
\end{equation}
% WPM-BLOCK:B0123
The limit $2/3$ in Eq.~\eqref{eq:twothirds-derivation} equals the
infrared envelope ratio. This limit is fixed by the $\xk\sim k^{-2}$ response
and the NS odd-momentum grid. Lattice-bandwidth details do not change the
limit provided they preserve this infrared structure. Figure~\ref{fig:twothirds} shows the
exact sequence approaching the limit with the displayed leading correction.

The value remains specific to this channel resolution and sector. Coarse
graining can change this concentration. For the lowest state in the fixed
odd-parity Ramond sector, the response instead samples the nonzero integer
ladder. Its envelope concentration is
\(\KF^{\mathrm{R},\mathrm{env}}=\zeta(4)/\zeta(2)^2=2/5\)
[Appendix~\ref{app:ramond}].

This establishes the first row of Eq.~\eqref{eq:paired-theorem-summary}.
We next turn to the half-filled XX anisotropy response, whose two-sided soft
structure supplies the remaining two rows.

% WPM-BLOCK:B0125
\subsection{XX pairing weights and the half-size identity}
% WPM-BLOCK:B0126
\label{sec:xy:xx-theorem}

% WPM-BLOCK:B0127
\emph{Paired XX sequence.} Having established the TFIM row of
Eq.~\eqref{eq:paired-theorem-summary}, we now derive its XX row and the
half-size identity.  Exact formulas for the field and anisotropy
susceptibilities of the \(XY\) model are already known~\cite{LuoZhaoWang2018}.
Here we resolve the response into positive-momentum block weights in the BdG
convention of Eq.~\eqref{eq:xy-bdg-xx-theorem}. We then normalize these weights
and evaluate the second moment of their distribution.
% WPM-BLOCK:B0129
On the XX line \(\gam=0\), the anisotropy direction is the gap-opening
pairing perturbation.  We denote the positive soft momentum by
\(k_{\star}=\arccos h\); at half filling, \(h=0\) and
\(k_{\star}=\pi/2\).  The NS momentum-block weight is then
% WPM-BLOCK:B0130
\begin{align}
  x_k^{(\gam)}(0)
  &=\left.\frac14\left(\partial_\gam\theta_k\right)^2\right|_{\gam=0,h=0}
  =\frac14\tan^2 k,
  \nonumber\\
  k_n&=\frac{(2n+1)\pi}{L}.
  \label{eq:xx-pairing-weight-main}
\end{align}
% WPM-BLOCK:B0131
For \(L\equiv0\pmod4\), the NS grid does not contain \(k_{\star}\), and the
finite-size moments are exactly
% WPM-BLOCK:B0132
\begin{align}
  \Ptwo^{\mathrm{XX},(\gam)}
  &=\frac{L(L-2)}{16},\nonumber\\
  \Pfour^{\mathrm{XX},(\gam)}
  &=\frac{L(L-2)(L^2+2L-12)}{768}.
  \label{eq:xx-pairing-moments-main}
\end{align}
% WPM-BLOCK:B0133
Consequently,
% WPM-BLOCK:B0134
\begin{equation}
  \begin{aligned}
  \KF^{\mathrm{XX},(\gam)}(0,L)
  &=\frac13\frac{L^2+2L-12}{L(L-2)}\\
  &=\frac13+\frac{4}{3L}+O(L^{-2}).
  \end{aligned}
  \label{eq:xx-pairing-KF-main}
\end{equation}
% WPM-BLOCK:B0135
Comparing this result with the TFIM sequence gives the exact half-size identity
% WPM-BLOCK:B0136
\begin{equation}
  \KF^{\mathrm{XX},(\gam)}(0,L)
  =\frac12\KF^{\mathrm{TFIM}}(\hc,L/2).
  \label{eq:xx-half-size-identity}
\end{equation}
% WPM-BLOCK:B0137
More generally, the normalized finite-size XX distribution consists of two
equally weighted copies of the half-size TFIM distribution. The channel
R\'enyi entropies therefore obey the exact product relation for every
\(\alpha\in(0,\infty]\)
% WPM-BLOCK:B0603
\begin{equation}
  S_\alpha^{\mathrm{ch},XX,(\gam)}(0,L)
  =S_\alpha^{\mathrm{ch},\rm TFIM}(h_c,L/2)+\ln2.
  \label{eq:xx-half-size-all-alpha}
\end{equation}
% WPM-BLOCK:B0604
At \(\alpha=2\), this relation is equivalent to Eq.~\eqref{eq:xx-half-size-identity}.
Appendix~\ref{app:multicone} gives the general product and entropy-domain conditions.

% WPM-BLOCK:B0138
Appendix~\ref{app:xx-folding-proof} proves the finite-size folding at the
level of individual weights.  Appendix~\ref{app:xx-pairing-zero-mode}
treats the singular \(L\equiv2\pmod4\) sequence. Its zero-mode-excluded ratio
is retained only as a regularized diagnostic.
Appendix~\ref{app:xx-pairing-offset} treats incommensurate offset arithmetic.

The duplication is consistent with the known zero-field XY decomposition
into two commuting Ising Hamiltonians~\cite{IgloiJuhasz2008}.  For uniform
\(J_x=1+\gam\), \(J_y=1-\gam\), their field-to-bond ratios are
\(g_\sigma=(1-\gam)/(1+\gam)\) and \(g_\tau=g_\sigma^{-1}\).
At \(\gam=0\), the induced Ising tangents are \(-2\partial_g\) and
\(+2\partial_g\). Squaring these tangent factors multiplies each response
weight by four. A common Hamiltonian energy scale does not change its eigenvectors.
Commuting Hamiltonians alone do not identify independent spin tensor factors
or fix periodic-sector projections.  The NS folding proof in
Appendix~\ref{app:xx-pairing-theorem} supplies those finite-size momentum
and projector conditions explicitly.  Thus the new comparison concerns the
resolved tangent weights and their concentration, while the two-Ising
Hamiltonian structure is prior input.

% WPM-BLOCK:B0101
\emph{Carrier interpretation.}
Equations~\eqref{eq:twothirds-derivation},
\eqref{eq:xx-pairing-KF-main}, and \eqref{eq:xx-half-size-identity}
establish the three rows of Eq.~\eqref{eq:paired-theorem-summary}. In the
infrared limit, both responses have the same inverse-square soft envelope.
The TFIM NS response occupies one odd soft-momentum ladder. For the XX
response, the two sides of the interior Dirac soft point form two equally
weighted resolved ladders. These ladders are not two independent topological
cones. Equation~\eqref{eq:xx-half-size-all-alpha}
shows that the responses share one within-ladder distribution and differ only
by a binary ladder label.  At order two, the resulting ladder-participation
factor halves the single-ladder limit \(2/3\) to \(1/3\).

% WPM-BLOCK:B0152
\subsection{NS-sector scaling function}
% WPM-BLOCK:B0153
\label{sec:tfim:scaling}

% WPM-BLOCK:B0154
\emph{Finite-size crossover.} This subsection asks how the exact critical sequence is rounded away from \(\hc\). The soft-mode structure is cut off by the mass \(m=h-\hc\). Expanding the dispersion near the band bottom gives \(\Lk^{2}\simeq k^{2}+m^{2}\), so that \(\xk\simeq k^{2}/[4(k^{2}+m^{2})^{2}]\). With the rescaled momentum \(\kp=2n+1\) and the single scaling variable
% WPM-BLOCK:B0155
\begin{equation}
  \mu=\frac{(h-\hc)\,L}{\pi},
  \label{eq:scaling-var}
\end{equation}
% WPM-BLOCK:B0156
we take \(L\to\infty\) and \(h\to\hc\) with \(\mu\) fixed. The weights then
take the form $\xk\propto \kp^{2}/(\kp^{2}+\mu^{2})^{2}$. Their common
prefactor cancels in \(\KF\), leaving the parameter-free NS-sector scaling function,
% WPM-BLOCK:B0157
\begin{equation}
  \KF(h,L)\to\Phisc(\mu)
  =\frac{\sum_{\kp=1,3,5,\ldots}\kp^{4}/(\kp^{2}+\mu^{2})^{4}}
        {\Bigl[\sum_{\kp=1,3,5,\ldots}\kp^{2}/(\kp^{2}+\mu^{2})^{2}\Bigr]^{2}},
  \label{eq:Phi}
\end{equation}
% WPM-BLOCK:B0158
with $\Phisc(0)=2/3$ reproducing the critical limit in Eq.~\eqref{eq:twothirds-derivation}. Lattice data for
several sizes collapse onto $\Phisc(\mu)$ without adjustable parameters
[Fig.~\ref{fig:collapse}(a)]. The small-$k$ derivation of
Eqs.~\eqref{eq:scaling-var}--\eqref{eq:Phi} appears in
Appendix~\ref{app:tfim-scaling-window}.

% WPM-BLOCK:B0159
\begin{figure*}[t]
  \centering
  \includegraphics[width=\linewidth]{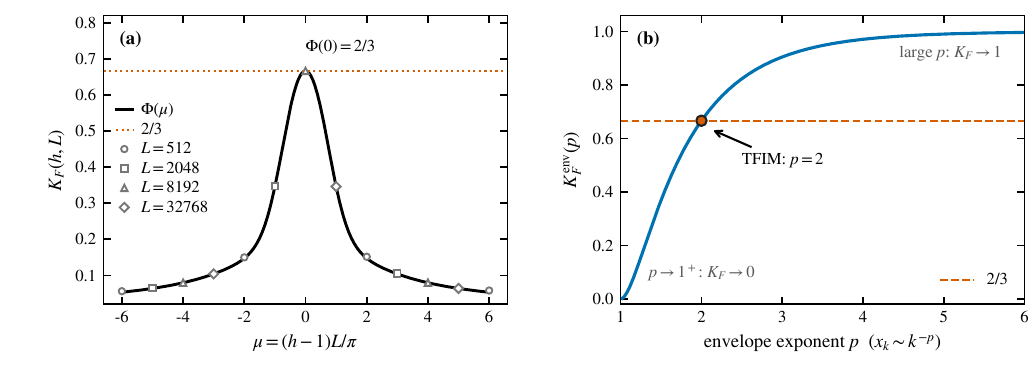}
  \caption{Finite-size scaling and the ideal NS odd-ladder envelope
    underlying the Ising concentration $2/3$. Here $\KF=\sum_k\pi_k^2$.
    (a) For the TFIM field response along $\gam=1$ near $h=\hc$, we use the
    even-parity NS positive-momentum-block resolution. Exact lattice data for
    $L=512,\,2048,\,8192,\,32768$ collapse onto $\Phisc(\mu)$ of Eq.~\eqref{eq:Phi}
    when plotted against $\mu=(h-\hc)L/\pi$. The scaling function reaches its maximum at the critical point,
    where $\Phisc(0)=2/3$.  Finite-size markers are thinned and staggered for legibility.
    (b) The concentration of an ideal NS odd ladder depends on its envelope exponent $p$.
    The Ising mass response has $p=2$ and gives
    $\KF^{\mathrm{env}}(2)=2/3$.  For this single ideal NS ladder with
    $p>1$, increasing $p$ concentrates more response in the softest channel.}
  \label{fig:collapse}
\end{figure*}

% WPM-BLOCK:B0160
\subsection{Conditional single-ladder envelope relation}
% WPM-BLOCK:B0161
\label{sec:tfim:envelope}

% WPM-BLOCK:B0162
\emph{Conditional RG extension.} We first evaluate the discrete soft-ladder sum,
then ask under what conditions RG scaling determines its exponent.
An asymptotic tail \(x_n\propto n^{-p}\) alone does not fix the concentration.
Here we assume that, after removing a common prefactor, the entire limiting
NS odd ladder has weights proportional to \((2n+1)^{-p}\). Normalization
requires \(p>1\). The moments of this ideal envelope are Dirichlet lambda values,
% WPM-BLOCK:B0163
\begin{equation}
  \lambda_{\mathrm{D}}(s)
  =\sum_{n=0}^{\infty}(2n+1)^{-s}
  =(1-2^{-s})\zeta(s),\qquad s>1,
  \label{eq:dirichlet-lambda-def}
\end{equation}
% WPM-BLOCK:B0164
and the critical concentration is
% WPM-BLOCK:B0165
\begin{equation}
  \KF^{\mathrm{env}}(p)
  =\frac{\lambda_{\mathrm{D}}(2p)}{\lambda_{\mathrm{D}}(p)^{2}}
  =\frac{(1-2^{-2p})\,\zeta(2p)}{\bigl[(1-2^{-p})\,\zeta(p)\bigr]^{2}}.
  \label{eq:envelope}
\end{equation}
% WPM-BLOCK:B0166
The Ising mass response has \(p=2\), giving
% WPM-BLOCK:B0167
\begin{equation}
  \KF^{\mathrm{env}}(2)
  =\frac{\lambda_{\mathrm{D}}(4)}{\lambda_{\mathrm{D}}(2)^2}
  =\frac{\pi^4/96}{(\pi^2/8)^2}
  =\frac{2}{3}.
\end{equation}
% WPM-BLOCK:B0168
The envelope curve in Fig.~\ref{fig:collapse}(b) interpolates from
\(\KF^{\mathrm{env}}\to0\) as \(p\to1^+\), where the normalized response is
broad, to \(\KF^{\mathrm{env}}\to1\) as \(p\to\infty\), where the softest
channel dominates.

% WPM-BLOCK:B0169
To relate the envelope exponent \(p\) to RG scaling, consider a local
perturbing density \(\mathcal O(x)\) with scaling dimension
\(\Delta_{\mathcal O}\). Its global probe is
\(V_\lambda=\partial_\lambda H=\int\d^d x\,\mathcal O(x)\).
Its infrared-singular response scales as
\cite{VenutiZanardi2007,Gu2010,RamsDamski2011A}
\begin{equation}
 \Ptwo\sim L^{2(d+z-\Delta_{\mathcal O})}
 \equiv L^{2y_{\mathcal O}},
 \label{eq:P2-RG-scaling-main}
\end{equation}
where \(d\) and \(z\) are the spatial dimension and dynamical exponent,
and \(y_{\mathcal O}=d+z-\Delta_{\mathcal O}\).
Now restrict the response to one participating NS odd ladder in one dimension.
Assume that the same exponent \(p\) controls both the finite-size prefactor
and the entire limiting envelope, with \(p>1\) for normalization.
Under this assumption, matching the total response gives
\begin{equation}
 x_n\sim\frac{L^p}{(2n+1)^p},
 \qquad p=2y_{\mathcal O},
 \label{eq:p-yo-main}
\end{equation}
and a tuning perturbation with \(y_{\mathcal O}=1/\nu\) gives, provided
\(p=2/\nu>1\),
\begin{equation}
 {\KF}^{\mathrm{env}}_\nu
 =\frac{\lambda_{\mathrm D}(4/\nu)}
 {\lambda_{\mathrm D}(2/\nu)^2}.
 \label{eq:nu-envelope}
\end{equation}
Appendix~\ref{app:rg-envelope} derives the singular scaling and separates
the analytic background. The identification \(p=2y_{\mathcal O}\) does not
apply to Lifshitz two-scale responses. Other boundary sectors, zero modes,
and multiple soft ladders change the discrete sum. Interacting level resolutions
also require the full normalized form-factor distribution, which the
response exponent alone does not determine. SM~\cite{SupplementalMaterial}, Sec.~\ref*{app:operator-dependence},
compares operator-dependent total-weight exponents at small sizes.

% WPM-BLOCK:B0176
\subsection{Multiple soft ladders}
\label{sec:tfim:multiple-ladders}
Equally weighted copies of one ladder spread the response over an
additional ladder label. To obtain this product structure asymptotically,
suppose a finite number \(n_{\rm lad}\) of equally coupled resolved soft ladders
share one normalizable leading envelope. Require the remainder to be \(o(L^p)\) in
\(\ell^1\) over all declared channels; pointwise control is insufficient.
The normalized distribution then approaches the product of a uniform distribution
over ladder labels and the common within-ladder distribution. Its limiting
order-two concentration is the single-ladder value divided by
\(n_{\rm lad}\).

The extension to other R\'enyi indices requires a finite defining moment.
For indices below one, uniform power-tail control is also needed.
At index one, uniform entropy-tail control is needed instead
[Appendix~\ref{app:multicone}]. Under these conditions, the R\'enyi entropies
converge to the corresponding single-ladder values plus \(\ln n_{\rm lad}\),
wherever those entropies are finite. The paired TFIM--XX
theorem realizes the one- versus two-ladder case exactly at finite size,
without an asymptotic remainder assumption.

The exact pair has now isolated carrier multiplicity while holding the
within-ladder shape fixed.  A distinct question remains: if the limiting
Hamiltonian and its dispersion are held fixed instead, can changing the
perturbing tangent alter the carrier envelope?  The Lifshitz analysis answers
this question.
% WPM-BLOCK:B0191
% ======================================================================
% WPM-BLOCK:B0192
\section{Lifshitz multicriticality and directional selectivity}
% WPM-BLOCK:B0193
\label{sec:lifshitz}
% WPM-BLOCK:B0194
% ======================================================================

% WPM-BLOCK:B0195
At a fixed gapless Hamiltonian, the dispersion and dynamical exponent control
the leading energy denominators. The perturbing operator supplies the form
factor in the spectral numerator. A common dispersion therefore need not fix
the normalized response distribution. The anisotropic \(XY\)-chain Lifshitz
point provides a controlled example. As the same \(z=2\) point is approached,
the field and anisotropy directions select inequivalent infrared envelopes.

% WPM-BLOCK:B0196
\subsection{XY-chain conventions}
% WPM-BLOCK:B0197
\label{sec:lifshitz:xy-conventions}

% WPM-BLOCK:B0198
We study the Lifshitz point and the later directional geometry using the anisotropic
$XY$ chain~\cite{LiebSchultzMattis1961,BaroukhMcCoy1971,Sachdev2011},
% WPM-BLOCK:B0199
\begin{equation}
  H=-\sum_{j}\Bigl[\tfrac{1+\gam}{2}\sigma^{x}_{j}\sigma^{x}_{j+1}
                  +\tfrac{1-\gam}{2}\sigma^{y}_{j}\sigma^{y}_{j+1}\Bigr]
    -h\sum_{j}\sigma^{z}_{j},
  \label{eq:xy}
\end{equation}
% WPM-BLOCK:B0200
whose positive-momentum block can be written as
% WPM-BLOCK:B0201
\begin{align}
  H_k&=(h-\cos k)\rho_z+\gam\sin k\,\rho_x,
  \nonumber\\
  \Lk^2&=(h-\cos k)^2+\gam^2\sin^2k,
  \label{eq:xy-bdg}
\end{align}
% WPM-BLOCK:B0202
with parameter point $\bm{\lambda}=(h,\gam)$. Here \(\rho_a\) denote the BdG Pauli
matrices, whereas \(\sigma^a\) denote the real-space spin Pauli matrices.
We use a real BdG convention,
related to the common $\sigma_y$ pairing convention by a parameter-independent
unitary rotation in Nambu space.  On each gapped branch, we choose the
continuous planar Bogoliubov angle
% WPM-BLOCK:B0203
\begin{equation}
  \thetak=\operatorname{atan2}(\gam\sin k,h-\cos k),
  \label{eq:theta-xy}
\end{equation}
% WPM-BLOCK:B0204
so that the following local derivatives are branch independent:
% WPM-BLOCK:B0205
\begin{equation}
  \partial_h\thetak=-\frac{\gam\sin k}{\Lk^{2}},
  \qquad
  \partial_{\gam}\thetak=\frac{(h-\cos k)\sin k}{\Lk^{2}}.
  \label{eq:xy-grads}
\end{equation}
% WPM-BLOCK:B0206
For a coordinate direction $a=h,\gam$, the corresponding free-fermion channel
weight is $x_k^{(a)}=\tfrac14(\partial_a\theta_k)^2$.  Appendix~\ref{app:mode}
derives these two-level metric conventions from the planar Bloch vector.

% WPM-BLOCK:B0207
\subsection{Small-momentum structure}
% WPM-BLOCK:B0208
\label{sec:lifshitz:smallk}

% WPM-BLOCK:B0209
% WPM-BLOCK:B0210
The anisotropic $h=1$ and isotropic $\gam=0$ lines meet at the $XY$-chain
Lifshitz point $(1,0)$. Earlier work at this point established trajectory dependence,
two correlation-length regimes, and quasi-critical oscillations of the total
QGT~\cite{RamsDamski2011A,MukherjeeDutta2011,MukherjeePolkovnikovDutta2011,LuoZhaoWang2018}.
Here we use CC to quantify the directional contrast and resolved mode
weights to explain its origin, keeping the punctured, noncommuting limits explicit.  At the point itself,
$\Lk^{2}\sim k^{4}$ and hence the quasiparticle gap scales as
$\Lk\sim k^{2}$. For small $\gam$ along $h=1$, the small-$k$ expansion gives
$\Lk^{2}\simeq k^{2}\bigl(k^{2}/4+\gam^{2}\bigr)$. The derivatives in
Eq.~\eqref{eq:xy-grads} then yield two sharply different
small-$k$ weights,
% WPM-BLOCK:B0211
\begin{equation}
  \xk^{(\gam)}\sim\frac{k^{2}}{(k^{2}+4\gam^{2})^{2}},
  \qquad
  \xk^{(h)}\sim\frac{4\gam^{2}}{k^{2}(k^{2}+4\gam^{2})^{2}}.
  \label{eq:lifshitz-weights}
\end{equation}
% WPM-BLOCK:B0212
At fixed finite size and \(h=1\), we first form the normalized weights at
\(\gam>0\) and then take \(\gam\to0^+\). The anisotropy response
approaches the Ising soft envelope, \(\xk^{(\gam)}\sim k^{-2}\).
For the field response, Eq.~\eqref{eq:lifshitz-weights} instead gives a
common factor \(\gam^2\) multiplying a \(k^{-6}\) envelope. This factor
cancels upon normalization at nonzero \(\gam\), leaving a punctured-limit
distribution dominated by the softest available channel. At exactly
\(\gam=0\), the field response vanishes and its concentration is undefined.

The same limiting \(z=2\) dispersion thus supports two different normalized
form-factor distributions.  We denote their tail exponents by
\(p_{\mathrm{env}}\).  The anisotropy direction has \(p_{\mathrm{env}}=2\), consistent
with the ideal NS odd-ladder relation under the single-ladder assumption.  The
field direction has \(p_{\mathrm{env}}=6\). As shown in the finite-size scaling
analysis below, this envelope exponent differs from the power of \(L\)
governing the total field-response weight. Within the stated NS momentum-block resolution,
the value \(2/3\) is therefore tied to the envelope exponent
\(p_{\mathrm{env}}=2\), not to a dispersion exponent by itself.

% WPM-BLOCK:B0213
\subsection{Anisotropic scaling functions}
% WPM-BLOCK:B0214
\label{sec:lifshitz:scaling}

% WPM-BLOCK:B0215
We write the NS momenta as $k=\kp\pi/L$, with $\kp=1,3,5,\ldots$.
The finite-size Lifshitz scaling limit takes $L\to\infty$ and $\gam\to0$
with the multicritical scaling variable $w=\gam L/\pi$ held fixed. The two
directions then yield two distinct scaling functions,
% WPM-BLOCK:B0216
\begin{equation}
  \KF^{(\gam)}(w)=\Phisc(2w),
  \qquad
  \KF^{(h)}(w)=\Psisc(w),
  \label{eq:lifshitz-scaling}
\end{equation}
% WPM-BLOCK:B0217
where $\Phisc$ is the Ising scaling function of Eq.~\eqref{eq:Phi} and
% WPM-BLOCK:B0218
\begin{equation}
  \Psisc(w)=
  \frac{\sum_{\kp=1,3,5,\ldots}\kp^{-4}(\kp^{2}+4w^{2})^{-4}}
       {\bigl[\sum_{\kp=1,3,5,\ldots}\kp^{-2}(\kp^{2}+4w^{2})^{-2}\bigr]^{2}}.
  \label{eq:Psi}
\end{equation}
% WPM-BLOCK:B0219
The scaling forms also show why the field-direction envelope exponent
\(p_{\mathrm{env}}=6\) does not imply \(y_{\mathcal O}=3\).
Fix \(w=\gam L/\pi>0\). To compare the total-weight scaling with the envelope,
we retain the common prefactors and write the small-momentum weights
schematically as
% WPM-BLOCK:B0220
\begin{equation}
  x_{\kappa}^{(\gam)}\propto
  L^2\frac{\kappa^2}{(\kappa^2+4w^2)^2},
  \qquad
  x_{\kappa}^{(h)}\propto
  L^4\frac{4w^2}{\kappa^2(\kappa^2+4w^2)^2}.
  \label{eq:lifshitz-two-scale-weights}
\end{equation}
% WPM-BLOCK:B0221
The common factor \(4w^2\) in the field-direction weights cancels in
\(\KF^{(h)}(w)\) for fixed \(w>0\). At exactly \(\gam=0\), however,
\(\Ptwo^{(h)}=0\), so the normalized ratio is undefined.
Under the single-ladder assumption, the anisotropy
direction matches the ideal NS odd-ladder relation between the
\(L\)-prefactor and the infrared envelope.  The field direction requires two
exponents.  At fixed nonzero \(w\), its total metric weight scales as
\(\Ptwo^{(h)}\sim L^4\). Its normalized small-\(w\) envelope instead has tail
\(\kappa^{-6}\), giving
\(p_{\mathrm{env}}=6\).  This envelope exponent is read directly from the
normalized form factor; it is not inferred from the single-ladder identity
\(p=2y_{\mathcal O}\).

Keeping the prefactor makes the order of limits explicit:
\begin{align}
 \Ptwo^{(h)}&=L^4F_h(w)+o(L^4),\qquad w>0\ \text{fixed},\notag\\
 F_h(w)&=\frac{4w^2}{\pi^4}
 \sum_{\kappa=1,3,5,\ldots}
 \frac{1}{\kappa^2(\kappa^2+4w^2)^2},\notag\\
 F_h(w)&\sim\frac{\pi^2}{240}w^2\quad(w\to0^+).
 \label{eq:lifshitz-field-prefactor}
\end{align}
The last coefficient follows from
\(\sum_{\kappa\ {\rm odd}>0}\kappa^{-6}=\pi^6/960\).
Along \(\gam=L^{-2}\), we have \(\gam/k_{\min}\to0\).
Expanding the exact lattice weights on this path gives
\(\Ptwo^{(h)}\sim\gam^2 L^6/240=L^2/240\).
The normalized \(\kappa^{-6}\) endpoint nevertheless remains the same.
This path statement follows from the lattice weights; it does not assume
that the fixed-\(w\) remainder is uniform for moving \(w(L)\).

\begin{figure*}[t]
  \centering
  \includegraphics[width=\linewidth]{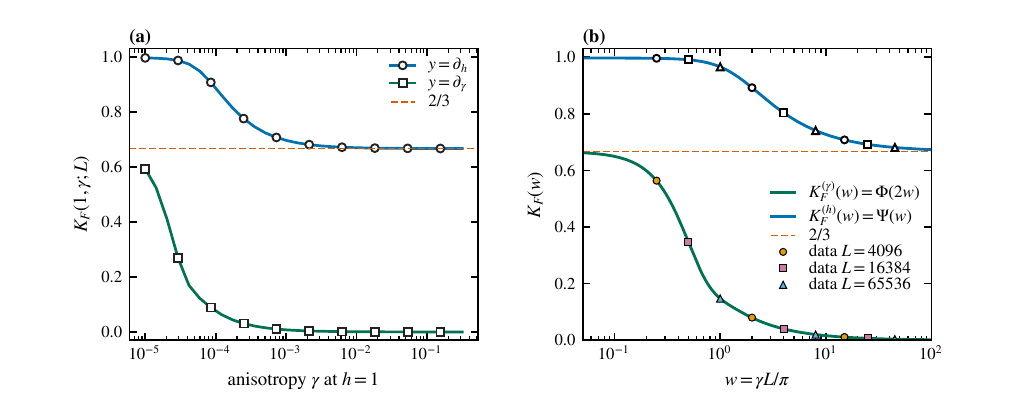}
  \caption{Directional anisotropy near the Lifshitz multicritical point
    $(h,\gam)=(1,0)$.  Both panels use the even-parity NS resolution into
    positive-momentum $(k,-k)$ Bogoliubov blocks, with normalized
    response weights $\pi_k$ and concentration $\KF=\sum_k\pi_k^2$.
    (a) At fixed large $L=65536$, the field and anisotropy directions show
    sharply different concentrations as $\gam$ is reduced along $h=1$ on a
    logarithmic axis.
    (b) In the finite-size scaling variable $w=\gam L/\pi$ (logarithmic
    axis), exact lattice data collapse onto the analytic functions
    $\KF^{(h)}(w)=\Psisc(w)$ and $\KF^{(\gam)}(w)=\Phisc(2w)$.
    As $w\to0^+$, the field curve approaches $\Psisc(0)\approx0.997$
    and the anisotropy curve approaches $2/3$. The anisotropy curve decreases
    toward zero as $w$ increases. The $w\to0^+$ endpoint is obtained by
    first taking the scaling limit at fixed $w>0$ and then sending $w$ to
    zero; its field value is the first-row result in
    Table~\ref{tab:lifshitz-order-limits}. At exactly $\gam=0$, the field
    response vanishes and its normalized concentration is undefined.
    For the same limiting Hamiltonian, the field and anisotropy tangents select
    different carrier distributions. The tangent and order of limits are
    therefore part of the observable.
    Markers are thinned and staggered for legibility; lines in panel (a)
    retain all evaluated points.}
  \label{fig:lifshitz}
\end{figure*}

% WPM-BLOCK:B0222
The limits of the two scaling functions encode the directional anisotropy.
The anisotropy response interpolates
\emph{downward} from the Ising value, $\KF^{(\gam)}\!:\,2/3\to0$ as
$w:0\to\infty$. The field response instead interpolates between near-unit
concentration and the Ising value,
% WPM-BLOCK:B0223
\begin{equation}
  \Psisc(0)=\frac{\lambda_{\mathrm{D}}(12)}{\lambda_{\mathrm{D}}(6)^{2}}\approx0.997,
  \qquad
  \Psisc(\infty)=\frac{2}{3}.
  \label{eq:Psi-limits}
\end{equation}
% WPM-BLOCK:B0224
The small-\(w\) endpoint in Eq.~\eqref{eq:Psi-limits} is the
\(p_{\mathrm{env}}=6\) member of the envelope in Eq.~\eqref{eq:envelope}.  It
describes the limiting normalized field-response distribution as
\(\gam\to0^+\) inside the scaling window.  At exactly \(\gam=0\), the
field-direction metric weight has \(\Ptwo^{(h)}=0\), so \(\KF^{(h)}\) is not
itself defined there.  Figure~\ref{fig:lifshitz} displays both the fixed-size
approach to the multicritical point and the scaling collapse in \(w\).  The
quantitative punctured-limit value \(\Psisc(0)\simeq0.997\) means that the
field response is dominated by the softest available channel in the small-\(w\)
regime.

% WPM-BLOCK:B0225

% WPM-BLOCK:B0226
\subsection{Order of limits}
% WPM-BLOCK:B0227
\label{sec:lifshitz:order}

% WPM-BLOCK:B0228
Equations~\eqref{eq:lifshitz-scaling}--\eqref{eq:Psi-limits} make explicit
that there is no unique concentration obtained by writing only
$(h,\gam)=(1,0)$.  The scaling variable is $w=\gam L/\pi$, and the two iterated
limits in Table~\ref{tab:lifshitz-order-limits} select different normalized
distributions.  The limits therefore do not commute and define distinct
finite-size trajectories toward the anisotropic multicritical point.

% WPM-BLOCK:B0229
\begin{table}[t]
  \caption{Iterated limits of the Lifshitz concentration ratio.  The inner
  limit is taken first.  In the first row the field ratio is formed for
  \(\gam>0\) before \(\gam\to0^+\); this punctured limit is necessary because
  \(\Ptwo^{(h)}=0\) and \(\KF^{(h)}\) is undefined at exactly \(\gam=0\).
  The closed form of the first-row field value is given in
  Eq.~\eqref{eq:Psi-limits}.}
  \label{tab:lifshitz-order-limits}
  \begin{ruledtabular}
  \begin{tabular}{lcc}
  Limit & \(\KF^{(\gam)}\) & \(\KF^{(h)}\) \\
  \hline
  \(\displaystyle\lim_{L\to\infty}\lim_{\gam\to0^+}\) & \(2/3\) & \(0.997\) \\
  \(\displaystyle\lim_{\gam\to0^+}\lim_{L\to\infty}\) & \(0\) & \(2/3\)
  \end{tabular}
  \end{ruledtabular}
\end{table}

% WPM-BLOCK:B0230
The exact and asymptotic analytic cases are compared in
Table~\ref{tab:operator-selective}.  The table keeps the paired sector theorems
separate from the exploratory interacting calculations of
Sec.~\ref{sec:interacting-benchmarks}.

% WPM-BLOCK:B0231
\begin{table*}[t]
  \caption{Analytic channel-concentration results for the response-structure comparison.  The first two rows show thermodynamic limits of
  exact finite-size NS-sector theorems; the last two are asymptotic Lifshitz
  distributions.  The XX value includes the participation factor of its two
  equal soft ladders.  For the Lifshitz field row, \(L^4F_h(w)\) refers to
  fixed \(w>0\); \(F_h(w)\sim\pi^2w^2/240\) vanishes at the punctured endpoint.}
  \label{tab:operator-selective}
  \footnotesize
  \setlength{\tabcolsep}{4pt}
  \begin{ruledtabular}
  \begin{tabular}{@{}lccccc@{}}
  case and tangent & resolution or limit & soft structure & \(\Ptwo\) scaling & status & \(\KF\) limit \\
  \hline
  TFIM \(\partial_h\) at \((1,1)\) & NS momentum blocks & one odd ladder, \(p_{\mathrm{env}}=2\) & \(L^2\) & exact theorem & \(2/3\) \\
  XX \(\partial_\gam\) at \((0,0)\) & NS, \(L\equiv0\pmod4\) & two equal ladders, \(p_{\mathrm{env}}=2\) & \(L^2\) & exact theorem & \(1/3\) \\
  Lifshitz anisotropy & punctured endpoint \(w=\gam L/\pi\to0^+\) & \(p_{\mathrm{env}}=2\) & \(\Ptwo^{(\gam)}\sim L^2\) & asymptotic & \(2/3\) \\
  Lifshitz field & \(w=\gam L/\pi\to0^+\) & \(p_{\mathrm{env}}=6\) & \(L^4F_h(w)\) & asymptotic & \(0.997\) \\
  \end{tabular}
  \end{ruledtabular}
\end{table*}

% WPM-BLOCK:B0232
Table~\ref{tab:operator-selective} fixes the coordinate directions and their
limiting procedures.  Together with the TFIM--XX theorem, these results
separate two kinds of carrier information. Ladder multiplicity changes the
number of equally loaded soft structures. Operator selectivity instead changes
the within-ladder envelope despite a shared \(z=2\) dispersion.
Appendix~\ref{sec:directional} extends the construction to arbitrary tangent
rays, the associated quartic tensor, and polar profiles. When momentum blocks
are no longer the relevant resolution, the next question is whether continuum
operator data can instead determine the weights of complete physical energy
levels. The conformal construction below answers that question in a controlled
single-primary domain.

% WPM-BLOCK:B0233
% ======================================================================
% WPM-BLOCK:B0278
\section{Conformal construction of channel concentration}
% WPM-BLOCK:B0677
\label{sec:cft-response}

The free-chain analysis relates differences in CC to ladder
multiplicity and tangent-dependent momentum-block weights. We now ask how continuum
operator and descendant data determine this concentration at a relativistic fixed point.
The construction proceeds through a continuum response ray and normalized weights over
complete physical energy levels, using a vacuum scalar primary in the convergence domain
specified below.

% WPM-BLOCK:B0678
\subsection{Euclidean response vector and two-replica identity}
% WPM-BLOCK:B0679
\label{sec:cft-bridge}

% WPM-BLOCK:B0680
The field-theory bridge begins from the same spectral object used on the
lattice. CFT supplies the continuum response state and physical level
structure. For a specified perturbation and projector resolution, the channel
construction extracts the corresponding normalized channel weights.
For a nondegenerate
finite-size ground state, let
\(Q=1-\ketbra{0}{0}\) and \(V_y=\partial_yH\).
The reduced resolvent then defines the unnormalized response vector
% WPM-BLOCK:B0681
\begin{equation}
  \ket{\chi_y}=Q(H-E_0)^{-1}QV_y\ket{0}.
  \label{eq:cft-response-state}
\end{equation}
% WPM-BLOCK:B0682
For a temporary lower Euclidean-time cutoff \(\epsilon>0\), its
half-cylinder representation is
% WPM-BLOCK:B0683
\begin{equation}
  \begin{aligned}
  \ket{\chi_y^{(\epsilon)}}
  &=\int_\epsilon^\infty d\tau\,
    e^{-\tau(H-E_0)}QV_y\ket{0},\\
  \ket{\chi_y}&=\lim_{\epsilon\to0^+}\ket{\chi_y^{(\epsilon)}}.
  \end{aligned}
  \label{eq:euclidean-response-state}
\end{equation}
% WPM-BLOCK:B0684
This Euclidean representation prepares the ground-state deformation spectrally.
The reduced resolvent in Eq.~\eqref{eq:cft-response-state} is the primary definition.
Comparing it with Eq.~\eqref{eq:agp-state-response} gives
\(\ket{\chi_y}=-D_y\ket{0}\), so
\(\braket{\chi_y}{\chi_y}=\Ptwo\).  At finite \(L\), \(Q\) removes
only the ground-state contribution. We first fix the vacuum, sector, tangent,
and projectors. When the spectral sum converges, we remove \(\epsilon\)
before normalizing. Appendix~\ref{app:cft-cylinder-derivation} derives the
Laplace-moment correlator identity and convention factor; the latter cancels
from every normalized weight.  Appendix~\ref{app:rg-envelope} gives the
ultraviolet--infrared and RG reduction.

% WPM-BLOCK:B0687
For \(\Ptwo(\bm{\lambda},y)>0\), choose any complete orthogonal projector family
\(\sum_a\Pi_a=Q\) and set
% WPM-BLOCK:B0750
\[
  x_a^\Pi=\bra{\chi_y}\Pi_a\ket{\chi_y},
  \qquad
  \pi_a^\Pi=\frac{x_a^\Pi}{\Ptwo(\bm{\lambda},y)}
  =\bra{\widehat\chi_y}\Pi_a\ket{\widehat\chi_y},
\]
% WPM-BLOCK:B0751
where \(\ket{\widehat\chi_y}=\ket{\chi_y}/\sqrt{\Ptwo(\bm{\lambda},y)}\)
is the normalized response ray.  Introduce the same-channel operator
% WPM-BLOCK:B0688
\begin{equation}
  \mathbb D_\Pi=\sum_a\Pi_a\otimes\Pi_a .
\end{equation}
% WPM-BLOCK:B0689
Tensor-product factorization gives the exact identity
% WPM-BLOCK:B0690
\begin{equation}
  \KF^\Pi
  \equiv\sum_a(\pi_a^\Pi)^2
  =\bra{\widehat\chi_y}^{\otimes2}\mathbb D_\Pi
   \ket{\widehat\chi_y}^{\otimes2} .
  \label{eq:replica-same-channel}
\end{equation}
% WPM-BLOCK:B0691
Thus \(\KF^\Pi\) is the probability that two independently prepared response
rays occupy the same declared channel.  In the Euclidean construction,
\(\mathbb D_\Pi\) enforces the same channel label on the two copies at the
readout surface. The relevant fourth-order quantity \(\Pfour\) is the sum of
squared channel weights. An ordinary connected fourth cumulant is a different
single-copy observable. Equating it with the concentration in
Eq.~\eqref{eq:replica-same-channel} requires an additional model-specific
measurement identity.

% WPM-BLOCK:B0692
\subsection{Vacuum scalar cylinder theorem for carrier weights}
\label{sec:cft-cylinder}

% WPM-BLOCK:B0693
Consider a unitary \(1+1\)-dimensional CFT on a circle of circumference
\(L\). We work in a fixed sector with a nondegenerate vacuum. The
perturbation is the spatial integral of one scalar primary \(\mathcal O\),
with equal holomorphic and antiholomorphic weights
\(h_{\mathcal O}=\bar h_{\mathcal O}=\Delta/2\).

The channels are complete physical conformal-energy levels at zero momentum.
Each projector groups the full level, with null descendants removed by the
physical Hilbert-space quotient. The spectral coefficient of each physical
level is therefore read from the physical two-point function. We restrict to
\(0<\Delta<3/2\), where the positive level sums converge. Within this range,
we remove the Euclidean cutoff, normalize the weights, and then compare with
the leading singular lattice response.

Expanding the plane-to-cylinder two-point function by conformal level gives
% WPM-BLOCK:B0694
\begin{equation}
  A_n(\Delta)=\left[\frac{(\Delta)_n}{n!}\right]^2,
  \qquad n=0,1,\ldots ,
  \label{eq:cft-level-coefficients}
\end{equation}
% WPM-BLOCK:B0695
where \((\Delta)_n=\Delta(\Delta+1)\cdots(\Delta+n-1)\) is the rising
Pochhammer symbol, with \((\Delta)_0=1\). Define
\(u_\epsilon=2\pi v\epsilon/L\), where \(v\) is the CFT velocity.
Up to one common factor, the response weights at finite cutoff and their
spectral limits are
% WPM-BLOCK:B0696
\begin{align}
  w_n^{(\epsilon)}(\Delta)
  &=\frac{A_n(\Delta)}{(\Delta+2n)^2}
    e^{-2u_\epsilon(\Delta+2n)},\notag\\
  w_n(\Delta)
  &=\lim_{\epsilon\to0^+}w_n^{(\epsilon)}(\Delta)
    =\frac{A_n(\Delta)}{(\Delta+2n)^2}.
  \label{eq:cft-response-weights}
\end{align}
% WPM-BLOCK:B0697
Appendix~\ref{app:cft-cylinder-levels} supplies the binomial expansion,
zero-momentum projection, resolvent integral, and complete-level degeneracy
grouping. Appendix~\ref{app:cft-uv-boundary} gives the ultraviolet convergence criterion.  The conformal map and descendant
organization follow Ref.~\cite[Chaps.~5 and 9]{DiFrancesco1997}; the general
finite-volume momentum-space organization is compared with
Ref.~\cite[Secs.~2 and 3]{Nishikawa2023}.

% WPM-BLOCK:B0698
The common operator normalization, velocity, cylinder radius, and microscopic
matching amplitude cancel from the normalized moment.  For
\(0<\Delta<3/2\), the weights have the tail \(w_n\sim n^{2\Delta-4}\).
The positive sums defining the norm and normalized distribution therefore
converge, allowing \(\epsilon\to0^+\) before normalization. In the same range, the singular CFT
response \(L^{4-2\Delta}\) dominates an analytic extensive lattice background.
Normalizing these complete-level weights gives
% WPM-BLOCK:B0699
\begin{equation}
  \KF^{\rm CFT}(\Delta)
  =\frac{\sum_{n=0}^\infty w_n(\Delta)^2}
  {\left[\sum_{n=0}^\infty w_n(\Delta)\right]^2},
  \qquad 0<\Delta<\frac32 .
  \label{eq:cft-KF-delta}
\end{equation}
% WPM-BLOCK:B0701
At \(\Delta=1\), \(A_n=1\) and \(w_n=(2n+1)^{-2}\), so
\(\KF^{\rm CFT}(1)=2/3\).  This is the continuum cylinder expression of the
NS Ising ladder derived microscopically in
Sec.~\ref{sec:tfim:twothirds}.  For the three-state Potts thermal field,
\(\Delta_\varepsilon=4/5\)~\cite{Wu1982,DiFrancesco1997,MongEtAl2014}, the
same complete-level prescription gives
% WPM-BLOCK:B0702
\begin{equation}
  K_{F,\mathrm{Potts}}^{\mathrm{CFT}}
  \simeq0.8515 .
  \label{eq:potts-cft-target}
\end{equation}
% WPM-BLOCK:B0703
Keeping only the exponent-only odd-ladder
envelope with \(p=12/5\) in Eq.~\eqref{eq:nu-envelope} instead gives
\(K_F^{\mathrm{ideal}}\simeq0.800\). CC therefore retains
the effect of the complete level weights, which the asymptotic envelope
approximation omits.
Appendices~\ref{app:cft-cylinder-levels}, \ref{app:cft-uv-boundary}, and
\ref{app:cft-potts-evaluation}
give the level-weight series, convergence conditions, and Potts evaluation supporting
Eq.~\eqref{eq:potts-cft-target}. Sec.~\ref*{supp:cft-cross-check} of the SM~\cite{SupplementalMaterial}
records two independent high-precision evaluation routes.

% WPM-BLOCK:B0700
Equation~\eqref{eq:cft-KF-delta} applies to a single primary in the
stated convergence domain. At \(\Delta=3/2\), the normalization sum
diverges logarithmically; for \(\Delta\ge3/2\), the regulated
concentration tends to zero without defining a normalized distribution
over the unregulated levels. Co-leading primaries or operator mixing
require the combined response and its actual projectors. For separately
resolved ladders, matching amplitudes enter
Eq.~\eqref{eq:multi-ladder-general} in Appendix~\ref{app:multicone};
merged projectors must be evaluated in the combined response state.

Matching a lattice response to this continuum comparator requires an
additional condition on the channel weights. First group the lattice
response into complete conformal levels, as required by
Eq.~\eqref{eq:potts-cft-target}.  Let \(q_N(L)\) be their
normalized weights, including the full tail. A sufficient condition for the
grouped concentration to converge to the CFT value is
\(\sum_N|q_N(L)-\pi_N^{\rm CFT}|\to0\), since
\begin{equation}
 \left|\sum_Nq_N(L)^2-\sum_N(\pi_N^{\rm CFT})^2\right|
 \le 2\sum_N|q_N(L)-\pi_N^{\rm CFT}|.
 \label{eq:cft-grouped-matching}
\end{equation}
This condition is an assumption for a grouped lattice observable; it is not
established for the fine-level Potts data below.  Grouping preserves \(\Ptwo\)
but changes \(\Pfour\) and \(\KF\).

For such a grouped observable, two mechanisms suggest candidate finite-size
correction powers: the analytic ultraviolet background and a symmetry-allowed
irrelevant field. The Potts thermal field has a singular response
\(\Ptwo^{\rm sing}\sim L^{12/5}\).
Because the global thermal probe is the spatial integral of a local density,
translation invariance extracts a center-of-mass factor \(L\). At fixed
lattice regulator, the remaining short-distance integral is finite, yielding
\(\Ptwo^{\rm UV}\sim L\).  The regular-to-singular ratio therefore scales as
\(L^{-7/5}\).  Appendix~\ref{app:rg-envelope} derives this ultraviolet--
infrared split.  The leading symmetry-allowed scalar irrelevant field has
\(\Delta_{\mathrm{irr}}=14/5\) and hence correction exponent
\(\omega_{\mathrm{irr}}=4/5\)
\cite{Wu1982,DiFrancesco1997,MongEtAl2014}.  If the grouped moments separately admit expansions in these two relative
powers, normalizing their ratio gives
% WPM-BLOCK:POTTS75-03
\begin{align}
  K_F^{\rm group}(L)
  &=K_{F,\mathrm{Potts}}^{\mathrm{CFT}}
    \left[1+d_{\mathrm{irr}}L^{-4/5}
            +d_{7/5}L^{-7/5}+\cdots\right]
  \notag\\
  &=K_{F,\mathrm{Potts}}^{\mathrm{CFT}}
    +a_{\mathrm{irr}}L^{-4/5}
    +b_{\mathrm{UV}}L^{-7/5}+\cdots ,
  \label{eq:potts-KF-correction-hierarchy}
\end{align}
Appendix~\ref{app:potts-correction-algebra} gives the assumed moment expansions
and the coefficient combinations produced by this normalization.
% WPM-BLOCK:POTTS75-04
Equation~\eqref{eq:potts-KF-correction-hierarchy} requires both grouped
matching and the stated moment expansions. Its nonuniversal coefficients
are moment combinations, so either displayed amplitude can vanish through a
selection rule or cancellation. Matching errors can introduce slower
corrections. SM~\cite{SupplementalMaterial},
Sec.~\ref*{supp:correction-diagnostics}, compares correction forms and
fit windows.

For the fine-level observable, grouped response matching and a bound on
the number of channels carrying each group's weight give the one-sided
interval derived in Appendix~\ref{app:resolution-stability}.
With at most \(d_N^2\) asymptotically responding fine channels per group,
where \(d_N\) is the physical holomorphic level multiplicity,
the reference budgets are below \(1.68\times10^{-4}\) for Ising and
\(4.04\times10^{-5}\) for Potts
[Appendix~\ref{app:thermal-refinement-budgets}]. Applying these budgets to
the interacting sequences requires control of the grouped weights and
the fine-channel support. The fixed-CFT fits below therefore remain
diagnostics against a continuum reference.

The cylinder construction supplies a fixed continuum reference for CC
and ranked weights. We next compare finite interacting responses with this reference and
an exact finite-size lattice sequence.

% WPM-BLOCK:B0704
\section{Interacting finite-size evidence}
% WPM-BLOCK:B0705
\label{sec:interacting-benchmarks}

We use two finite-size benchmarks with different reference distributions.
The critical Potts chain is compared with fixed descendant-resolved CFT and
exponent-only references. The NNN-TFIM is compared with the exact same-size
TFIM lattice sequence. Both calculations use many-body level projectors to compare CC and its underlying weight distribution at accessible sizes.

\paragraph{Lattice models and probes.}
The interacting calculations use periodic rings with \(j=1,\ldots,L\), with
all site indices understood modulo \(L\).

\emph{Potts comparison.}
We write the three-state quantum Potts chain as
\begin{equation}
  \begin{split}
    H_{\mathrm{Potts}}(g)
    ={}&-\sum_{j=1}^{L}(\tau_j+\tau_j^\dagger)\\
       &-g\sum_{j=1}^{L}
       (\sigma_j^\dagger\sigma_{j+1}
       +\sigma_{j+1}^\dagger\sigma_j).
  \end{split}
  \label{eq:interacting-potts-hamiltonian}
\end{equation}
Here the three-state clock and shift operators obey
\(\sigma_j^3=\tau_j^3=\mathbf{1}\) and
\(\sigma_j\tau_j=e^{2\pi i/3}\tau_j\sigma_j\).  The Potts calculation is
performed at the self-dual critical point \(g=1\) with thermal probe
\(V_g=\partial_g H_{\mathrm{Potts}}\).  It uses the complete translation-invariant \(k=0\) block without a charge
projection. The Hamiltonian, ground state, and thermal probe preserve the
global \(\mathbb Z_3\) color shift, so the response has \(q=0\) support.

\emph{Ising-family calculation.}
The nonintegrable next-nearest-neighbor transverse-field Ising (NNN-TFIM)
calculation uses the Hamiltonian
\begin{equation}
  \begin{split}
    H_{\mathrm{NNN}}(h,J_2)
    ={}&-\sum_{j=1}^{L}\sigma_j^z\sigma_{j+1}^z
       -J_2\sum_{j=1}^{L}\sigma_j^z\sigma_{j+2}^z\\
       &-h\sum_{j=1}^{L}\sigma_j^x,
  \end{split}
  \label{eq:interacting-nnn-tfim-hamiltonian}
\end{equation}
with field probe
\(V_h=\partial_h H_{\mathrm{NNN}}=-\sum_{j=1}^{L}\sigma_j^x\).
At \(J_2=0\), Eq.~\eqref{eq:interacting-nnn-tfim-hamiltonian} reduces to
the TFIM convention in Eq.~\eqref{eq:tfim}.

\emph{Operator conventions.}
In
Eq.~\eqref{eq:interacting-potts-hamiltonian}, \(\sigma_j\) and \(\tau_j\)
denote three-state operators; the superscripted \(\sigma_j^{x,z}\) in
Eq.~\eqref{eq:interacting-nnn-tfim-hamiltonian} are Pauli matrices.

% WPM-BLOCK:B0281
\paragraph{Matched level-projector observable.}
% WPM-BLOCK:B0282
The exact free-chain results use momentum blocks. The interacting finite-size
calculations instead use projectors onto many-body energy levels, grouping
only tolerance-stable numerical degeneracies. This change of channel
resolution applies the same distributional construction to interacting
spectra. The momentum-block results do not by themselves determine the
interacting level-projector values. At \(J_2=0\), the matched-projector
control below compares \(\Ptwo\), \(\Pfour\), and \(\KF\) with the
momentum-block results at the tested sizes.

% WPM-BLOCK:B0283
For either \(V=V_g\) or \(V_h\), use the reduced-resolvent response
\(\ket{\chi}=Q(H-E_0)^{-1}QV\ket{0}\), as in
Eq.~\eqref{eq:cft-response-state}. Here \(Q=1-\ketbra{0}{0}\), and the inverse
is restricted to the excited subspace.
Let \(\Pi_a\) project onto the \(a\)th tolerance-stable numerical
energy group. Its exact response mass is
\begin{equation}
  r_a=\norm{\Pi_a\ket{\chi}}^2
  =\sum_{n\in a}\frac{|\langle n|V|0\rangle|^2}{(E_n-E_0)^2}.
  \label{eq:level-projector-benchmark}
\end{equation}
For mutually orthogonal groups complete on the response support,
\(\Ptwo=\sum_a r_a>0\), and the level-projector concentration is
\begin{equation}
  K_F^{\mathrm{proj}}
  =\frac{\sum_a r_a^2}{\Ptwo^2},
  \qquad \pi_a=\frac{r_a}{\Ptwo}.
\end{equation}
The plotted numerical estimates use the cluster mean gap
\(\bar\Delta_a\) to form
\(x_a=\norm{\Pi_a V\ket{0}}^2/\bar\Delta_a^2\).
For an exactly degenerate group, \(x_a=r_a\). For a nearly degenerate
group, \(x_a\) is a mean-gap approximation to the exact response mass.
SM~\cite{SupplementalMaterial}, Sec.~\ref*{supp:conditional-tail},
gives the same-quantity enclosure and conversion bounds.

% WPM-BLOCK:B0287
Numerically degenerate eigenvectors must be grouped into one projector; otherwise
\(\Pfour\) would depend on an arbitrary basis within the degenerate subspace.
This tolerance grouping restores a numerical eigenspace; it does not merge
finite-size splittings into putative conformal levels.

The denominator depends on the dataset. The Potts $L=6$ full-sector point
self-normalizes the mean-gap weights. The sparse scalar and ranked diagnostics use retained mean-gap numerators.
Their denominators are the independently solved total $\Ptwo$ for ranked
weights and its square for concentrations. These plotted estimators remain
distinct from the exact target \(K_F^{\mathrm{proj}}\).

For Potts \(L=12,13,14\), the stored resolvent-mass intervals also yield
conditional enclosures for \(K_F^{\mathrm{proj}}\). These enclosures lie fully
inside the archived intervals. Other sizes, the NNN comparisons, and the operator
diagnostics retain their documented approximation and truncation status.

\paragraph{Finite-size numerical controls.}
The Potts spectral-tail intervals below are conditional numerical enclosures.
They propagate residual and tail bounds, assuming correct identification of
the computed ground state, retained clusters, and first omitted level.
Residuals alone do not prove this spectral completeness.
As a calibration, we perform exact diagonalization at \(J_2=0\) using matched
level projectors. We compare the resulting \(\Ptwo\), \(\Pfour\), and \(\KF\)
with the momentum-block results.
For \(L=6,8,10,12\) and the grouping tolerances in this control,
the maximum absolute discrepancies are approximately
\(1.217\times10^{-13}\), \(7.230\times10^{-13}\), and
\(4.885\times10^{-15}\), respectively.  This is a representation and
floating-point check, not a physical uncertainty estimate.

Repeating that calibration with grouping tolerances from \(10^{-10}\) to
\(10^{-6}\) leaves \(\KF\) unchanged at the displayed precision. For Potts, the same-response \(L=12,13,14\) conditional enclosure widths
are below \(10^{-8}\). For the \(L=20,J_2=0.20\) NNN-TFIM point, shifting the
specified gap-crossing field by \(\pm10^{-4}\) changes \(\KF\) by about
\(10^{-6}\).
Section~\ref*{supp:retained-checks} of the SM~\cite{SupplementalMaterial}
gives the tolerance, retained-spectrum, and field-sensitivity records.

\begin{figure*}[tbp]
  \centering
  \includegraphics[width=\linewidth]{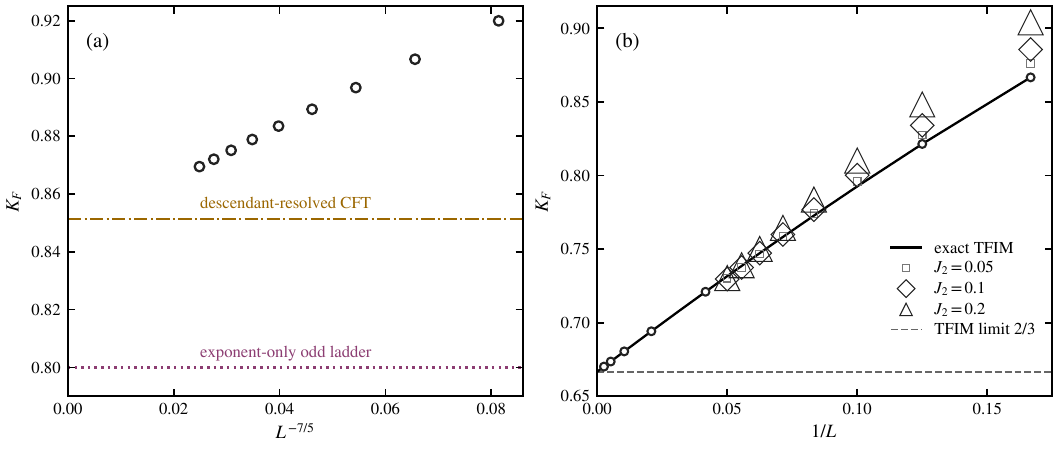}
  \caption{Finite-size interacting concentrations against fixed theory
  references. The target observable is the spectral-projector concentration
  \(K_F^{\mathrm{proj}}\); sparse interacting markers use the approximate
  mean-gap diagnostic \(\widetilde K_F^{\mathrm{low}}\), while the full-sector
  Potts \(L=6\) point self-normalizes its mean-gap weights. (a) At \(g=1\),
  the critical three-state Potts thermal response is resolved into
  tolerance-stable energy-level projectors in the translation-invariant
  \(k=0\) block, with color-neutral response support. Sizes \(L=7,\ldots,14\)
  use the calibrated retained count (Sec.~\ref*{supp:retained-checks} of
  the SM~\cite{SupplementalMaterial}). The fixed references are dash-dotted
  descendant-resolved CFT \(K_F^{\rm CFT}\simeq0.8515\) and the dotted
  exponent-only odd ladder \(K_F^{\mathrm{ideal}}\simeq0.800\). On the
  diagnostic \(L^{-7/5}\) axis motivated by the relative analytic background,
  the sequence decreases with increasing \(L\) and remains above the CFT
  reference. No fitted extrapolation is plotted because free-intercept
  estimates depend on correction form and size window (SM
  Sec.~\ref*{supp:correction-diagnostics}). For \(L=12,13,14\), separate
  deterministic same-response enclosures of \(K_F^{\mathrm{proj}}\),
  conditional on the documented spectral premises, have widths below graphical
  resolution; hence no error bars are visible (SM
  Sec.~\ref*{supp:conditional-tail}). (b) The solid black curve and
  selected open circles show the exact finite-size NS TFIM sequence; the gray
  dashed line marks its \(K_F=2/3\) limit at \(1/L=0\). Open squares,
  diamonds, and triangles show NNN-TFIM field responses for
  \(J_2=0.05,\,0.10,\,0.20\), respectively, in the \(k=0\), spin-flip-even
  sector, at even \(L=6,\ldots,20\) and scaled-gap-crossing fields
  \(h_\times(L;J_2)\). Their cross-\(J_2\) spread contracts; at \(L=20\),
  their values lie near the exact TFIM value at the same size. These
  finite-size data do not establish thermodynamic convergence.}
  \label{fig:interacting-benchmarks}
\end{figure*}

% WPM-BLOCK:B0288
\paragraph{Scalar finite-size evidence.}
% WPM-BLOCK:B0289
Figure~\ref{fig:interacting-benchmarks} summarizes the finite-size
observations.
% WPM-BLOCK:B0292

The regular-to-singular response ratio motivates the \(L^{-7/5}\) diagnostic axis in
Fig.~\ref{fig:interacting-benchmarks}(a), which shows only finite-size data and
two fixed theory comparators.  The plotted Potts diagnostic decreases from \(0.920\) at
\(L=6\) to \(0.869\) at \(L=14\), remaining above
\(K_F^{\rm CFT}\simeq0.8515\). The fixed-CFT two-amplitude diagnostic
uses all nine sizes \(L=6,7,\ldots,14\), with \(K_F^{\rm CFT}\) held fixed
and seven residual degrees of freedom. It fits the plotted fine-level diagnostic sequence;
the grouped matching assumed in Eq.~\eqref{eq:potts-KF-correction-hierarchy}
is not established for these data. Section~\ref*{supp:correction-diagnostics}
of the SM~\cite{SupplementalMaterial} compares this diagnostic with
\(L^{-4/5}\), \(L^{-7/5}\), free-intercept, fit-window, and polynomial forms.
The free-intercept estimates depend on the correction form and size window,
so these data do not determine a correction-model-independent limit.  The archived checks cover retained-subspace, level-grouping and residual
stability. For \(L=12,13,14\), the same-response tail replay yields
enclosures below graphical resolution.

% WPM-BLOCK:B0290
\paragraph{Ising-family spread.}
% WPM-BLOCK:B0291
For the nonintegrable NNN-TFIM, the three sequences
\(J_2=0.05,0.10,0.20\) are evaluated at finite-size pseudocritical fields
\(h_\times(L;J_2)\) through \(L=20\).  We define these fields by the phenomenological-renormalization scaled-gap
crossing~\cite{Nightingale1976}
\((L-2)\Delta_{L-2}(h_\times,J_2)=L\Delta_L(h_\times,J_2)\). Here
\(\Delta_L=E_{0,-}^{(L)}-E_{0,+}^{(L)}\) is the lowest spin-flip odd--even gap
in the \(k=0\) momentum sector.  The response is then evaluated in the
translation-invariant, spin-flip-even sector.
Figure~\ref{fig:interacting-benchmarks}(b) uses the closed-form sequence in
Eq.~\eqref{eq:twothirds-derivation} as the finite-size reference.
The spread across the three \(J_2\) sequences shrinks from
\(2.81\times10^{-2}\) at \(L=6\), through \(9.34\times10^{-3}\) at \(L=12\),
to \(6.63\times10^{-4}\) at \(L=20\).
Relative to the exact same-size TFIM sequence, the \(J_2=0.05,0.10,0.20\)
sequences cross from above to below between \(L=14\) and \(16\), \(16\) and
\(18\), and \(18\) and \(20\), respectively. The largest-size values lie near \(0.730\), within
\(1.27\times10^{-3}\)--\(1.93\times10^{-3}\) of the exact finite-size Ising
value \(0.732\).  Together, the shrinking spread and largest-size proximity support alignment
of the \(J_2\) sequences with the exact same-size TFIM reference under the
specified response conditions.  Sections~\ref*{supp:correction-diagnostics} and
\ref*{supp:retained-checks} of the SM~\cite{SupplementalMaterial}
report fit-form, field, and implementation sensitivity.

% WPM-BLOCK:B0708
\begin{figure*}[tbp]
  \centering
  \includegraphics[width=\linewidth]{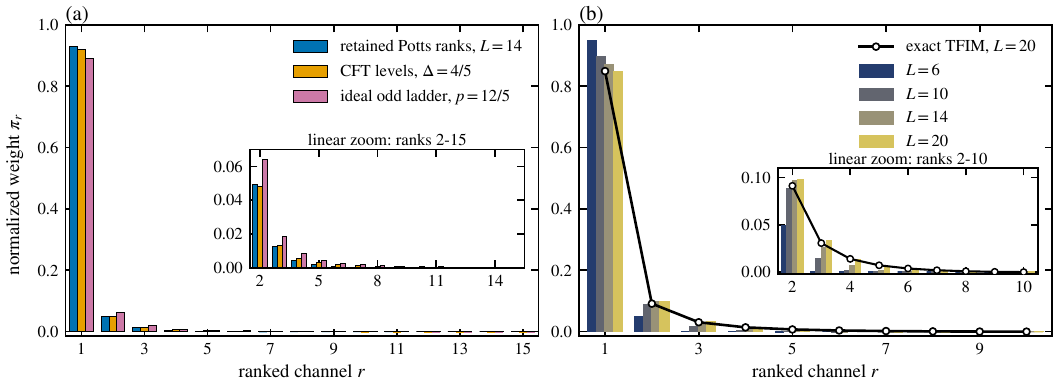}
  \caption{Ranked response distributions reveal the channel rearrangements
  behind the scalar interacting comparisons. (a) Blue bars show the leading 15
  retained weights of the critical \(L=14\) three-state Potts thermal
  response, resolved by tolerance-stable many-body energy-level projectors in
  the translation-invariant \(k=0\), color-neutral response sector. Under the
  spectral premises of Sec.~\ref*{supp:conditional-tail} of the
  SM~\cite{SupplementalMaterial}, the omitted-mass bound protects only the
  first ten ranks against intrusion by an omitted channel; ranks
  \(11\)--\(15\) retain only their ordering among retained channels. Orange
  and purple bars show the descendant-resolved CFT and exponent-only
  odd-ladder comparators; the inset resolves ranks \(2\)--\(15\). The ratio of
  the second-largest Potts weight to the largest is \(0.0531\), closer to the
  CFT value \(0.0522\) than to the exponent-only value \(0.0716\). At
  \(L=14\), the leading normalized Potts weight exceeds its CFT counterpart.
  The Potts bars use $x_a/\Ptwo$. Under the same spectral premises,
  same-response intervals for $r_a/\Ptwo$ preserve the displayed precision and
  ordering among the retained weights. The comparison orders weights by rank
  within each resolution. Rank labels do not identify Potts projectors with
  conformal levels, and the unresolved tail remains uninterpreted. (b) Grouped
  bars show the leading ten level-projector weights of the NNN-TFIM field
  response at \(J_2=0.20\) in the \(k=0\), spin-flip-even sector. Results are
  shown for \(L=6,\,10,\,14,\,20\) at the corresponding odd--even
  scaled-gap-crossing fields \(h_\times(L;J_2)\). The black open-circle line
  is the exact \(L=20\) TFIM distribution, and the inset resolves ranks
  \(2\)--\(10\). From \(L=6\) to \(20\), leading-weight depletion and
  subleading-weight buildup bring the profile near that exact finite-size
  reference; this is an accessible-size trajectory comparison. These NNN bars
  are mean-gap weights $x_a/\Ptwo$, using an independently solved denominator,
  without a same-response enclosure. The displayed ranks contain more than
  \(99.97\%\) of that normalization at every shown size. Construction details
  are in Sec.~\ref*{supp:distribution-construction} of the
  SM~\cite{SupplementalMaterial}.}
  \label{fig:potts-weight-comparators}
\end{figure*}

% WPM-BLOCK:B0294
\paragraph{Distribution-level evidence.}
% WPM-BLOCK:B0706
To explain the CC comparisons, Fig.~\ref{fig:potts-weight-comparators}
resolves the ranked response weights of both interacting models.  In panel (a), the descendant coefficients
\(A_n(4/5)\) [Eq.~\eqref{eq:cft-level-coefficients}] distinguish the CFT
construction from the exponent-only odd ladder.  Here the second-to-leading ratio means the second-largest weight divided
by the largest, \(\pi_{(2)}/\pi_{(1)}\); parentheses denote sorted ranks,
not conformal-level labels. At \(L=14\), the Potts ratio \(0.0531\) is
closer to the CFT value \(0.0522\) than to the exponent-only value
\(0.0716\). At this size, the leading normalized Potts weight exceeds its
CFT counterpart.

% WPM-BLOCK:B0707
Panel (b) shows that for \(J_2=0.20\) the leading NNN-TFIM weight falls from
\(0.9496\) at \(L=6\) to \(0.8482\) at \(L=20\), within \(1.52\times10^{-3}\)
of the exact \(L=20\) TFIM value \(0.8497\). Subleading weights build up under
the matched field-response, symmetry-sector, and level-projector specification.

% WPM-BLOCK:B0709
Under these finite-size specifications, the NNN-TFIM evidence is an
accessible-size trajectory toward the exact finite-size lattice reference. For
Potts, the evidence combines the \(L=6,\ldots,14\) scalar sequence with
conditionally bounded \(L=14\) retained-rank structure against theoretical
comparators.
These data establish scalar and ranked response-structure comparisons beyond
independent momentum modes.  Thermodynamic convergence remains unresolved for
both, and the Potts comparison does not identify lattice projectors with
conformal levels.

The following discussion connects these finite-size comparisons with the
analytic results, the role of channel resolution, and the counting
interpretation of CC.

\section{Discussion}
\label{sec:physical-discussion}

\subsection{Response scale, resolved composition, and critical data}
% WPM-BLOCK:B0619
\label{sec:discussion:response-shape}

% WPM-BLOCK:B0620
The exact TFIM--XX theorem shows that leading susceptibility scaling does
not determine resolved response composition. Both total responses scale as \(L^2\),
whereas one versus two equal soft ladders give concentrations tending to
\(2/3\) and \(1/3\). The resolved distributions identify the source of this
difference: a common within-ladder shape and an additional binary ladder
label in the XX response. For factorized ladder weights,
\(S_2^{\mathrm{ch}}[\pi]=S_2^{\mathrm{ch}}[q]+S_2^{\mathrm{ch}}[\phi]\),
where \(q\) is the ladder-label distribution and \(\phi\) the common
within-ladder distribution. Equal ladders contribute \(\ln n_{\rm lad}\).
The TFIM--XX R\'enyi family realizes this factorization exactly for the
declared response weights.

Beyond this multiplicity effect, the Lifshitz tangents and XX grid offsets
show that the perturbation direction and momentum grid can each change CC.
The response distribution therefore describes a specified perturbation and
channel resolution, together with its state sector and limiting procedure.
The exact lattice relations provide benchmarks for comparisons under these
conditions.

For the spatially integrated response of a single scalar primary in a unitary
\(1+1\)-dimensional CFT on a circle with a nondegenerate vacuum in a fixed sector,
Eq.~\eqref{eq:cft-KF-delta} determines CC from complete
zero-momentum level weights for \(0<\Delta<3/2\). The cylinder coefficients
retain descendant contributions omitted by the exponent-only envelope.
In this domain, the scaling dimension fixes both the leading metric exponent
and the complete-level distribution. Normalization removes the common
response amplitude, yielding a universal concentration function of \(\Delta\)
for this level resolution.
The boundary \(\Delta=3/2\) identifies where this continuum level
distribution ceases to be normalizable without a cutoff. The regulated
concentration can still tend to zero; the boundary concerns the normalized
distribution, not the absence of a scalar limit
[Appendix~\ref{app:cft-uv-boundary}]. Within its domain, the function and
ranked weights supply fixed continuum benchmarks for studying response
composition beyond the exactly solvable chains.

% WPM-BLOCK:B0622
\subsection{Resolution and interacting comparisons}
% WPM-BLOCK:B0623
\label{sec:discussion:probe-contract}

% WPM-BLOCK:B0624
Resolution and spin structure specify the physical probe. Momentum blocks
follow the \((k,-k)\) factorization of the paired Hamiltonian and its ideal
final-Bogoliubov readout; the NS sector fixes the nondegenerate ground-state
sequence and odd soft grid. Coarse graining increases \(\KF\), while a
different spin structure changes the grid. The fixed odd-parity Ramond
lowest state has fixed unpaired occupations with no state derivative; its
\(2/5\) limit concerns a different state and leaves the NS ground-state
theorem unchanged.

The XX point gives an exact example of what the resolution retains.
For \(h=\gam=0\) and \(L\equiv0\pmod4\), the blocks at \(k\) and
\(\pi-k\) have equal response weights and equal pair-excitation energies.
Momentum-block resolution keeps the two labels, whereas a measurement
resolving only excitation energy merges each such pair.  The merged weight
is \(2x_k\): its contribution to \(\Ptwo\) stays \(2x_k\), while its
contribution to \(\Pfour\) changes from \(2x_k^2\) to \(4x_k^2\).
Consequently,
\begin{equation}
 K_F^{\rm XX,energy}(L)=2K_F^{\rm XX,block}(L)
 =K_F^{\rm TFIM}(L/2)\longrightarrow\frac23.
 \label{eq:xx-energy-coarsening}
\end{equation}
The block value still tends to \(1/3\).  The difference is the binary ladder
label retained by the block projectors and lost upon energy grouping; it
requires no change of state or perturbation.  Hence the same total response
can support different concentrations for two explicitly specified readouts.
This finite-size example also explains why energy proximity alone cannot
justify identifying differently resolved interacting observables.

% WPM-BLOCK:B0625
Unlike the free-chain momentum-block probe, the interacting calculations use
many-body level projectors with numerical degeneracies grouped. These
projectors need not match complete CFT levels: irrelevant operators may split
one conformal level into several finite-size lattice levels with weights
\(x_{N,a}\). Resolving these levels separately or merging them gives
\begin{equation}
  K_{F,N}^{\mathrm{coarse}}-K_{F,N}^{\mathrm{fine}}
  =\frac{2}{\Ptwo^2}\sum_{a<b}x_{N,a}x_{N,b}\ge0.
  \label{eq:cft-lattice-resolution-gap}
\end{equation}
Appendix~\ref{app:resolution-stability} bounds this refinement loss for a
fixed response using group weights and counts of responding fine channels.
Its finite-size error bound separately tracks grouped-response matching and
weight outside the specified fine-channel support. With both errors and the
channel counts controlled, it confines fine-resolution concentration to an
interval relative to the reference. A vanishing energy splitting does not by
itself make the two projector observables equal. Numerical tolerance grouping
restores degenerate eigenspaces but does not implement a scale-dependent
conformal grouping. The Potts analysis therefore compares finite-size CC
values and ranked lattice weights with fixed conformal and exponent-only
references; it neither identifies individual projectors nor asserts equality
of continuum concentrations.

% WPM-BLOCK:B0626
\paragraph{Finite-size comparisons.}
\label{sec:discussion:interacting}
The interacting benchmarks in Sec.~\ref{sec:interacting-benchmarks}
therefore answer complementary finite-size questions. The NNN-TFIM follows
an exact same-size lattice reference, whereas Potts is compared with fixed
continuum distributions. Together, CC and the ranked weights compare response
composition beyond independent momentum modes, resolving how changes in
concentration accompany redistribution among many-body levels.
Thermodynamic convergence remains unresolved for both sequences.
Establishing a continuum identification would require joint control of the scalar concentration,
ranked distribution, and correction hierarchy under one probe specification;
agreement of a fitted intercept alone would not supply that test.

% WPM-BLOCK:B0629
\subsection{Operational access and outlook}
% WPM-BLOCK:B0630
\label{sec:discussion:operational}

% WPM-BLOCK:B0631
% WPM-BLOCK:B0270
\label{sec:operational-access}

Consider pair occupations resolved in final-Hamiltonian Bogoliubov blocks.
A weak one-sided displacement gives binary excitation probabilities
\(p_k=\delta^2x_k+O(\delta^3)\). Assume that these occupations are
independent Bernoulli events, and let \(N\) count excited pairs. Then
\begin{align*}
  \langle N\rangle&=\sum_{k>0}p_k,\\
  \Var(N)&=\sum_{k>0}p_k(1-p_k),\\
  \langle N\rangle-\Var(N)&=\sum_{k>0}p_k^2.
\end{align*}
At fixed finite \(L\), the weak-displacement expansion therefore gives
\(\langle N\rangle-\Var(N)=\delta^4\Pfour+O(\delta^5)\) and
\(\langle N\rangle^2=\delta^4\Ptwo^2+O(\delta^5)\), so
\begin{equation}
  \lim_{\delta\to0}
  \frac{\langle N\rangle-\Var(N)}{\langle N\rangle^2}
  =\KF.
  \label{eq:quench-KF}
\end{equation}
Under these counting assumptions, the identity provides a readout of CC
in the zero-amplitude limit. The normalized variance deficit is not a Fano factor,
and the construction is not an interacting level-projector protocol.  SM~\cite{SupplementalMaterial} presents the counting benchmark, final-basis
projector contract, and expansion
(Secs.~\ref*{supp:sec:operational-access} and
\ref*{app:weak-quench}). It also details the finite-amplitude error
budget, fixed-amplitude bias, and shot-cost scaling
(Secs.~\ref*{app:weak-quench-budget} and
\ref*{app:weak-feasibility}).

Three directions follow from these results. Continuum--lattice matching
requires joint control of grouped response weights, unresolved tails, and
fine-channel support, using the CFT concentration and full distribution as
fixed targets for a lattice perturbation governed by a single scalar primary.

The imaginary part of the channel QGT in Eq.~\eqref{eq:channel-qgt} also
motivates an analysis of Berry-curvature composition.
Contributions associated with two parameter directions can have opposite
signs, so a small net curvature may arise from weak channels or strong
cancellation.
Which channels carry the metric and curvature responses? How does grouping
change concentration and cancellation? These are distinct questions, and the
positive-weight refinement rules do not transfer directly to signed
curvature contributions. Studying curvature composition requires models with
nonzero curvature, since the smooth gapped branches of the present
\((h,\gam)\) plane have zero curvature.

A third direction concerns readout design.
The present construction requires zero-amplitude extrapolation and large
ideal sample counts at the analyzed finite amplitudes
[SM~\cite{SupplementalMaterial}].
Its bias and shot-cost analysis provides a starting point for comparing
alternative counting models and detector partitions, each requiring its own
relation to the target response weights.

% WPM-BLOCK:B0632

% WPM-BLOCK:B0633
\newpage
\section{Summary}
% WPM-BLOCK:B0710
\label{sec:summary}

% WPM-BLOCK:B0711
Channel concentration (CC) distinguishes how excitations share a
quantum-geometric response even when its leading scaling is unchanged.
At the specified even-parity NS momentum-block resolution, the exact
TFIM--XX theorem gives thermodynamic concentrations of \(2/3\) and \(1/3\)
for the field and pairing tangents, respectively. One versus two equal soft
ladders with a common shape produce this contrast. The XX sequence has
\(L\equiv0\pmod4\).
The punctured \(\gamma>0\) Lifshitz approaches isolate a different
distinction: field and anisotropy tangents select different response
envelopes at the same limiting quadratic-dispersion Hamiltonian.
A shared leading metric scaling or dispersion therefore does not fix the
resolved response composition.

For one spatially integrated scalar primary in a unitary
\(1+1\)-dimensional CFT on a circle with a nondegenerate vacuum in a fixed sector,
the complete zero-momentum level weights determine a universal
concentration function of \(\Delta\) for \(0<\Delta<3/2\).
The exact analytic expression reproduces Ising \(2/3\) and gives
Potts \(\simeq0.8515\), compared with \(\simeq0.800\) in the
exponent-only approximation.
The concentration and full continuum distribution provide fixed benchmarks
for testing response composition in the declared resolution.
The interacting calculations compare scalar and ranked finite-size
responses with fixed references; the Potts comparison does not identify
lattice projectors with conformal levels or establish convergence to the
CFT value.

The full distribution can distinguish responses even when their CC values
coincide. Refinement bounds constrain changes in concentration when a fixed
response is regrouped between specified resolutions. At fixed finite size,
independent final-Hamiltonian Bogoliubov pair counting gives a conditional
zero-amplitude readout through the variance deficit under the stated
sampling assumptions. Together, these results make the composition of
critical response beyond total metric scaling calculable for declared probes.

% WPM-BLOCK:B0310
\begin{acknowledgments}
We thank Chen-Chih Wang, Yi-Ping Huang, and Po-Yao Chang for helpful discussions and valuable comments.
This work is supported by the Higher Education Sprout Project funded by the National Science and Technology Council, and the Ministry of Education in Taiwan. DWW is supported under the Grant NSTC 113–2112-M-007-036.

OpenAI ChatGPT (GPT-5.6 Sol) assisted with language editing for clarity and
concision under author instructions to preserve scientific content. The
authors exclusively made all scientific judgments and conclusions, manually
reviewed and edited every AI-assisted revision, and take full responsibility
for the manuscript.
\end{acknowledgments}

% WPM-BLOCK:B0311
% ======================================================================
% WPM-BLOCK:B0312
% Let the appendices follow acknowledgments without forcing a sparse column.
% WPM-BLOCK:B0313
\appendix
% WPM-BLOCK:B0314
\section*{Appendices}
% WPM-BLOCK:B0315
Appendices~\ref{app:spectral-projector} and
\ref{app:cft-cylinder-derivation} establish the general projector and
conformal-response formulas; the remaining appendices supply the two-level
directional construction and the exact-chain, RG, sector, and multi-ladder
derivations.  Weak-quench details, feasibility analysis, numerical fits,
comparators, and retained-level audits are in the SM~\cite{SupplementalMaterial}.

% WPM-BLOCK:B0316
\section{Channel-decomposition formalism}
% WPM-BLOCK:B0317
\label{app:spectral-projector}

The spectral derivation supports Sec.~\ref{sec:channel}; its final positive-weight sum establishes the coarse-graining comparison used in Sec.~\ref{sec:channel:scope}.

% WPM-BLOCK:B0318
\subsection{Spectral and projector decompositions}

% WPM-BLOCK:B0319
Let \(n=0\) denote the normalized nondegenerate ground state and \(n>0\) the
excited states.  Each eigenstate obeys
% WPM-BLOCK:B0320
\begin{equation}
  H(\bm{\lambda})\ket{n(\bm{\lambda})}=E_n(\bm{\lambda})\ket{n(\bm{\lambda})}.
\end{equation}
% WPM-BLOCK:B0321
Differentiate the eigenvalue equation and project onto $\bra{m}$ with
$m\ne0$:
% WPM-BLOCK:B0322
\begin{equation}
  \bra{m}\partial_{\mu}H\ket{0}+E_m\braket{m}{\partial_{\mu}0}
  =E_0\braket{m}{\partial_{\mu}0}.
\end{equation}
% WPM-BLOCK:B0323
This gives
% WPM-BLOCK:B0324
\begin{equation}
  \braket{m}{\partial_{\mu}0}
  =-\frac{\bra{m}\partial_{\mu}H\ket{0}}{E_m-E_0}.
  \label{eq:perturb-state}
\end{equation}
% WPM-BLOCK:B0325
The sign drops out of diagonal channel weights.  With
$D_{\mu}\ket{0}\equiv(1-\ketbra{0}{0})\partial_\mu\ket{0}
=\sum_{m>0}\ket{m}\braket{m}{\partial_{\mu}0}$, the QGT is
% WPM-BLOCK:B0326
\begin{equation}
  \Qgt_{\mu\nu}
  =\sum_{m>0}
   \frac{\bra{0}\partial_{\mu}H\ket{m}\bra{m}\partial_{\nu}H\ket{0}}
        {(E_m-E_0)^2}.
  \label{eq:spectral-qgt}
\end{equation}
% WPM-BLOCK:B0327
Contracting with a real tangent vector $y$ gives
% WPM-BLOCK:B0328
\begin{equation}
  \Ptwo(\bm{\lambda},y)
  =\sum_{m>0}x_m(\bm{\lambda},y),
  \qquad
  x_m(\bm{\lambda},y)
  =\frac{\abs{\bra{m}\partial_yH\ket{0}}^2}{(E_m-E_0)^2}.
  \label{eq:spectral-channel-weight-app}
\end{equation}
% WPM-BLOCK:B0329
Equation~\eqref{eq:spectral-channel-weight-app} supplies the spectral-channel
formula for a many-body resolution of the decomposition.

% WPM-BLOCK:B0330
The projector-resolved form is invariant under basis changes within each
resolved projector subspace.  At the chosen base point set
$\ket{\psi}=\ket{0}$ and $P_0=\ketbra{\psi}{\psi}$, and let
$\{\Pi_{\chann}\}$ satisfy
Eq.~\eqref{eq:projector-resolution}.  Define
% WPM-BLOCK:B0331
\begin{equation}
  q^{\chann}_{\mu\nu}
  =\bra{D_{\mu}\psi}\Pi_{\chann}\ket{D_{\nu}\psi}.
\end{equation}
% WPM-BLOCK:B0332
Then
% WPM-BLOCK:B0333
\begin{equation}
  \sum_{\chann}q^{\chann}_{\mu\nu}
  =\bra{D_{\mu}\psi}(1-P_0)\ket{D_{\nu}\psi}
  =\braket{D_{\mu}\psi}{D_{\nu}\psi}
  =\Qgt_{\mu\nu},
\end{equation}
% WPM-BLOCK:B0334
because $D_{\nu}\ket{\psi}$ is horizontal.  For real $y^{\mu}$,
% WPM-BLOCK:B0335
\begin{equation}
  x_{\chann}=y^{\mu}y^{\nu}\Real\!\left(q^{\chann}_{\mu\nu}\right)
  =\bra{D_y\psi}\Pi_{\chann}\ket{D_y\psi}
  =\norm{\Pi_{\chann}D_y\psi}^{2}\ge0.
\end{equation}
% WPM-BLOCK:B0336
Together with $\sum_{\chann}x_{\chann}=\Ptwo$, this proves that, for
$\Ptwo>0$, the normalized weights
$\pi_{\chann}=x_{\chann}/\Ptwo$ form a probability distribution over the
declared channels.
% WPM-BLOCK:B0673
The resolution dependence then obeys a monotonic ordering rule.  If a coarse
channel groups fine probabilities over disjoint sets \(G_a\), then
% WPM-BLOCK:B0674
\begin{align}
  \KF^{\mathrm{coarse}}
  &=\sum_a \bar\pi_a^2
    =\sum_a\left(\sum_{c\in G_a}\pi_c\right)^2
  \nonumber\\
  &=\KF^{\mathrm{fine}}
    +2\sum_a\sum_{c<d\in G_a}\pi_c\pi_d
  \ge \KF^{\mathrm{fine}}.
  \label{eq:coarse-monotonicity}
\end{align}
% WPM-BLOCK:B0675
Coarse graining can therefore only increase \(\KF\), or equivalently reduce
the order-two effective channel number.  Equation~\eqref{eq:coarse-monotonicity} gives the coarse-graining comparison
rule used in Sec.~\ref{sec:channel:scope}: numerical comparisons between
declared probes must use the same grouping on both sides.

% WPM-BLOCK:R3-BOUND-A
\subsection{Concentration bounds under channel refinement}
\label{app:resolution-stability}

We quantify the refinement dependence in
Eq.~\eqref{eq:coarse-monotonicity} and the conformal-level comparison in
Sec.~\ref{sec:discussion:probe-contract}. All weights below are exact
projector masses of the same normalized response vector. At fixed finite
size, partition the fine channels into disjoint groups $N$ and write
\begin{equation}
 \begin{aligned}
 q_N&=\sum_{a\in N}\pi_{N,a},\\
 K_F^{\mathrm{group}}&=\sum_Nq_N^2,\\
 K_F^{\mathrm{fine}}&=\sum_{N,a}\pi_{N,a}^2.
 \end{aligned}
 \label{eq:resolution-group-definitions}
\end{equation}
For $q_N>0$, let $m_N$ count the fine channels with positive weight.

\emph{Finite-size refinement bound.} Every such partition satisfies
\begin{align}
 0&\le K_F^{\mathrm{group}}-K_F^{\mathrm{fine}}\notag\\
 &\le\sum_{N:q_N>0}\left(1-\frac1{m_N}\right)q_N^2.
 \label{eq:resolution-finite-bound}
\end{align}
Indeed, $\sum_a\pi_{N,a}^2\ge q_N^2/m_N$ by Cauchy--Schwarz.
The upper bound is attained when the responding channels within each
group have equal weights. A group supported on one channel contributes
zero. Thus the effect of a refinement is controlled by both the group
weight and the number of channels over which it can be shared.

To compare with a reference distribution $\pi_N^\star$, let
$K_F^\star=\sum_N(\pi_N^\star)^2$. Pad absent groups by zeros and use
one common countable set of group labels at all sizes. Fix positive
integer budgets $M_N$, independent of $L$. For each group and size,
choose a subset $\mathcal A_N(L)$ of at most $M_N$ actual fine channels.
Define the weight outside these subsets and the grouped matching error by
\begin{align}
 e_N(L)&=\sum_{a\in N\setminus\mathcal A_N(L)}\pi_{N,a}(L),\notag\\
 \delta_{\mathrm{out}}(L)&=\sum_N e_N(L),\notag\\
 \delta_{\mathrm{grp}}(L)&=\sum_N|q_N(L)-\pi_N^\star|.
 \label{eq:resolution-error-definitions}
\end{align}
These definitions refer to finite-size projector channels, rather than
components of the response vector in an arbitrarily chosen basis of a
limiting degenerate subspace.

\emph{Reference bound with finite-size errors.} Set
\begin{equation}
 B_M=\sum_N\left(1-\frac1{M_N}\right)(\pi_N^\star)^2.
 \label{eq:resolution-reference-budget}
\end{equation}
Then
\begin{align}
 K_F^{\mathrm{fine}}(L)&\ge K_F^\star-B_M\notag\\
 &\quad-2[\delta_{\mathrm{grp}}(L)+\delta_{\mathrm{out}}(L)],\notag\\
 K_F^{\mathrm{fine}}(L)&\le K_F^\star+2\delta_{\mathrm{grp}}(L).
 \label{eq:resolution-reference-bound}
\end{align}
For the lower bound, put $t_N=q_N-e_N$. Cauchy--Schwarz on the selected
channels gives $K_F^{\mathrm{fine}}\ge\sum_Nt_N^2/M_N$.
Since $0\le t_N,\pi_N^\star\le1$ and
$\sum_N|t_N-\pi_N^\star|\le\delta_{\mathrm{grp}}+\delta_{\mathrm{out}}$,
\[
 \begin{aligned}
 &\left|\sum_N\frac{t_N^2-(\pi_N^\star)^2}{M_N}\right|\\
 &\qquad\le2(\delta_{\mathrm{grp}}+\delta_{\mathrm{out}}).
 \end{aligned}
\]
The upper bound follows from coarse-graining monotonicity and the
same squared-probability estimate applied to $q_N$ and $\pi_N^\star$.
Both estimates control the complete sum, so no interchange of a
limit superior with an uncontrolled tail is required.

If $\delta_{\mathrm{grp}}\to0$ and $\delta_{\mathrm{out}}\to0$, the
finite-size bound yields
\begin{equation}
 \begin{aligned}
 K_F^\star-B_M
 &\le\liminf_{L\to\infty}K_F^{\mathrm{fine}}(L)\\
 &\le\limsup_{L\to\infty}K_F^{\mathrm{fine}}(L)
 \le K_F^\star.
 \end{aligned}
 \label{eq:resolution-limit-band}
\end{equation}
This confines all subsequential limits to a one-sided interval. A unique
limit or equality with $K_F^\star$ requires additional information about
the allocation of weight within the groups.

% WPM-BLOCK:B0713
\section{Euclidean response vector and conformal-cylinder level weights}
% WPM-BLOCK:B0714
\label{app:cft-cylinder-derivation}

% WPM-BLOCK:B0715
This appendix derives the CFT formulas of Sec.~\ref{sec:cft-response} under the
normalization, projector, and ultraviolet prescriptions used for the
comparison.  We follow the Euclidean information-metric construction of
Ref.~\cite[Sec.~2]{BakTrivella2017}; the conformal map and descendant expansion
use the conventions of Ref.~\cite[Chaps.~5 and 9]{DiFrancesco1997}.
In particular, the resolvent calculation below supplies the Laplace-moment
representation and convention cancellation used with
Eq.~\eqref{eq:cft-response-state}.

% WPM-BLOCK:B0716
\subsection{Resolvent and Euclidean normalization}

% WPM-BLOCK:B0717
For \(Q=1-\ketbra{0}{0}\), the spectral theorem gives
% WPM-BLOCK:B0718
\begin{equation}
  Q(H-E_0)^{-1}Q
  =\int_0^\infty d\tau\,Qe^{-\tau(H-E_0)}Q,
  \label{eq:app-cft-resolvent-laplace}
\end{equation}
% WPM-BLOCK:B0719
on the excited-state subspace.  Truncating its lower limit at
\(\epsilon>0\) gives Eq.~\eqref{eq:euclidean-response-state} and
\(\ket{\chi_y^{(\epsilon)}}=e^{-\epsilon(H-E_0)}\ket{\chi_y}\).  Consequently,
% WPM-BLOCK:CND219-01
\begin{equation}
  \braket{\chi_y^{(\epsilon)}}{\chi_y^{(\epsilon)}}
  =\sum_{m>0}
  \frac{e^{-2\epsilon(E_m-E_0)}
  \abs{\bra{m}V_y\ket{0}}^2}{(E_m-E_0)^2}.
  \label{eq:app-cft-regulated-norm}
\end{equation}
Applying the untruncated representation to the bra and ket gives the
\(\epsilon\to0^+\) double Euclidean-time integral for the norm.
% WPM-BLOCK:B0720
\begin{equation}
  \begin{aligned}
  \braket{\chi_y}{\chi_y}
  &=\int_0^\infty d\tau_1\int_0^\infty d\tau_2
    \bra{0}V_yQe^{-(\tau_1+\tau_2)(H-E_0)}QV_y\ket{0}\\
  &=\int_0^\infty ds\int_0^s d\tau_1\,C_y(s)
   =\int_0^\infty ds\,s\,C_y(s)\\
  &=\sum_{m>0}\frac{\abs{\bra{m}V_y\ket{0}}^2}{(E_m-E_0)^2}.
  \end{aligned}
  \label{eq:app-cft-response-norm}
\end{equation}
% WPM-BLOCK:B0721
Here \(C_y(s)=\langle V_y(s)V_y(0)\rangle_c\), and the two \(Q\)
projections remove only the ground-state contribution.  At fixed
\(s=\tau_1+\tau_2\), the interval \(0\le\tau_1\le s\) produces the
first-moment factor.  The spectral resolvent is defined first at finite size
with a nondegenerate ground state and positive gaps.  No additional local
contact counterterm enters this spectral observable.  For the CFT level sum,
the cutoff is removed only in the convergent range established below, before
normalization and large-size lattice comparison.  Ref.~\cite[Sec.~2]{BakTrivella2017} uses
\(G_{yy}=\Ptwo/2\), a common normalization that cancels from \(\pi_a^\Pi\)
and \(\KF^\Pi\).

% WPM-BLOCK:B0722
For a complete orthogonal family \(\sum_a\Pi_a=Q\), set
\(x_a^\Pi=\bra{\chi_y}\Pi_a\ket{\chi_y}\),
\(\pi_a^\Pi=x_a^\Pi/\braket{\chi_y}{\chi_y}\), and
\(\ket{\widehat\chi_y}=\ket{\chi_y}/\sqrt{\braket{\chi_y}{\chi_y}}\).
Direct multiplication on two copies gives
% WPM-BLOCK:B0723
\begin{align}
  &\bra{\widehat\chi_y}^{\otimes2}
  \left(\sum_a\Pi_a\otimes\Pi_a\right)
  \ket{\widehat\chi_y}^{\otimes2} \notag\\
  &\qquad=\sum_a
  \left(\bra{\widehat\chi_y}\Pi_a\ket{\widehat\chi_y}\right)^2 \notag\\
  &\qquad=\sum_a(\pi_a^\Pi)^2 .
  \label{eq:app-cft-replica-proof}
\end{align}
% WPM-BLOCK:B0724
This proves Eq.~\eqref{eq:replica-same-channel} without identifying it with a
connected single-copy four-point cumulant.

% WPM-BLOCK:B0725
\subsection{Cylinder expansion and complete level projectors}\label{app:cft-cylinder-levels}

We derive the complete-level coefficients and inverse-gap factors in Eq.~\eqref{eq:cft-response-weights}, which also supply the Potts specialization in Eq.~\eqref{eq:potts-cft-target}.

% WPM-BLOCK:B0726
Under the vacuum single-primary assumptions of
Sec.~\ref{sec:cft-cylinder}, start from the plane two-point function of a
Hermitian scalar primary with
\(h_{\mathcal O}=\bar h_{\mathcal O}=\Delta/2\):
\[
  \bigl\langle\mathcal O_{\rm pl}(z_1)
  \mathcal O_{\rm pl}(z_2)\bigr\rangle
  =\frac{C_{\mathcal O}}{|z_1-z_2|^{2\Delta}}.
\]
Here \(C_{\mathcal O}>0\) fixes the operator normalization, and the
antiholomorphic arguments are suppressed.  At finite circumference \(L\),
the cylinder coordinate and conformal map are
\[
  w=v\tau+ix,\qquad x\sim x+L,\qquad z=e^{2\pi w/L}.
\]
Using cylinder translation invariance, choose the insertions at
\(w_1=v\tau+ix\) and \(w_2=0\), with \(\tau>0\).  Then
\[
\begin{aligned}
  u&=2\pi v\tau/L, & \theta&=2\pi x/L,\\
  z_1&=e^{u+i\theta}, & z_2&=1.
\end{aligned}
\]
The conjugate coordinates are \(\bar z_1=e^{u-i\theta}\) and
\(\bar z_2=1\).  Thus the insertion at \(w_2=0\) maps to \(z_2=1\);
the plane origin instead corresponds to \(\tau\to-\infty\).

The primary transformation supplies the Jacobian factors:
\[
\begin{aligned}
  \mathcal O_{\rm cyl}(w,\bar w)
  & =\left(\frac{dz}{dw}\right)^{h_{\mathcal O}}
     \left(\frac{d\bar z}{d\bar w}\right)^{\bar h_{\mathcal O}}\\
  &\quad\times\mathcal O_{\rm pl}(z,\bar z).
\end{aligned}
\]
Since \(dz/dw=(2\pi/L)z\), their product at the two insertions gives
\[
\begin{aligned}
  &\bigl\langle\mathcal O_{\rm cyl}(w_1)
    \mathcal O_{\rm cyl}(0)\bigr\rangle\\
  &\qquad=C_{\mathcal O}\left(\frac{2\pi}{L}\right)^{2\Delta}
    \frac{e^{\Delta u}}{|e^{u+i\theta}-1|^{2\Delta}}.
\end{aligned}
\]
The distance identity
\[
\begin{aligned}
  |e^{u+i\theta}-1|^2
  &=e^{2u}+1-2e^u\cos\theta\\
  &=2e^u(\cosh u-\cos\theta)
\end{aligned}
\]
cancels the factor \(e^{\Delta u}\) against the denominator.  Defining
\(G(u,\theta)\) by removing the common factor
\(C_{\mathcal O}(2\pi/L)^{2\Delta}\) from this cylinder correlator gives
% WPM-BLOCK:B0727
\begin{equation}
  \begin{split}
  G(u,\theta)
  &=\left[2\bigl(\cosh u-\cos\theta\bigr)\right]^{-\Delta} \\
  &=e^{-\Delta u}
    (1-e^{-u+i\theta})^{-\Delta}
    (1-e^{-u-i\theta})^{-\Delta} .
  \end{split}
  \label{eq:app-cylinder-correlator}
\end{equation}
% WPM-BLOCK:B0728
For \(u>0\), both factors admit the convergent binomial expansion
\((1-z)^{-\Delta}=\sum_{n\ge0}(\Delta)_n z^n/n!\).  Hence
\[
\begin{aligned}
  G(u,\theta)
  &=\sum_{m,n\ge0}\frac{(\Delta)_m(\Delta)_n}{m!\,n!}\\
  &\quad\times e^{-(\Delta+m+n)u}e^{i(m-n)\theta}.
\end{aligned}
\]
The zero-momentum projection enforces equal left- and right-moving levels
through \(\int_0^{2\pi}d\theta\,e^{i(m-n)\theta}/(2\pi)=\delta_{mn}\),
so that
% WPM-BLOCK:B0729
\begin{equation}
  \int_0^{2\pi}\frac{d\theta}{2\pi}\,G(u,\theta)
  =\sum_{n=0}^\infty
   \left[\frac{(\Delta)_n}{n!}\right]^2
   e^{-(\Delta+2n)u} .
  \label{eq:app-cylinder-zero-momentum}
\end{equation}
% WPM-BLOCK:B0730
The zero-momentum projector groups the full degeneracy at each common level;
individual descendant basis vectors inside that block are not separate
channels.  Equation~\eqref{eq:app-cylinder-zero-momentum} therefore identifies
\(A_n(\Delta)=[(\Delta)_n/n!]^2\) as the total squared-matrix-element
weight in the complete physical level, after removing the common factor,
rather than the number of states in that level.  Null descendants are
absent in the physical Hilbert-space quotient.  The cylinder gap is
% WPM-BLOCK:B0731
\begin{equation}
  E_n-E_0=\frac{2\pi v}{L}(\Delta+2n).
  \label{eq:app-cylinder-gap}
\end{equation}
% WPM-BLOCK:B0732
The lower cutoff in Eq.~\eqref{eq:app-cft-regulated-norm} multiplies a level
amplitude by \(e^{-\epsilon(E_n-E_0)}\); squaring it gives
\(e^{-2u_\epsilon(\Delta+2n)}\), where
\(u_\epsilon=2\pi v\epsilon/L\).  The two inverse resolvents therefore produce
the regulated weight in Eq.~\eqref{eq:cft-response-weights}, up to one positive
level-independent factor.
Operator normalization, microscopic matching amplitude, velocity, and
cylinder radius multiply every \(x_n\) equally and cancel from \(K_F\).
These level coefficients also supply the Potts series used in
Eq.~\eqref{eq:potts-cft-target}.

% WPM-BLOCK:B0733
\subsection{Ultraviolet boundary and order of limits}\label{app:cft-uv-boundary}

We establish the convergence domain and cutoff order of Eq.~\eqref{eq:cft-response-weights}, including the Potts evaluation in Eq.~\eqref{eq:potts-cft-target}.

% WPM-BLOCK:B0734
Stirling asymptotics gives
% WPM-BLOCK:B0735
\begin{equation}
  A_n(\Delta)\sim\frac{n^{2\Delta-2}}{\Gamma(\Delta)^2},
  \qquad
  w_n(\Delta)\sim
  \frac{n^{2\Delta-4}}{4\Gamma(\Delta)^2} .
  \label{eq:app-cft-weight-asymptotic}
\end{equation}
% WPM-BLOCK:B0736
Write the response-weight moments as
\(W_r(\Delta)=\sum_{n\ge0}w_n(\Delta)^r\). Then \(W_1\) converges for
\(0<\Delta<3/2\), while
\(\sum_nw_n^2\) converges over the larger range \(0<\Delta<7/4\).  At
\(\Delta=3/2\), the first moment grows logarithmically and the normalized
concentration tends to zero.  The same boundary follows from finite-size
scaling: the singular scalar response is \(L^{4-2\Delta}\), whereas a generic
analytic ultraviolet background is extensive, \(O(L)\).  Because the regulated
positive terms are bounded by the summable \(w_n\) for
\(0<\Delta<3/2\), dominated convergence permits
\(\epsilon\to0^+\) in \(W_1\) and \(W_2\).  The construction first fixes finite \(L\), the nondegenerate vacuum,
\(Q\), the tangent, and the complete level projectors. We then remove
\(\epsilon\), normalize the level weights, and form \(K_F\). Only after
these steps do we compare the singular response with the large-\(L\)
lattice result at fixed microscopic matching.  No regulator-free claim is made for
\(\Delta\ge3/2\).
This establishes the convergence domain of the response weights in
Eq.~\eqref{eq:cft-response-weights} and of the Potts evaluation in
Eq.~\eqref{eq:potts-cft-target}.

% WPM-BLOCK:B0737
\subsection{Ising specialization and Potts evaluation}\label{app:cft-potts-evaluation}

We evaluate the Ising and Potts specializations of Eq.~\eqref{eq:cft-KF-delta}, completing the value quoted in Eq.~\eqref{eq:potts-cft-target}.

% WPM-BLOCK:B0738
The concentration is \(K_F=W_2/W_1^2\) in terms of the moments defined
above. For \(\Delta=1\), \(A_n=1\), and the
odd-integer sums give
% WPM-BLOCK:B0739
\begin{equation}
  \begin{split}
  W_1(1)&=\sum_{n=0}^\infty\frac{1}{(2n+1)^2}=\frac{\pi^2}{8}, \\
  W_2(1)&=\sum_{n=0}^\infty\frac{1}{(2n+1)^4}=\frac{\pi^4}{96}, \\
  \frac{W_2(1)}{W_1(1)^2}&=\frac23 .
  \end{split}
  \label{eq:app-cft-ising-specialization}
\end{equation}
% WPM-BLOCK:B0740
For general \(\Delta\), one evaluation route iterates
\(A_{n+1}/A_n=[(n+\Delta)/(n+1)]^2\) and encloses the tail by the gamma-ratio
asymptotic series.  An independent route evaluates both \(W_1\) and \(W_2\)
from their unit-argument generalized-hypergeometric representations.  SM~\cite{SupplementalMaterial} records the two moment values and their
high-precision cross-check.  At \(\Delta=4/5\), the routes agree and give
\(K_F\simeq0.8515\).  Derived without a microscopic Potts fit, this specified
conformal observable supplies the level-weight series and
convergence-supported theory comparator for the finite-size lattice sequence in
main-text Eq.~\eqref{eq:potts-cft-target}.

% WPM-BLOCK:R3-BOUND-B
\subsection{Conditional thermal-family refinement budgets}
\label{app:thermal-refinement-budgets}

For the complete-level reference of Sec.~\ref{sec:cft-cylinder}, take
$\pi_N^\star=\pi_N^{\mathrm{CFT}}=w_N/W_1$ and
$K_F^\star=K_F^{\mathrm{CFT}}(\Delta)$. If the grouped weights obey
Eq.~\eqref{eq:cft-grouped-matching} and their response is asymptotically
carried by at most $d_N^2$ fine channels per group, the choice $M_N=d_N^2$
in Eq.~\eqref{eq:resolution-limit-band} gives a thermal-family budget
\begin{equation}
 B_{\mathrm{fam}}=
 \sum_N\left(1-\frac1{d_N^2}\right)
             (\pi_N^{\mathrm{CFT}})^2.
 \label{eq:thermal-family-budget}
\end{equation}
Here $d_N$ is the holomorphic multiplicity in the physical Virasoro
module after null states are removed. The support assumption means that
the subsets in Eq.~\eqref{eq:resolution-error-definitions} can be chosen
with $M_N=d_N^2$ and $\delta_{\mathrm{out}}\to0$.

For the thermal modules, the physical characters give
\cite{DiFrancesco1997}
\begin{align*}
 d_N^{\mathrm{Ising}}&=1,1,1,1,2,2,3,4,5,6,\ldots,\\
 d_N^{\mathrm{Potts}}&=1,1,1,2,3,4,6,8,11,15,\ldots.
\end{align*}
The values can be evaluated with the Rocha--Caridi character:
\begin{align}
 d_n&=\sum_{k\in\mathbb Z}
        [p(n-\ell_k^-)-p(n-\ell_k^+)],\notag\\
 \ell_k^\pm&=
 \frac{(2pqk+qr\pm ps)^2-(qr-ps)^2}{4pq},
 \label{eq:thermal-character-count}
\end{align}
where $p(n)$ denotes the partition number, with $p(n)=0$ for $n<0$.
Use $(p,q,r,s)=(3,4,2,1)$ for Ising and $(5,6,2,1)$ for Potts.

Outward-rounded interval arithmetic and analytic tail estimates give
\begin{align}
 B_{\mathrm{fam}}^{\mathrm{Ising}}&<1.68\times10^{-4},\notag\\
 B_{\mathrm{fam}}^{\mathrm{Potts}}&<4.04\times10^{-5}.
 \label{eq:thermal-budget-values}
\end{align}
These are upper bounds on the reference budgets, conditional on the
stated support model when used for a lattice observable. For both
thermal dimensions $0<\Delta\le1$, the coefficients $A_n$ are nonincreasing with
$n$. If the partial sums stop at $n=N$, integral tests therefore give
\begin{align}
 \sum_{n>N}w_n&\le\frac{A_N}{2(2N+\Delta)},\notag\\
 \sum_{n>N}w_n^2&\le\frac{A_N^2}{6(2N+\Delta)^3}.
 \label{eq:thermal-budget-tails}
\end{align}
The omitted numerator of Eq.~\eqref{eq:thermal-family-budget} is bounded
by the second tail. These estimates enclose both the numerator and its
normalization without assigning zero weight to the omitted levels.

The first four Ising levels are single states and carry about $94.96\%$
of the conformal response; the first three Potts levels are single states
within the thermal Virasoro family and carry about $98.28\%$.
Their large weights explain why the family budgets are small.

Application to a lattice spectrum requires the support statement above.
In the specified even, zero-momentum Ising CFT sector, identity and
thermal levels have even and odd scaled energies, respectively, so there
is no cross-family degeneracy at a thermal level. Lattice spectral
multiplicity and grouped response matching still need to be controlled.
In Potts, other Virasoro families have states at the same scaled energies
as thermal descendants. The continuum vacuum selection rule alone does
not prevent finite-size eigenvectors from mixing those states. The small
Potts family budget therefore requires an additional symmetry or
projector-support argument for the lattice model under study.

With only an asymptotically unsplit leading response group and grouped
matching, a weaker budget is
\begin{equation}
 B_{\mathrm{tail}}=\sum_{N\ge1}(\pi_N^{\mathrm{CFT}})^2
 \simeq2.5433\times10^{-3}\quad\text{(Potts)}.
 \label{eq:potts-leading-group-budget}
\end{equation}
It follows by bounding the refinement loss of every other group by
$q_N^2$. This alternative avoids thermal-family multiplicity assumptions
for the excited groups, while retaining the leading-group and matching
conditions. Neither budget supplies the finite-size rates of
$\delta_{\mathrm{grp}}$ and $\delta_{\mathrm{out}}$ for the interacting
sequences in Sec.~\ref{sec:interacting-benchmarks}.

% WPM-BLOCK:B0353
\section{Two-level planar mode metric and XY derivatives}
% WPM-BLOCK:B0354
\label{app:mode}

The two-level QGT in Eq.~\eqref{eq:twolevel-qgt-app} fixes the metric convention
used in Secs.~\ref{sec:lifshitz:xy-conventions} and
\ref{sec:directional:weights}; its angle contraction gives the main mode-weight
identity, Eq.~\eqref{eq:xk-angle}.  The resulting \(XY\) angle derivatives and
directional weights give Eqs.~\eqref{eq:xy-grads} and
\eqref{eq:P4-quartic}.

% WPM-BLOCK:B0355
For a two-level Hamiltonian
$H=\bm d\cdot\bm\sigma$, the ray depends only on
$\hat{\bm d}=\bm d/\abs{\bm d}$.  Parametrize the unit vector by spherical
angles $(\theta,\phi)$.  Here $\phi_k$ is the Bloch-sphere azimuth, not the
tangent-space direction angle $\varphi$ of
Sec.~\ref{sec:directional:direction}.  For the lower eigenstate of a gapped
block, the rank-one projector is
\begin{equation*}
  P_{\rm low}=\frac12\left(1-\hat{\bm d}\cdot\bm\sigma\right).
\end{equation*}
The pure-state metric can be written in terms of this projector.  Using
\(\operatorname{Tr}(\sigma_i\sigma_j)=2\delta_{ij}\) gives
\begin{equation*}
  \begin{aligned}
  g_{\mu\nu}
  &=\frac12\operatorname{Tr}
    \left[(\partial_\mu P_{\rm low})(\partial_\nu P_{\rm low})\right]\\
  &=\frac14\partial_\mu\hat{\bm d}\cdot
    \partial_\nu\hat{\bm d}.
  \end{aligned}
\end{equation*}
Thus changes in \(\abs{\bm d}\) alone do not change the ray.  Substituting
\(\hat{\bm d}=(\sin\theta\cos\phi,\sin\theta\sin\phi,\cos\theta)\)
yields the Fubini--Study line element
% WPM-BLOCK:B0356
\begin{equation}
  \d s^2=\frac14\left(\d\theta^2+\sin^2\theta\,\d\phi^2\right),
\end{equation}
% WPM-BLOCK:B0357
and the corresponding QGT convention is
% WPM-BLOCK:B0358
\begin{widetext}
\begin{equation}
  \Qgt_{\mu\nu}^{(k)}
  =\frac14\left(\partial_{\mu}\theta_k\partial_{\nu}\theta_k
  +\sin^2\theta_k\,\partial_{\mu}\phi_k\partial_{\nu}\phi_k\right)
  -\frac{\mathrm{i}}{4}\sin\theta_k
  \left(\partial_{\mu}\theta_k\partial_{\nu}\phi_k
       -\partial_{\nu}\theta_k\partial_{\mu}\phi_k\right).
  \label{eq:twolevel-qgt-app}
\end{equation}
\end{widetext}
% WPM-BLOCK:B0359
The minus sign in the imaginary term corresponds to the lower eigenstate
with the stated spherical coordinates and the standard Pauli orientation.
A local phase change of the eigenstate leaves the entire QGT unchanged.
In the Ising and $XY$ chains
used here, $\phi_k$ is locally constant on each gapped branch in $(h,\gam)$,
hence
% WPM-BLOCK:B0360
\begin{equation}
  g_{\mu\nu}^{(k)}=\frac14\partial_{\mu}\theta_k\partial_{\nu}\theta_k,
  \qquad
  \Berry_{\mu\nu}^{(k)}=0.
\end{equation}
Contracting this metric with the tangent direction gives the main-text mode
weight in Eq.~\eqref{eq:xk-angle}.

% WPM-BLOCK:B0361
For the $XY$ block, define
% WPM-BLOCK:B0362
\begin{align}
  X_k&=h-\cos k,
  &
  Y_k&=\gam\sin k,
  \nonumber\\
  \Lk^2&=X_k^2+Y_k^2,
  &
  \theta_k&=\operatorname{atan2}(Y_k,X_k).
\end{align}
% WPM-BLOCK:B0363
On a continuous branch away from a gap closing,
% WPM-BLOCK:B0364
\begin{equation}
  \partial_a\theta_k
  =\frac{X_k\partial_aY_k-Y_k\partial_aX_k}{X_k^2+Y_k^2},
\end{equation}
% WPM-BLOCK:B0365
which yields Eq.~\eqref{eq:xy-grads}:
% WPM-BLOCK:B0366
\begin{equation}
  \partial_h\theta_k=-\frac{\gam\sin k}{\Lk^2},
  \qquad
  \partial_{\gam}\theta_k=\frac{(h-\cos k)\sin k}{\Lk^2}.
\end{equation}
% WPM-BLOCK:B0367
The TFIM is the slice $\gam=1$ with tangent direction $y=\partial_h$.

% Moved atomically from the main Lifshitz sequence under
% CC-2.1.11-CND-NARRATIVE-CLOSURE-R1 / D04-A.
% WPM-BLOCK:B0234
\subsection{Directional fourth moments in the \texorpdfstring{$XY$}{XY} plane}
% WPM-BLOCK:B0235
\label{sec:directional}
% WPM-BLOCK:B0236
% ======================================================================

This subsection extends the coordinate-tangent Lifshitz result of
Sec.~\ref{sec:lifshitz} to arbitrary tangent rays and supplies the quartic
directional structure invoked there.

% WPM-BLOCK:B0237
% WPM-BLOCK:B0238
\subsubsection{Two-parameter weights}
% WPM-BLOCK:B0239
\label{sec:directional:weights}

% WPM-BLOCK:B0240
With the $XY$-chain conventions of Sec.~\ref{sec:lifshitz:xy-conventions},
define the mode covector
% WPM-BLOCK:B0241
\begin{equation}
  v_{k,\mu}=\partial_{\mu}\thetak,
  \qquad
  v_k(y)=v_{k,\mu}y^{\mu}
  =y^h\partial_h\thetak+y^{\gam}\partial_{\gam}\thetak.
  \label{eq:vk-covector}
\end{equation}
% WPM-BLOCK:B0242
The directional weight of mode $k$ is
% WPM-BLOCK:B0243
\begin{equation}
  \xk(y)=\frac14\,v_k(y)^2,
  \label{eq:xk-directional}
\end{equation}
% WPM-BLOCK:B0244
and the first two channel moments are
% WPM-BLOCK:B0245
\begin{equation}
  \Ptwo(y)=\sum_{k>0}\xk(y)
  =g_{\mu\nu}y^{\mu}y^{\nu},
  \qquad
  g_{\mu\nu}=\frac14\sum_{k>0}v_{k,\mu}v_{k,\nu},
  \label{eq:P2-directional}
\end{equation}
% WPM-BLOCK:B0246
\begin{align}
  \Pfour(y)
  &=\sum_{k>0}\xk(y)^2
    =T_{\mu\nu\rho\sigma}y^{\mu}y^{\nu}y^{\rho}y^{\sigma},
  \nonumber\\
  T_{\mu\nu\rho\sigma}
  &=\frac{1}{16}\sum_{k>0}
    v_{k,\mu}v_{k,\nu}v_{k,\rho}v_{k,\sigma}.
  \label{eq:P4-directional-tensor}
\end{align}
% WPM-BLOCK:B0247
Equivalently,
% WPM-BLOCK:B0248
\begin{equation}
  \Pfour(y)=\frac{1}{16}\sum_{k>0}v_k(y)^4.
  \label{eq:P4-quartic}
\end{equation}
% WPM-BLOCK:B0249
The two-level metric derivation in Appendix~\ref{app:mode} yields these weights.

% WPM-BLOCK:B0250
A one-parameter TFIM has a single
quench direction, so $\KF$ is a single number and no directional structure
exists.  Directional structure requires a parameter space of dimension at
least two, which the $XY$ plane provides.  Moreover, since $H_k$ lies in the
$\rho_z$--$\rho_x$ plane for all $(h,\gam)$, the Hamiltonian is real in
this basis.  On a smooth gapped branch its eigenstates can be chosen
locally real, so the Berry curvature
$\Berry_{h\gam}=-2\Imag\!\left(\Qgt_{h\gam}\right)$ vanishes there.  This local
statement does not extend the nondegenerate QGT across gap closings.  The directional response in this plane is therefore entirely metric;
no Berry-curvature channel enters the analysis.

% WPM-BLOCK:B0251
\subsubsection{Direction dependence and the two-dimensional equivalence}
% WPM-BLOCK:B0252
\label{sec:directional:direction}

% WPM-BLOCK:B0253
The quadratic moment \eqref{eq:P2-directional} is the squared Riemannian
length.  The quartic moment \eqref{eq:P4-quartic}, however, is not generally the
square of any quadratic form.  The directional concentration,
% WPM-BLOCK:B0254
\begin{equation}
  \KF(y)=\frac{\Pfour(y)}{\Ptwo(y)^2},
  \label{eq:KF-directional}
\end{equation}
% WPM-BLOCK:B0255
is a function on tangent directions.  It is direction independent only in
special cases where $\Pfour(y)=C(\bm{\lambda})\Ptwo(y)^2$ for all $y$.
The mode covectors need not be collinear, so direction dependence is possible;
the explicit sums at the two points used in
Fig.~\ref{fig:xy-directional-profiles} show that the proportionality condition
fails there.

% WPM-BLOCK:B0256
We parameterize rays in the tangent space of the $XY$ plane by the Euclidean
angle \(\varphi\) in the interval \(0\le\varphi<2\pi\).  The coordinates
\(h\) and \(\gam\) are the dimensionless bare coefficients of the Hamiltonian
in Eq.~\eqref{eq:xy}; a unit vector therefore represents equal displacement
amplitude in these two bare coordinates.  This is the quench convention used
for the polar profiles, not a coordinate-independent unit circle.  A rescaling
or nonlinear reparameterization of the couplings would define a different
equal-amplitude protocol.  With this convention,
% WPM-BLOCK:B0257
\begin{equation}
  \hat y(\varphi)
  =\cos\varphi\,\partial_h+
   \sin\varphi\,\partial_{\gam}.
  \label{eq:tangent-angle-varphi}
\end{equation}
% WPM-BLOCK:B0258
We define $\KF(\varphi)=\KF(\hat y(\varphi))$.  Because
$\Ptwo(s\,y)=s^2\Ptwo(y)$ and $\Pfour(s\,y)=s^4\Pfour(y)$, the ratio is
homogeneous of degree zero in the tangent vector.  Moreover
$\KF(-\hat y(\varphi))=\KF(\hat y(\varphi))$, so the
profile is $\pi$-periodic even when displayed over $0\le\varphi<2\pi$.
Figure~\ref{fig:xy-directional-profiles}(a) shows this polar profile at a representative
point.  Figure~\ref{fig:xy-directional-profiles}(b) repeats the same display at the
finite-size-regulated point $(h,\gam)=(1+s,s)$ with $s=0.05$ on the gapped
approach to the Lifshitz point.  The exact point $(1,0)$ is not used for this
polar comparison: as in Sec.~\ref{sec:lifshitz}, at $\gam=0$ the
field-direction metric weight has $\Ptwo^{(h)}=0$, so the directional ratio is
not defined in every tangent direction.

% WPM-BLOCK:B0259
In two dimensions the fully symmetric
quartic tensor has only five independent components,
% WPM-BLOCK:B0260
\begin{equation}
  T_{hhhh},\quad T_{hhh\gam},\quad T_{hh\gam\gam},\quad
  T_{h\gam\gam\gam},\quad T_{\gam\gam\gam\gam}.
\end{equation}
% WPM-BLOCK:B0261
At a single base point, $\Pfour(\varphi)$ is a quartic trigonometric polynomial
whose five independent Fourier coefficients determine the five components of
$T$, and vice versa.  Where \(\Ptwo>0\), the pair \(\Ptwo(\varphi)\) and
\(\KF(\varphi)\) recovers the same fourth-moment curve.  Thus $T$ is a
coordinate representation of these local data, not a second observable.

% WPM-BLOCK:B0262
\subsubsection{Directional fourth-moment structure}
% WPM-BLOCK:B0263
\label{sec:directional:fourth-moment}

% WPM-BLOCK:B0264
The quartic data are expressed through the real mode-response
amplitudes \(v_k(\hat y)=v_{k,\mu}\hat y^\mu\).  For
\(N_k=L/2\) positive NS momenta, define the channel average
\(\langle f\rangle_k=N_k^{-1}\sum_{k>0}f_k\).  Then
% WPM-BLOCK:B0265
\begin{equation}
  \KF(\hat y)
  =\frac{\sum_{k>0}v_k(\hat y)^4}{\left[\sum_{k>0}v_k(\hat y)^2\right]^2}
  =\frac{1}{N_k}
   \frac{\langle v(\hat y)^4\rangle_k}{\langle v(\hat y)^2\rangle_k^2}.
  \label{eq:KF-kurtosis}
\end{equation}
% WPM-BLOCK:B0266
Thus \(N_k\KF\) is the fourth-to-second-moment ratio of this finite set of
resolved amplitudes.  The channel average is deterministic and assumes no
probability distribution over the amplitudes.  The QGT fixes the summed
quadratic response, whereas \(\Pfour\) retains fourth-moment information about
how that response is distributed among the chosen channels.  A large value of
\(\KF\) means that a few modes carry most of the response, while a small value
means that the same total metric weight is spread over more modes.

% WPM-BLOCK:B0267
\begin{figure*}[t]
  \centering
  \includegraphics[width=\linewidth]{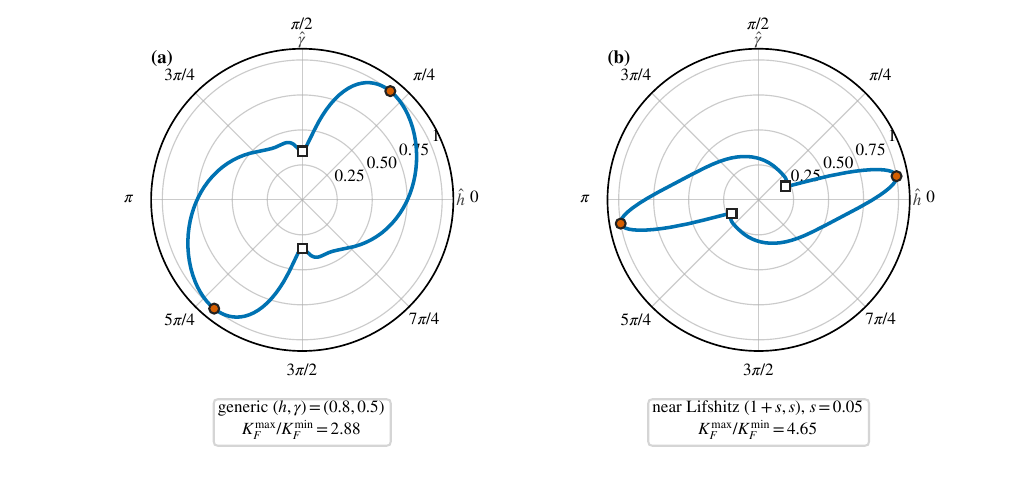}
  \caption{Directional channel-concentration profiles in the $XY$ parameter
    plane at $L=2000$, using even-parity NS momentum blocks.
    Here $\KF=\sum_k\pi_k^2$ is the channel concentration and
    $\Neff=1/\KF$ its effective number of channels. The radial coordinate is
    $\KF(\hat y(\varphi))/\KF^{\max}$, with
    $\hat y(\varphi)=(\cos\varphi,\sin\varphi)$ in $(h,\gamma)$ coordinates.
    Circles and squares mark the maximum and minimum, respectively.
    The unit direction uses equal
    displacement amplitude in the dimensionless bare couplings $h$ and
    $\gam$ of Eq.~\eqref{eq:xy}; it is a quench-protocol convention rather
    than a coordinate-invariant circle.  Panel (a), at the generic gapped
    point $(h,\gam)=(0.8,0.5)$, has
    $0.00163\le\KF\le0.00467$ and
    $214\lesssim\Neff\lesssim615$.  Panel (b), at the finite-size-regulated
    point $(h,\gam)=(1+s,s)$ with $s=0.05$, has
    $0.00189\le\KF\le0.00877$ and
    $114\lesssim\Neff\lesssim530$.  Thus the normalized curves compare
    directional shape, while the quoted ranges restore their absolute
    concentration scale. The displayed maximum-to-minimum ratios are about
    \(2.88\) in panel (a) and \(4.65\) in panel (b). Both panels retain
    hundreds of effective channels. Their directional variation shows that
    changing the tangent redistributes response within the fixed resolution
    rather than merely rescaling its total magnitude.
    The exact Lifshitz point is avoided because the field-direction metric
    weight vanishes at $\gam=0$.}
  \label{fig:xy-directional-profiles}
\end{figure*}

% WPM-BLOCK:B0268
The quartic tensor, together with \(\Ptwo(\varphi)\), determines the
\(\KF(\varphi)\) profile plotted in
Fig.~\ref{fig:xy-directional-profiles}.  Together with the Lifshitz
coordinate-direction results, the full tangent profiles establish operator
selectivity at infrared and finite sizes: the approach direction changes
carrier composition.  The tensor represents the same local fourth-moment data,
not a separate observable.

% WPM-BLOCK:B0368
\section{TFIM critical and scaling derivations}
% WPM-BLOCK:B0369
\label{app:tfim-scaling}

% WPM-BLOCK:B0370
This appendix collects the exact half-grid power sums and the scaling-window
derivation used in Sec.~\ref{sec:tfim}.

% WPM-BLOCK:B0371
\subsection{Exact NS half-grid sums}

% WPM-BLOCK:B0372
The exact sums in this subsection complete the finite-size NS sequence in
Sec.~\ref{sec:tfim:twothirds}.  The following scaling-window derivation instead
supports the crossover analysis in Sec.~\ref{sec:tfim:scaling}.

Differentiating the standard cotangent root identity
$\sum_{m=0}^{L-1}\cot(z+m\pi/L)=L\cot(Lz)$ gives
% WPM-BLOCK:B0373
\begin{equation}
  \sum_{m=0}^{L-1}\csc^2\left(z+\frac{m\pi}{L}\right)
  =L^2\csc^2(Lz),
  \label{eq:csc-root-identity}
\end{equation}
% WPM-BLOCK:B0374
and generates all required critical sums.  The coefficients in its higher
derivatives follow from
\begin{equation*}
  \begin{aligned}
  \frac{\d^2}{\d t^2}\csc^{2r}t
  ={}&2r(2r+1)\csc^{2r+2}t\\
     &-4r^2\csc^{2r}t.
  \end{aligned}
\end{equation*}
Using this identity first at \(r=1\), then at \(r=1,2\), generates the
second and fourth derivatives.  Let
$\mathcal C_{2r}(z)=\sum_{m=0}^{L-1}\csc^{2r}(z+m\pi/L)$.  At
$z=\pi/(2L)$, Eq.~\eqref{eq:csc-root-identity} and its second and fourth
derivatives give
% WPM-BLOCK:B0375
\begin{align}
  \mathcal C_2&=L^2,
  \nonumber\\
  6\mathcal C_4-4\mathcal C_2&=2L^4,
  \nonumber\\
  120\mathcal C_6-120\mathcal C_4+16\mathcal C_2&=16L^6.
\end{align}
% WPM-BLOCK:B0376
Thus
% WPM-BLOCK:B0377
\begin{equation}
  \mathcal C_4=\frac{L^4+2L^2}{3},
  \qquad
  \mathcal C_6=\frac{2L^6+5L^4+8L^2}{15}.
\end{equation}
% WPM-BLOCK:B0378
Reflection symmetry halves the full grid, and
$\cot^2t=\csc^2t-1$ converts these results into
\begin{align}
  \sum_{n=0}^{L/2-1}\cot^{2}\frac{(2n+1)\pi}{2L}
  &=\frac{L(L-1)}{2},
  \nonumber\\
  \sum_{n=0}^{L/2-1}\cot^{4}\frac{(2n+1)\pi}{2L}
  &=\frac{L(L-1)(L^{2}+L-3)}{6}.
  \label{eq:tfim-critical-cot-sums}
\end{align}
The same construction at sixth order gives
% WPM-BLOCK:B0379
\begin{equation}
  \begin{aligned}
  \sum_{n=0}^{L/2-1}
  \cot^6\frac{(2n+1)\pi}{2L}
  &=\frac{L(L-1)}{30}\\
  &\quad\times
  (2L^4+2L^3-8L^2-8L+15).
  \end{aligned}
  \label{eq:tfim-critical-cot6-sum}
\end{equation}

For later use on the nonzero Ramond integer ladder, take \(z\to0\) in
Eq.~\eqref{eq:csc-root-identity} and subtract the singular \(m=0\) term.
The identity and its second derivative give
\begin{align*}
  \sum_{m=1}^{L-1}\csc^2\frac{m\pi}{L}
  &=\frac{L^2-1}{3},\\
  \sum_{m=1}^{L-1}\csc^4\frac{m\pi}{L}
  &=\frac{(L^2-1)(L^2+11)}{45}.
\end{align*}
Using \(\cot^2t=\csc^2t-1\), expanding \(\cot^4t\), and applying reflection
symmetry then yields
\begin{align*}
  \sum_{n=1}^{L/2-1}\cot^2\frac{\pi n}{L}
  &=\frac{(L-1)(L-2)}{6},\\
  \sum_{n=1}^{L/2-1}\cot^4\frac{\pi n}{L}
  &=\frac{(L-2)(L-1)(L^2+3L-13)}{90}.
\end{align*}

% WPM-BLOCK:B0380
\subsection{Critical scaling window}\label{app:tfim-scaling-window}

% WPM-BLOCK:B0381
To derive the finite-size rounding in Sec.~\ref{sec:tfim:scaling}, set the
local TFIM detuning to $m=h-\hc=h-1$ and take
\begin{equation*}
  L\to\infty,\qquad m\to0,\qquad
  \mu=\frac{mL}{\pi}\ \text{fixed}.
\end{equation*}
This keeps the detuning relative to the lowest NS momentum fixed.  The
low-momentum expansion is
% WPM-BLOCK:B0391
\begin{equation}
  \Lk^2=(h-\cos k)^2+\sin^2k
  =m^2+k^2+O(mk^2,k^4).
\end{equation}
% WPM-BLOCK:B0392
The low-momentum weight becomes
% WPM-BLOCK:B0393
\begin{equation}
  \xk(h)\simeq\frac{k^2}{4(k^2+m^2)^2}.
\end{equation}
% WPM-BLOCK:B0394
With $k_n=\kappa\pi/L$, $\kappa=1,3,5,\ldots$, and
$\mu=mL/\pi$, this becomes
% WPM-BLOCK:B0395
\begin{equation}
  x_{\kappa}\simeq\frac{L^2}{4\pi^2}
  \frac{\kappa^2}{(\kappa^2+\mu^2)^2}.
\end{equation}
% WPM-BLOCK:B0396
To control the sum over the growing grid, define the scaled exact weights
\(b_\kappa^{(L)}=4\pi^2x_\kappa/L^2\) for odd \(\kappa<L\), and set them
to zero for odd \(\kappa\ge L\).  The exact identity
\(\Lk^2=m^2+4h\sin^2(k/2)\), together with
\(\sin(k/2)\ge k/\pi=\kappa/L\), implies
\begin{equation*}
  0\le b_\kappa^{(L)}
  \le\frac{\pi^2}{4h^2\kappa^2}
  \le\frac{\pi^2}{\kappa^2},
  \qquad h\ge\frac12.
\end{equation*}
At fixed \(\mu\), the last condition holds for all sufficiently large
\(L\).  For each fixed \(\kappa\), \(b_\kappa^{(L)}\) approaches
\(\kappa^2/(\kappa^2+\mu^2)^2\), while the bound and its square are
summable.  Dominated convergence therefore applies to both moments of the
scaled weights.  The discrete infrared ladder survives this limit, and
the common prefactor cancels in $\KF$, leaving
% WPM-BLOCK:B0397
\begin{equation}
  \KF\longrightarrow
  \frac{\sum_{\kappa=1,3,\ldots}
    \kappa^4/(\kappa^2+\mu^2)^4}
  {\left[\sum_{\kappa=1,3,\ldots}
    \kappa^2/(\kappa^2+\mu^2)^2\right]^2}
  =\Phisc(\mu),
\end{equation}
% WPM-BLOCK:B0398
which is Eq.~\eqref{eq:Phi}.

% WPM-BLOCK:B0399
\section{RG origin of the infrared envelope exponent}
% WPM-BLOCK:B0400
\label{app:rg-envelope}

% WPM-BLOCK:B0401
The following zero-temperature argument separates a general response-scale
result from a conditional envelope result.  Infrared dominance gives the
singular scaling of \(\Ptwo\).  Identifying \(p=2y_{\mathcal O}\) further assumes one participating soft
ladder in one dimension, resolved on the NS odd-momentum grid. The same
infrared power must control both the finite-size prefactor and the channel
envelope.  Lifshitz two-scale
responses, other spin structures, multiple cones, and interacting
level-projector resolutions lie outside that identification.  Under these
conditions the argument establishes the contributions behind
Eqs.~\eqref{eq:P2-RG-scaling-main} and \eqref{eq:p-yo-main} in
Sec.~\ref{sec:tfim:envelope}.  Let
\(H(\lambda)=H_c+\lambda\int \d^d x\,\mathcal O(x)\), and specialize the
general response identity to the coordinate tangent \(y=\partial_\lambda\):
% WPM-BLOCK:B0752
\begin{align*}
  V_\lambda\equiv\partial_\lambda H
  &=\int \d^d x\,\mathcal O(x),\\
  V_\lambda(\tau)
  &=e^{\tau(H-E_0)}V_\lambda e^{-\tau(H-E_0)}.
\end{align*}
Equation~\eqref{eq:app-cft-response-norm} gives the finite-size identity
\(\Ptwo=\int_0^\infty \d\tau\,\tau
\langle V_\lambda(\tau)V_\lambda(0)\rangle_c\).
Its projector subtraction, spectral insertion, and Laplace-moment derivation
are given in Sec.~\ref{sec:cft-bridge} and
Appendix~\ref{app:cft-cylinder-derivation}; here we use the identity only to
separate the translation-invariant response into ultraviolet and infrared
parts.
% WPM-BLOCK:B0403
Using \(V_\lambda=\int \d^d x\,\mathcal O(x)\), the global connected
correlator is
% WPM-BLOCK:B0758
\begin{align*}
  \langle V_\lambda(\tau)V_\lambda(0)\rangle_c
  &=\int \d^d x\int \d^d x'\,
    \langle\mathcal O(x,\tau)\mathcal O(x',0)\rangle_c\\
  &=L^d\int \d^d r\,
    \langle\mathcal O(r,\tau)\mathcal O(0,0)\rangle_c .
\end{align*}
% WPM-BLOCK:B0759
The second equality uses translation invariance: one spatial integral gives the
center-of-mass volume \(L^d\), while \(r=x-x'\) is the relative coordinate.
At fixed microscopic regulator, split the remaining relative-coordinate and
Euclidean-time integration into short- and long-distance domains.  The
short-distance domain has an \(L\)-independent range and gives a nonuniversal
constant per center-of-mass position.  It therefore contributes
\(\Ptwo^{\mathrm{UV}}=\mathcal A_{P_2}^{\rm UV}L^d\), which becomes
\(\mathcal A_{P_2}^{\rm UV}L\) in one dimension.
After separating this analytic background, the long-distance domain supplies
the infrared-singular part in the declared critical scaling limit.  It obeys
% WPM-BLOCK:B0404
\begin{equation}
  \Ptwo
  \sim L^d\int^{L^z} \d\tau\,\tau
       \int^{L} \d^d r\,\langle \mathcal O(r,\tau)\mathcal O(0,0)\rangle_c.
  \label{eq:app-P2-scaling-integral}
\end{equation}
% WPM-BLOCK:B0405
Under the anisotropic scaling \(r\to br\), \(\tau\to b^z\tau\), and
\(\mathcal O\to b^{-\Delta_{\mathcal O}}\mathcal O\), the integral scales as
% WPM-BLOCK:B0406
\begin{equation}
  \Ptwo\sim L^{d}\,L^{2z+d-2\Delta_{\mathcal O}}
  =L^{2(d+z-\Delta_{\mathcal O})}
  \equiv L^{2y_{\mathcal O}},
  \label{eq:app-P2-scaling}
\end{equation}
% WPM-BLOCK:B0407
where \(y_{\mathcal O}=d+z-\Delta_{\mathcal O}\).  This power describes the leading total response
when the infrared integral is divergent; otherwise analytic or ultraviolet
background terms can compete.  For a tuning perturbation, \(y_{\mathcal O}=1/\nu\), so
the usual infrared exponent \(2/\nu\) for the fidelity susceptibility is
recovered.

For the Potts thermal response, the extensive ultraviolet term is smaller
than the singular \(L^{12/5}\) response by \(L^{-7/5}\). Its propagation
through the normalized moments gives the denominator-side correction source
in Eq.~\eqref{eq:potts-KF-correction-hierarchy}, conditional on the grouped
matching and moment expansions stated there. This denominator scaling alone
does not determine the fine-level concentration or its corrections.

% WPM-BLOCK:B0408
Under the additional assumption of a one-dimensional NS odd-ladder resolution
with one participating soft ladder, represent the infrared channel weights by
% WPM-BLOCK:B0409
\begin{equation}
  x_n\sim \frac{L^p}{(2n+1)^p},
  \qquad p=2y_{\mathcal O}.
  \label{eq:app-soft-ladder-weight}
\end{equation}
% WPM-BLOCK:B0410
For a normalizable full limiting ladder, \(p>1\),
\(\Ptwo=\sum_nx_n\sim L^p\) and
\(\Pfour=\sum_nx_n^2\sim L^{2p}\), so the ratio \(\KF=\Pfour/\Ptwo^2\) has a
finite critical limit.  This identification requires the same power \(p\) to
control both the finite-size prefactor and the infrared momentum envelope.
It does not apply to two-scale responses such as the Lifshitz field direction,
where \(\Ptwo^{(h)}\sim L^4\) at fixed \(w>0\) but the limiting normalized envelope is
\(\kappa^{-6}\).  It also requires the NS odd-momentum ladder: different
boundary sectors, zero-mode structures, multiple inequivalent cones, or
interacting-system channel decompositions can change the discrete sum.

\subsection{Conditional Potts moment-correction algebra}
\label{app:potts-correction-algebra}
This algebra supports the grouped correction hierarchy in
Eq.~\eqref{eq:potts-KF-correction-hierarchy}, conditional on the matching
criterion of Eq.~\eqref{eq:cft-grouped-matching}.  In addition to that
matching, assume
\begin{align}
  \frac{\Ptwo(L)}{A_2L^{12/5}}
  &=1+c_{2,\mathrm{irr}}L^{-4/5}
     +c_{2,\mathrm{UV}}L^{-7/5}+\cdots,
  \notag\\
  \frac{P_4^{\rm group}(L)}{A_4^{\rm group}L^{24/5}}
  &=1+c_{4,\mathrm{irr}}L^{-4/5}
     +c_{4,7/5}L^{-7/5}+\cdots.
  \label{eq:potts-moment-correction-propagation}
\end{align}
Here \(c_{4,7/5}\) allows a fourth-moment contribution at the same relative
order if one is present.  Expanding \(K_F^{\rm group}=P_4^{\rm group}/\Ptwo^2\) gives
the form of Eq.~\eqref{eq:potts-KF-correction-hierarchy},
where \(K_{F,\mathrm{Potts}}^{\mathrm{CFT}}=A_4^{\rm group}/A_2^2\),
\(d_{\mathrm{irr}}=c_{4,\mathrm{irr}}-2c_{2,\mathrm{irr}}\), and
\(d_{7/5}=c_{4,7/5}-2c_{2,\mathrm{UV}}\), with
\(a_{\mathrm{irr}}=K_{F,\mathrm{Potts}}^{\mathrm{CFT}}d_{\mathrm{irr}}\)
and \(b_{\mathrm{UV}}=K_{F,\mathrm{Potts}}^{\mathrm{CFT}}d_{7/5}\).
Products of the \(L^{-4/5}\) correction first enter at relative order
\(L^{-8/5}\), beyond the two displayed powers.
The two relative powers are therefore candidate corrections with
nonuniversal amplitudes, either of which can cancel.  This algebra supplies
no intercept or correction theorem for the fine-level observable.

% WPM-BLOCK:B0424
\section{Boundary-sector dependence and fixed-parity Ramond response}
% WPM-BLOCK:B0425
\label{app:ramond}

% WPM-BLOCK:B0426
The periodic spin chain decomposes as
\(H=H_{\rm NS}P_+ + H_{\rm R}P_-\), where \(P_\pm\) project onto even and
odd fermion parity.  For even \(L\ge4\), \(\gam=1\), and positive \(h\)
near \(h_c=1\), the lowest state \(\lvert\psi_-\rangle\) of the restricted
Hamiltonian \(H_-=H|_{\mathcal H_-}\) is nondegenerate.  Its \(k=0\) mode
is occupied and its \(k=\pi\) mode is empty~\cite{DamskiRams2014}.
Changing the zero-mode occupation alone changes parity, so the degeneracy
of the unprojected Ramond Fock space is not a degeneracy within
\(\mathcal H_-\).  The nondegenerate-state response is well defined there:
\begin{equation}
 \lvert D_h\psi_-\rangle
 =-(H_--E_-)^{-1}_{\perp}
 (P_--\lvert\psi_-\rangle\langle\psi_-\rvert)
 (\partial_h H)\lvert\psi_-\rangle .
 \label{eq:ramond-restricted-response}
\end{equation}
The reduced inverse acts only on the complement of the ray within
\(\mathcal H_-\).  Both unpaired occupations remain fixed locally in \(h\)
and hence contribute no state derivative.  The responding channels are the
paired modes \(0<k<\pi\); no physical response channel has been discarded.
This sector lowest state is not the unrestricted periodic spin-chain ground
state, which lies in the even NS sector.  It supplies the boundary-sector
comparison in Secs.~\ref{sec:channel:scope}, \ref{sec:tfim:weights},
\ref{sec:tfim:majorana}, and \ref{sec:tfim:twothirds}, while the NS theorem remains unchanged.

% WPM-BLOCK:B0427
To see this explicitly, evaluate the critical TFIM lattice weight from
Eq.~\eqref{eq:xk-crit-exact},
% WPM-BLOCK:B0428
\begin{equation}
  x_k(\hc)=\frac{1}{16}\cot^2\frac{k}{2}
\end{equation}
% WPM-BLOCK:B0429
on the nonzero Ramond integer ladder
\(k_n=2\pi n/L\), \(n=1,\ldots,L/2-1\), of these responding pair blocks.  The
standard trigonometric power sums~\cite{ShevelevMoses2012} give
% WPM-BLOCK:B0430
\begin{align}
  \sum_{n=1}^{L/2-1}\cot^2\frac{\pi n}{L}
  &=\frac{(L-1)(L-2)}{6},\\
  \sum_{n=1}^{L/2-1}\cot^4\frac{\pi n}{L}
  &=\frac{(L-2)(L-1)(L^2+3L-13)}{90}.
\end{align}
% WPM-BLOCK:B0431
Appendix~\ref{app:tfim-scaling} derives these trigonometric identities from
Eq.~\eqref{eq:csc-root-identity} by taking $z\to0$, subtracting the
singular \(m=0\) term, converting \(\csc^2\) to \(\cot^2+1\), and using
reflection to retain the positive half-ladder.  The absent unpaired-mode
contribution follows from the fixed occupations in the physical sector, not
from a regularization of a divergent channel weight.  Since
$\Ptwo^{\mathrm{R}}=16^{-1}\sum_n\cot^2(\pi n/L)$ and
$\Pfour^{\mathrm{R}}=256^{-1}\sum_n\cot^4(\pi n/L)$, the channel moments are
% WPM-BLOCK:B0432
\begin{align}
  \Ptwo^{\mathrm{R}}(\hc,L)&=\frac{(L-1)(L-2)}{96},\\
  \Pfour^{\mathrm{R}}(\hc,L)&=\frac{(L-2)(L-1)(L^2+3L-13)}{23040},
\end{align}
% WPM-BLOCK:B0433
and
% WPM-BLOCK:B0434
\begin{equation}
  \KF^{\mathrm{R}}(\hc,L)
  =\frac{2(L^2+3L-13)}{5(L-1)(L-2)}
  =\frac{2}{5}+\frac{12}{5L}+O(L^{-2}).
  \label{eq:ramond-KF}
\end{equation}
% WPM-BLOCK:B0435
The thermodynamic value is equivalently
% WPM-BLOCK:B0436
\begin{equation}
  \KF^{\mathrm{R},\mathrm{env}}
  =\frac{\zeta(4)}{\zeta(2)^2}=\frac25,
\end{equation}
% WPM-BLOCK:B0437
because the nonzero Ramond ladder uses ordinary positive integers rather than
positive odd integers.  This diagnostic shows why the boundary sector is part
of the channel resolution; it does not replace the NS theorem in the main
text.

% WPM-BLOCK:B0438
\section{Exact XX pairing theorem and momentum-ladder arithmetic}
% WPM-BLOCK:B0439
\label{app:xx-pairing-theorem}

% WPM-BLOCK:B0440
The derivation below proves
Eqs.~\eqref{eq:xx-pairing-moments-main}--\eqref{eq:xx-half-size-identity}.
It also separates the nondegenerate
\(L\equiv0\pmod4\) theorem from the singular \(L\equiv2\pmod4\) sequence and
states the arithmetic conditions behind the incommensurate oscillations.

% WPM-BLOCK:B0441
\subsection{Pairing-direction weight on the XX line}

We derive the pairing-direction weight used in the XX row of
Eq.~\eqref{eq:paired-theorem-summary}, before the half-size folding in
Appendix~\ref{app:xx-folding-proof}.

% WPM-BLOCK:B0442
For the two-parameter block in Eq.~\eqref{eq:xy-bdg-xx-theorem}, differentiation
at \(\gam=0\) gives
% WPM-BLOCK:B0443
\begin{equation}
  x_k^{(\gam)}(h)
  =\left.\frac14\left(\partial_\gam\theta_k\right)^2\right|_{\gam=0}
  =\frac{\sin^2 k}{4(h-\cos k)^2}.
  \label{eq:xx-pairing-weight-app}
\end{equation}
% WPM-BLOCK:B0444
For $\abs{h}<1$, the gap closes at the interior soft momenta
\(\pm k_{\star}\), where \(k_{\star}=\arccos h\).  Near the positive soft
momentum,
% WPM-BLOCK:B0445
\begin{equation}
  x_k^{(\gam)}(h)
  =\frac{1}{4(k-k_{\star})^2}\left[1+O(\abs{k-k_{\star}})\right].
  \label{eq:xx-pairing-local-envelope}
\end{equation}
% WPM-BLOCK:B0446
Thus the two sides of the interior soft point carry equal \(p=2\) envelopes.
At half filling, \(h=0\) and \(k_{\star}=\pi/2\), Eq.~\eqref{eq:xx-pairing-weight-app}
reduces to Eq.~\eqref{eq:xx-pairing-weight-main}.

% WPM-BLOCK:B0447
\subsection{Complementary-angle folding and the half-size identity}\label{app:xx-folding-proof}

The folding below establishes the weight-level half-size identity in
Eq.~\eqref{eq:xx-half-size-identity}, its R\'enyi participation extension in
Eq.~\eqref{eq:xx-half-size-all-alpha}, and the corresponding relation in the
paired theorem, Eq.~\eqref{eq:paired-theorem-summary}.
% WPM-BLOCK:B0448
\label{app:xx-pairing-folding}

% WPM-BLOCK:B0449
Let \(L\equiv0\pmod4\) and write \(M=L/2\), so \(M\) is even.  Reflection about
\(\pi/2\) pairs the \(L/2\) positive NS momenta according to
\(k\leftrightarrow\pi-k\), with equal values of \(\tan^2 k\).  Each pair
contributes $2\times\tfrac14\tan^2k$ to $\Ptwo$ and
$2\times\tfrac1{16}\tan^4k$ to $\Pfour$.  Folding the sum onto
\(0<k<\pi/2\) therefore gives
% WPM-BLOCK:B0450
\begin{align}
  \Ptwo^{\mathrm{XX},(\gam)}
  &=\frac12\sum_{j=0}^{M/2-1}
    \tan^2\frac{(2j+1)\pi}{2M},
  \nonumber\\
  \Pfour^{\mathrm{XX},(\gam)}
  &=\frac18\sum_{j=0}^{M/2-1}
    \tan^4\frac{(2j+1)\pi}{2M}.
  \label{eq:xx-folded-moments}
\end{align}
% WPM-BLOCK:B0451
The complementary-angle map
% WPM-BLOCK:B0452
\begin{equation}
  \tan\frac{(2j+1)\pi}{2M}
  =\cot\frac{(M-2j-1)\pi}{2M}
  \label{eq:xx-complementary-map}
\end{equation}
% WPM-BLOCK:B0453
is a bijection of the positive odd numerators because \(M\) is even.  The
remaining sums are therefore precisely the critical TFIM cotangent sums in
Eq.~\eqref{eq:tfim-critical-cot-sums}, evaluated at size \(M\):
% WPM-BLOCK:B0454
\begin{align}
  \sum_{j=0}^{M/2-1}\tan^2\frac{(2j+1)\pi}{2M}
  &=\frac{M(M-1)}{2},
  \nonumber\\
  \sum_{j=0}^{M/2-1}\tan^4\frac{(2j+1)\pi}{2M}
  &=\frac{M(M-1)(M^2+M-3)}{6}.
  \label{eq:xx-tangent-power-sums}
\end{align}
% WPM-BLOCK:B0455
Substitution into Eq.~\eqref{eq:xx-folded-moments} yields
% WPM-BLOCK:B0456
\begin{align}
  \Ptwo^{\mathrm{XX},(\gam)}
  &=\frac{M(M-1)}{4}=\frac{L(L-2)}{16},
  \nonumber\\
  \Pfour^{\mathrm{XX},(\gam)}
  &=\frac{M(M-1)(M^2+M-3)}{48}
  \nonumber\\
  &=\frac{L(L-2)(L^2+2L-12)}{768}.
  \label{eq:xx-pairing-moments-app}
\end{align}
% WPM-BLOCK:B0457
Their ratio is
% WPM-BLOCK:B0458
\begin{equation}
  \KF^{\mathrm{XX},(\gam)}(0,L)
  =\frac{M^2+M-3}{3M(M-1)}
  =\frac13\frac{L^2+2L-12}{L(L-2)}.
  \label{eq:xx-pairing-KF-app}
\end{equation}
% WPM-BLOCK:B0459
Comparison with Eq.~\eqref{eq:twothirds-derivation} at size \(M\) proves the
finite-size identity
% WPM-BLOCK:B0460
\begin{equation}
  \KF^{\mathrm{XX},(\gam)}(0,L)
  =\frac12\KF^{\mathrm{TFIM}}(\hc,L/2).
  \label{eq:xx-half-size-identity-app}
\end{equation}
% WPM-BLOCK:B0461
It also fixes the leading correction without an additional fit:
\(\KF^{XX,(\gam)}(0,L)=1/3+4/(3L)+O(L^{-2})\).  To make the stronger
weight-level correspondence explicit, write the complementary-angle index as
\(j'=M/2-1-j\).  The two XX momenta labeled by \(s=\pm\) lie on opposite
sides of \(k_\star=\pi/2\); each is already a resolved \((k,-k)\) block.
Comparing their \(\tfrac14\tan^2k\) weights with the
\(\tfrac1{16}\cot^2\) TFIM weights in Eq.~\eqref{eq:xk-crit-exact} gives
\begin{equation*}
  \begin{aligned}
  x^{XX}_{s,j}(0,L)&=4x^{\rm TFIM}_{j'}(h_c,M),\\
  \Ptwo^{XX,(\gam)}(0,L)&=8\Ptwo^{\rm TFIM}(h_c,M).
  \end{aligned}
\end{equation*}
The factor eight counts the two copies of each fourfold raw weight.
Relabeling each XX branch by the TFIM index \(j'\), and then denoting that
index by \(j\), normalization gives
% WPM-BLOCK:B0639
\begin{equation}
  \pi^{XX}_{s,j}(0,L)
  =\frac12\pi^{\rm TFIM}_{j}(h_c,L/2),
  \qquad s=\pm.
  \label{eq:xx-distribution-duplication-app}
\end{equation}
% WPM-BLOCK:B0640
Equation~\eqref{eq:xx-distribution-duplication-app} proves the finite-size
product distribution used in Eq.~\eqref{eq:xx-half-size-all-alpha}; no
thermodynamic or power-law convergence assumption enters this step.

% WPM-BLOCK:B0462
\subsection{Singular \texorpdfstring{$L\equiv2\pmod4$}{L = 2 (mod 4)} sequence}
% WPM-BLOCK:B0463
\label{app:xx-pairing-zero-mode}

We derive the singular-sequence distinction in
Sec.~\ref{sec:xy:xx-theorem}.

% WPM-BLOCK:B0464
If \(L\equiv2\pmod4\), the NS grid contains \(k=\pi/2\).  At \(h=\gam=0\)
this block is gapless at finite size, the ground-state ray is degenerate, and
Eq.~\eqref{eq:xx-pairing-weight-app} diverges.  The nondegenerate-state QGT and
its CC are therefore not defined at that point.

% WPM-BLOCK:B0465
For comparison with the Ramond arithmetic only, one may remove the zero-gap
block before forming the moments.  With \(M=L/2\) now odd, the complementary
map sends the remaining tangent ladder to the nonzero integer cotangent ladder.
The corresponding zero-mode-excluded moments are
% WPM-BLOCK:B0466
\begin{align}
  \left.\Ptwo^{\mathrm{XX},(\gam)}\right|_{k\ne k_\star}
  &=\frac{(M-1)(M-2)}{12},
  \nonumber\\
  \left.\Pfour^{\mathrm{XX},(\gam)}\right|_{k\ne k_\star}
  &=\frac{(M-1)(M-2)(M^2+3M-13)}{720}.
\end{align}
% WPM-BLOCK:B0467
Their regularized ratio is therefore
% WPM-BLOCK:B0468
\begin{equation}
  \begin{aligned}
  \left.\KF^{\mathrm{XX},(\gam)}(0,L)\right|_{k\ne k_{\star}}
  &=\frac{M^2+3M-13}{5(M-1)(M-2)}\\
  &=\frac15+\frac{12}{5L}+O(L^{-2}).
  \end{aligned}
  \label{eq:xx-zero-mode-excluded}
\end{equation}
% WPM-BLOCK:B0469
Substituting $M=L/2$ into Eq.~\eqref{eq:ramond-KF} shows explicitly that this
is one half of the corresponding half-size Ramond polynomial.  Here \(M\)
is odd; this arithmetic substitution does not identify the physical
even-length odd-parity state defined in Appendix~\ref{app:ramond}.
The zero-gap XX block has two even-parity pair states, so fixing total parity
does not remove its degeneracy.
Equation~\eqref{eq:xx-zero-mode-excluded} is a
regularized projector convention, not a second value of the original QGT at a
nondegenerate ground state.

% WPM-BLOCK:B0470
\subsection{Commensurate and incommensurate offsets}
% WPM-BLOCK:B0471
\label{app:xx-pairing-offset}

We derive the offset qualification of the half-filled result in
Sec.~\ref{sec:xy:xx-theorem}.

% WPM-BLOCK:B0472
For the general interior soft point, with the half-filled case treated above,
define the NS grid spacing \(\Delta k=2\pi/L\) and the fractional offset
% WPM-BLOCK:B0473
\begin{equation}
  \eta_L
  =\left\{\frac{k_{\star}}{\Delta k}-\frac12\right\}
  =\left\{\frac{Lk_{\star}}{2\pi}-\frac12\right\},
  \qquad 0\le\eta_L<1,
  \label{eq:xx-offset-definition}
\end{equation}
% WPM-BLOCK:B0474
where braces denote the fractional part.  For $n=0,1,\ldots$, the two sides
of the soft point have distances \((n+\eta_L)\Delta k\) and
\((n+1-\eta_L)\Delta k\).  At sizes with \(\eta_L=0\), a grid point
coincides with \(k_{\star}\), so the nondegenerate-state QGT is singular.  For
nonsingular sizes, substituting these distances into
Eq.~\eqref{eq:xx-pairing-local-envelope} gives the two leading infrared
ladders and the common amplitude $C_L=1/(4\Delta k^2)$:
% WPM-BLOCK:B0475
\begin{equation}
  \begin{aligned}
  x_{n,+}^{\mathrm{IR}}&=C_L(n+\eta_L)^{-2},\\
  x_{n,-}^{\mathrm{IR}}&=C_L(n+1-\eta_L)^{-2},\\
  C_L&=\frac{1}{4\Delta k^2}.
  \end{aligned}
  \label{eq:xx-offset-ladder-weights}
\end{equation}
% WPM-BLOCK:B0476
For \(s>1\) and \(a>0\), define the Hurwitz zeta function by
% WPM-BLOCK:B0477
\begin{equation}
  \zeta(s,a)=\sum_{n=0}^{\infty}(n+a)^{-s}.
  \label{eq:xx-hurwitz-definition}
\end{equation}
% WPM-BLOCK:B0478
Summing both sides of the soft point then gives
% WPM-BLOCK:B0479
\begin{align}
  \Ptwo^{\mathrm{IR}}
  &=C_L\left[\zeta(2,\eta_L)+\zeta(2,1-\eta_L)\right],
  \nonumber\\
  \Pfour^{\mathrm{IR}}
  &=C_L^2\left[\zeta(4,\eta_L)+\zeta(4,1-\eta_L)\right].
  \label{eq:xx-offset-ir-moments}
\end{align}
% WPM-BLOCK:B0480
The common factor \(C_L\) cancels from \(\Pfour^{\mathrm{IR}}/(\Ptwo^{\mathrm{IR}})^2\).
For \(0<\eta_L<1\), the polygamma reflection formula gives
% WPM-BLOCK:B0481
\begin{align}
  \zeta(2,\eta_L)+\zeta(2,1-\eta_L)
  &=\frac{\pi^2}{\sin^2(\pi\eta_L)},
  \nonumber\\
  \zeta(4,\eta_L)+\zeta(4,1-\eta_L)
  &=\frac{\pi^4[2+\cos(2\pi\eta_L)]}
  {3\sin^4(\pi\eta_L)}.
  \label{eq:xx-hurwitz-reflection}
\end{align}
% WPM-BLOCK:B0482
Consequently, the leading \(p=2\) envelope is
% WPM-BLOCK:B0483
\begin{equation}
  \begin{aligned}
    \KF^{\mathrm{XX},(\gam),\mathrm{env}}(\eta_L)
    &=\frac{\zeta(4,\eta_L)+\zeta(4,1-\eta_L)}
    {\left[\zeta(2,\eta_L)+\zeta(2,1-\eta_L)\right]^2}\\
    &=\frac{2+\cos(2\pi\eta_L)}{3},
    \qquad 0<\eta_L<1.
  \end{aligned}
  \label{eq:xx-hurwitz-offset}
\end{equation}
% WPM-BLOCK:B0484
The centered ladder has \(\eta_L=1/2\) and gives \(1/3\), while
\(\KF^{\mathrm{XX},(\gam),\mathrm{env}}(\eta_L)\to1\) as
\(\eta_L\to0^+\) or \(1^-\), when
one near-resonant channel dominates.  The latter is a punctured envelope
limit, not a value assigned to the singular \(\eta_L=0\) finite-size QGT.

% WPM-BLOCK:B0485
Along even sizes $L=2m$, Eq.~\eqref{eq:xx-offset-definition} gives the
recurrence
$\eta_{L+2}=\{\eta_L+k_{\star}/\pi\}$, so the offset advances by
$k_{\star}/\pi$ modulo one.  If
\(k_{\star}/\pi\) is irrational, the offsets are dense in \([0,1]\), so the full
finite-size sequence generically has no unique limit and samples the continuous
range \(1/3\le\KF<1\) through subsequences.  If \(k_{\star}/\pi\) is rational, the
offset is periodic; special subsequences may be centered, may hit an exact zero
mode, or may cycle through a finite set of concentrations.  Thus the
oscillation is controlled momentum-ladder arithmetic rather than a failure of
the numerical sum.

% WPM-BLOCK:B0486
\section{Proofs for equivalent resolved ladders}
% WPM-BLOCK:B0641
\label{app:multicone}

This appendix establishes the product and asymptotic-factorization conditions
for the multiple-ladder result in Sec.~\ref{sec:tfim:multiple-ladders}.
Its entropy-domain arguments extend the exact XX relation,
Eq.~\eqref{eq:xx-half-size-all-alpha}, to the R\'enyi participation family
of Sec.~\ref{sec:channel:three}. The amplitude-weighted formula also treats
separately resolved co-leading primaries following
Eq.~\eqref{eq:cft-KF-delta}. Appendix~\ref{app:xx-folding-proof} supplies
the independent finite-size XX proof.

% WPM-BLOCK:B0642

% Moved from the main evidentiary sequence under
% CC-2.1.11-CND-NARRATIVE-CLOSURE-R1 / D06-A.
% WPM-BLOCK:B0176
\subsection{Multiple soft ladders}
% WPM-BLOCK:B0177
\label{sec:xy:multi-ladders}

% WPM-BLOCK:B0178
Suppose finitely many resolved soft ladders share the leading power \(L^p\)
and write
% WPM-BLOCK:B0605
\begin{align}
  x_{n,a}^{(L)}&=A_aL^pf_a(n)+r_{n,a}^{(L)},\notag\\
  A_a&>0,
  &\sum_{a,n}|r_{n,a}^{(L)}|&=o(L^p).
  \label{eq:multi-ladder-ansatz}
\end{align}
% WPM-BLOCK:B0606
The global remainder condition controls the growing ultraviolet tail; a
pointwise \(o(L^p)\) statement would not suffice.  With
% WPM-BLOCK:B0607
\begin{equation}
  \mathcal M_{r,a}=\sum_n f_a(n)^r,
  \qquad r=1,2,
\end{equation}
% WPM-BLOCK:B0608
the leading concentration is
% WPM-BLOCK:B0609
\begin{equation}
  \KF^{\rm IR}
  =\frac{\sum_aA_a^2\mathcal M_{2,a}}
  {\left(\sum_aA_a\mathcal M_{1,a}\right)^2}.
  \label{eq:multi-ladder-general}
\end{equation}
% WPM-BLOCK:B0610
Equation~\eqref{eq:multi-ladder-general} also supplies the combined-ladder
relation required by the co-leading-primary exception following
Eq.~\eqref{eq:cft-KF-delta}.
The envelope and grid data reside in \(f_a\), while \(A_a\) records how the
perturbation distributes response among the resolved soft branches.  If
different ladders have different leading powers, only the dominant set enters
this normalized infrared limit.

% WPM-BLOCK:B0611
For a common shape \(f_a=f\), the normalized distribution converges in
\(\ell^1\) to a product,
% WPM-BLOCK:B0612
\begin{equation}
  \pi_{a,n}=q_a\phi_n,
  \qquad
  q_a=\frac{A_a}{\sum_bA_b},
  \qquad
  \phi_n=\frac{f(n)}{\sum_mf(m)}.
  \label{eq:factorized-pi}
\end{equation}
% WPM-BLOCK:B0613
Whenever the relevant entropy is finite, product structure gives
% WPM-BLOCK:B0614
\begin{equation}
  S_\alpha^{\mathrm{ch}}[\pi]
  =S_\alpha^{\mathrm{ch}}[q]+S_\alpha^{\mathrm{ch}}[\phi].
  \label{eq:renyi-additive}
\end{equation}
% WPM-BLOCK:B0615
Thus \(\KF=(\sum_a q_a^2)(\sum_n\phi_n^2)\) at \(\alpha=2\).  For
\(n_{\rm lad}\) equal resolved ladders, \(q\) is uniform and
% WPM-BLOCK:B0616
\begin{equation}
  S_\alpha^{\mathrm{ch},(n_{\rm lad})}
  =S_\alpha^{\mathrm{ch},(1)}+\ln n_{\rm lad},
  \qquad
  N_{\rm eff}^{(\alpha,n_{\rm lad})}
  =n_{\rm lad}N_{\rm eff}^{(\alpha,1)}.
  \label{eq:multiplicity-additive}
\end{equation}
% WPM-BLOCK:B0617
In particular,
\(\KF^{(n_{\rm lad})}=\KF^{(1)}/n_{\rm lad}\).  This multiplicity belongs to
the declared resolution and equal-coupling product structure; it is not by
itself a central-charge theorem.

% WPM-BLOCK:B0618
For an infinite odd power-law ladder \(f(n)=(2n+1)^{-p}\), normalization
requires \(p>1\), and a finite R\'enyi moment below the Shannon index requires
\(\alpha p>1\).  The exact finite-size identity
Eq.~\eqref{eq:xx-half-size-all-alpha} is not subject to this infinite-ladder
threshold.  The following subsections prove the product and asymptotic-factorization
statements and record the entropy-domain conditions.  In the
paired TFIM--XX theorem
\(p_{\rm env}=2\) is fixed while the resolved ladder multiplicity changes;
the Lifshitz section instead varies the perturbing operator at one gapless
Hamiltonian.

% WPM-BLOCK:B0643
\subsection{Exact product distributions}
\label{app:ladder-product}

We establish the exact product and entropy-replication clauses
of Sec.~\ref{sec:xy:multi-ladders}, namely
Eqs.~\eqref{eq:factorized-pi} and \eqref{eq:multiplicity-additive}.

% WPM-BLOCK:B0644
Let \(\pi_{a,n}=q_a\phi_n\), where \(q\) is a normalized distribution over
resolved ladder labels and \(\phi\) is a normalized within-ladder
distribution.  For \(\alpha\in(0,\infty)\setminus\{1\}\),
% WPM-BLOCK:B0645
\begin{equation}
  \sum_{a,n}\pi_{a,n}^{\alpha}
  =\left(\sum_aq_a^\alpha\right)
   \left(\sum_n\phi_n^\alpha\right),
\end{equation}
% WPM-BLOCK:B0646
and hence, whenever the relevant moment is finite,
% WPM-BLOCK:B0647
\begin{equation}
  S_\alpha^{\mathrm{ch}}[\pi]
  =S_\alpha^{\mathrm{ch}}[q]+S_\alpha^{\mathrm{ch}}[\phi].
  \label{eq:renyi-product-proof}
\end{equation}
% WPM-BLOCK:B0648
At \(\alpha=1\), the result follows by expanding
\(-\sum_{a,n}q_a\phi_n\ln(q_a\phi_n)\); at \(\alpha=\infty\), it follows from
\(\max_{a,n}q_a\phi_n=(\max_aq_a)(\max_n\phi_n)\).  If \(q\) is uniform on
\(n_{\rm lad}\) labels, \(S_\alpha^{\mathrm{ch}}[q]=\ln n_{\rm lad}\), proving
Eq.~\eqref{eq:multiplicity-additive}.  In particular, the order-two member is
% WPM-BLOCK:B0649
\begin{equation}
  \KF=\left(\sum_aq_a^2\right)\left(\sum_n\phi_n^2\right)
  .
\end{equation}

% WPM-BLOCK:B0653
\subsection{Sufficient asymptotic factorization condition}
\label{app:ladder-asymptotic}

We prove the sufficient convergence condition for the
concentration and factorization statements of Sec.~\ref{sec:xy:multi-ladders},
Eqs.~\eqref{eq:multi-ladder-general} and \eqref{eq:factorized-pi}, and state
the additional tail conditions needed for Eq.~\eqref{eq:multiplicity-additive}.

% WPM-BLOCK:B0654
To make the leading-ladder argument uniform over a growing set of channels,
zero-pad every finite-size array onto one countable index set and assume
% WPM-BLOCK:B0655
\begin{equation}
  x_{a,n}^{(L)}=L^pA_af_n+r_{a,n}^{(L)},
  \qquad
  \sum_{a,n}|r_{a,n}^{(L)}|=o(L^p),
  \label{eq:l1-ladder-remainder}
\end{equation}
% WPM-BLOCK:B0656
with finitely many ladders, \(A_a>0\), \(f_n\ge0\), and
\(0<F=\sum_nf_n<\infty\).  Set
\(g_{a,n}=A_af_n\), \(G=\sum_{a,n}g_{a,n}\), and
\(X_L=\sum_{a,n}x_{a,n}^{(L)}\).  Then
\(|X_L-L^pG|\le\lVert r^{(L)}\rVert_1=o(L^p)\), and
% WPM-BLOCK:B0657
\begin{equation}
  \left\lVert\frac{x^{(L)}}{X_L}-\frac{g}{G}\right\rVert_1
  \le\frac{2\lVert r^{(L)}\rVert_1}{X_L}
  \longrightarrow0.
  \label{eq:l1-factorization-bound}
\end{equation}
% WPM-BLOCK:B0658
Because
\(g_{a,n}/G=[A_a/(\sum_bA_b)][f_n/(\sum_mf_m)]\), the normalized limit is
the product in Eq.~\eqref{eq:factorized-pi}.  Pointwise
\(r_{a,n}^{(L)}=o(L^p)\) is insufficient: an increasing number of individually
small ultraviolet terms can carry order-\(L^p\) total weight.

In particular, write \(\pi^{(L)}=x^{(L)}/X_L\) and \(\pi=g/G\).
Since every entry of these normalized distributions lies in \([0,1]\),
\begin{equation*}
  \begin{aligned}
  \bigl|\KF^{(L)}-\KF[\pi]\bigr|
  &=\left|\sum_{a,n}
    \left[(\pi_{a,n}^{(L)})^2-\pi_{a,n}^2\right]\right|\\
  &\le 2\lVert\pi^{(L)}-\pi\rVert_1\longrightarrow0.
  \end{aligned}
\end{equation*}
Thus the global remainder bound also guarantees convergence of the
concentration to that of the product distribution.

% WPM-BLOCK:B0659
On a countable support, Eq.~\eqref{eq:l1-factorization-bound} alone does not
imply convergence of every entropy.  It suffices for \(\alpha>1\) moments and
for \(S_\infty^{\mathrm{ch}}\); \(0<\alpha<1\) additionally requires uniform control of the
\(\alpha\)-power tail, while \(S_1^{\mathrm{ch}}\) requires uniform entropy-tail
integrability.  The exact product identity of the limiting distribution is
used only where its defining entropy is finite.
Together, the positive-weight limit and the product proof complete the
multiple-ladder result of Sec.~\ref{sec:xy:multi-ladders}: the concentration
is Eq.~\eqref{eq:multi-ladder-general}, the common-shape limit is
Eq.~\eqref{eq:factorized-pi}, and equivalent ladders give
Eq.~\eqref{eq:multiplicity-additive} on the stated entropy domains.
The global remainder and, where required, entropy-tail assumptions are part
of this conclusion.

% WPM-BLOCK:B0746
\section*{Data availability}

% WPM-BLOCK:B0572
The numerical data, computational parameters, and source code supporting this
work are available in the
\href{https://github.com/ToelUl/channel-concentration/tree/companion-2026-09-21-rc1}{Channel Concentration GitHub repository}.
The numerical results reported here are preserved in the versioned release
cited in Ref.~\cite{NumericalBaseline2026}. The repository also provides
figure-generation scripts, instructions for reproducing the numerical checks
and figures, and a verification script with a reviewed interval receipt for
the conditional channel-refinement budgets. Those budget checks concern the
stated conformal reference families and finite-channel examples; application
to lattice fine channels retains the grouped-matching and projector-support
assumptions stated in the appendices. An independent evaluation of the
conformal-cylinder moments is included in the numerical release, whose
documentation specifies the scope and limitations of the archived validation
records.
No external proprietary data were used.

% WPM-BLOCK:B0573
% ======================================================================
% WPM-BLOCK:B0574
\bibliography{references.rev1}

% BEGIN NUMERICAL SUPPLEMENT
\clearpage
\onecolumngrid
\setcounter{section}{0}
\setcounter{subsection}{0}
\setcounter{equation}{0}
\setcounter{figure}{0}
\setcounter{table}{0}
\counterwithout{equation}{section}
\renewcommand{\thesection}{S\arabic{section}}
\renewcommand{\thesubsection}{S\arabic{section}.\arabic{subsection}}
% The complete subsection number already contains its S-prefixed section.
% REVTeX's inherited reference prefix would otherwise produce "S4 S4.4".
\makeatletter
\def\p@subsection{}
\makeatother
\renewcommand{\theequation}{S\arabic{equation}}
\renewcommand{\thefigure}{S\arabic{figure}}
\renewcommand{\thetable}{S\arabic{table}}
\renewcommand{\theHsection}{supplement.\arabic{section}}
\renewcommand{\theHsubsection}{supplement.\arabic{section}.\arabic{subsection}}
\renewcommand{\theHequation}{supplement.\arabic{equation}}
\renewcommand{\theHfigure}{supplement.\arabic{figure}}
\renewcommand{\theHtable}{supplement.\arabic{table}}
\phantomsection
\pdfbookmark[0]{Supplemental Material}{supplemental-material}
\section*{Supplemental Material}
\begin{center}
{\large\bfseries Numerical Supplemental Material for\\
Channel concentration of critical quantum geometry}
\end{center}

% WPM-BLOCK:B0003
This document records the weak-quench benchmark and feasibility analysis,
independent conformal-moment cross-check, model contracts, finite-size
sequences and ranked distributions, correction diagnostics, convergence
checks, operator-dependence diagnostics, and implementation details supporting
the main-text comparisons. Channel concentration (CC) is denoted by \(\KF\),
as in the main text.

% BEGIN INLINED WEAK-QUENCH MODULE
% Moved atomically from the main article under

% CC-2.1.11-CND-NARRATIVE-CLOSURE-R1 / D05-A.

% WPM-BLOCK:B0269
\section{Operational weak-quench access}
% WPM-BLOCK:B0270
\label{supp:sec:operational-access}

This section benchmarks the conditional readout identity in main-text
Eq.~\textup{(\ref*{eq:quench-KF})} for independent modes.  The CC
definition applies to any declared orthogonal-projector resolution.  Exact
free-chain and conditional CFT results determine CC in their respective
regimes; interacting calculations provide finite-size applications, with
conditional bounds only where the stated spectral premises are available. Here,
final-Hamiltonian Bogoliubov pair counting accesses the free-chain distribution
in the zero-amplitude limit under the conditions below.

\begin{figure*}[t]
  \centering
  \includegraphics[width=\linewidth]{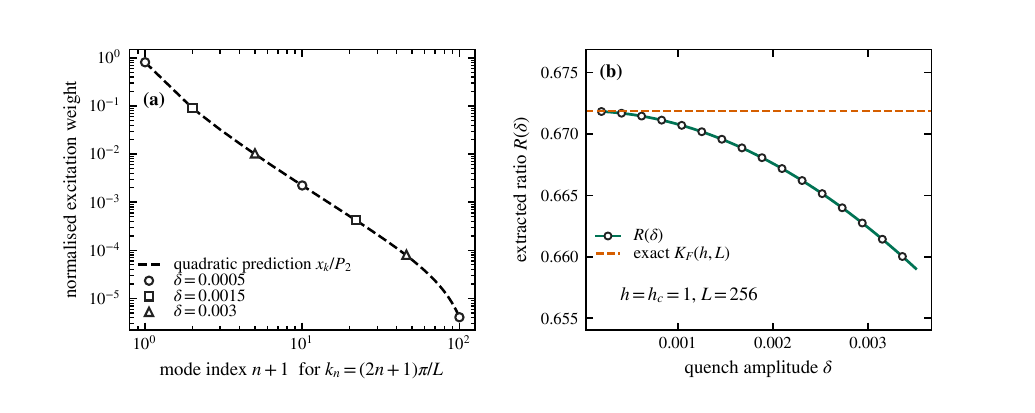}
  \caption{Ideal independent-mode counting after a weak one-sided field quench
    of the critical TFIM at $L=256$, with $N$ the number of pairs excited in the
    final-BdG basis.  The data use
    $p_k(\delta)=\sin^2\left[
    \frac{\theta_k(h+\delta)-\theta_k(h)}{2}\right]$ at $h=1$.
    (a) Normalized excitation weights approach the quadratic-response weights
    $x_k/\Ptwo$ as $\delta\to0$.
    (b) The counting ratio
    $R(\delta)=[\langle N\rangle-\Var(N)]/\langle N\rangle^2$
    recovers the channel concentration $\KF(h,L)=\sum_k(x_k/\Ptwo)^2$
    (orange dashed finite-size reference).  The common TFIM cubic factor
    cancels the generic $O(\delta)$ term, leaving $O(\delta^2)$ drift.
    Markers are thinned and staggered for legibility; all calculated samples
    are retained in the source data.}
  \label{fig:quench}
\end{figure*}

For the free paired chains, prepare the ground state, apply a small one-sided
displacement \(\bm\lambda\to\bm\lambda+\delta\hat y\), and resolve pair
occupations in the final-Hamiltonian Bogoliubov basis.  The excitation
probability of block \(k\) is
\(p_k=\delta^2x_k(\bm\lambda,\hat y)+O(\delta^3)\).  When the declared blocks
are independent Bernoulli events and \(N\) counts excited pairs, the
directional second and fourth moments appear as
% WPM-BLOCK:B0274
\begin{equation}
  \langle N\rangle=\delta^2\Ptwo+O(\delta^3),
  \qquad
  \langle N\rangle-\Var(N)
  =\delta^4\Pfour+O(\delta^5).
  \label{eq:FCS-P4-directional}
\end{equation}
Their normalized variance deficit therefore obeys
% WPM-BLOCK:B0271
% WPM-BLOCK:B0272
\begin{align}
  R(\delta;\hat y)
  &\equiv
  \frac{\langle N\rangle-\Var(N)}{\langle N\rangle^2},\notag\\
  R(\delta;\hat y)
  &=\KF(\bm{\lambda},\hat y)+O(\delta),\notag\\
  \KF(\bm{\lambda},\hat y)
  &=\lim_{\delta\to0}R(\delta;\hat y).
  \label{supp:eq:quench-KF}
\end{align}
\begingroup
\makeatletter
\edef\@currentlabel{\theequation}
\label{eq:directional-quench-ratio}
\makeatother
\endgroup
Here \(R\) is the normalized variance deficit, not the Fano factor.
% WPM-BLOCK:B0275
Under the stated independent-mode conditions, the zero-amplitude ratio can be
evaluated for each tangent direction with \(\Ptwo>0\) to obtain a directional
\(\KF\) profile.  Figure~\ref{fig:quench} tests one field direction in the
critical TFIM.  Sec.~\ref{app:weak-quench-feasibility} gives the finite-amplitude
bias and shot-cost analysis; Sec.~\ref{app:weak-quench} gives the final-basis,
channel-resolution, efficiency, and false-count conditions.  For this TFIM
benchmark, the common cubic correction cancels. At fixed \(L\),
\(R(\delta)-\KF=O(\delta^2)\); in the fixed-\(u=\delta L\) scaling limit,
\(R_{\mathrm{sc}}(u)-2/3=O(u^2)\).

% WPM-BLOCK:B0276

% WPM-BLOCK:B0277
Under the stated independent-mode, final-Bogoliubov-basis conditions, the
low-order variance deficit recovers \(\KF\) without reconstructing excitation
gaps.  For interacting level projectors,
Main-text Eq.~\textup{(\ref*{eq:replica-same-channel})} remains an exact formal collision identity
for resolved many-body labels; no interacting counting protocol is implemented
here.

% WPM-BLOCK:B0337
\subsection{Weak one-sided quench and counting statistics}
% WPM-BLOCK:B0338
\label{app:weak-quench}

This component supplies the final-basis projector contract and finite-amplitude expansion for the conditional
readout identity in main-text Eq.~\textup{(\ref*{eq:quench-KF})}.

% WPM-BLOCK:B0339
For the free-Bogoliubov realization used in the main article, fix the positive
momentum set \(\mathcal K_+\).  In block \(k\), the final-Hamiltonian ground
and pair-excited projectors are
% WPM-BLOCK:B0634
\begin{equation}
  \Pi_{k,0}^{f}=\ket{g_k^f}\bra{g_k^f},\qquad
  \Pi_{k,1}^{f}=\ket{e_k^f}\bra{e_k^f},
  \label{eq:final-bdg-block-povm}
\end{equation}
% WPM-BLOCK:B0635
and the joint readout is
% WPM-BLOCK:B0636
\begin{equation}
  \Pi_{\boldsymbol n}^{f}
  =\bigotimes_{k\in\mathcal K_+}\Pi_{k,n_k}^{f},
  \qquad n_k\in\{0,1\}.
  \label{eq:final-bdg-joint-povm}
\end{equation}
% WPM-BLOCK:B0637
This projector choice identifies a geometric momentum block with one binary
pair event; a coarse projector or local detector need not have this property.
The identification requires a Gaussian product over the declared blocks with
no interchannel preparation or readout covariance, with occupations resolved
in the final-Hamiltonian Bogoliubov basis.  The readout must also avoid channel
merging or pair splitting and have unit, mode-independent efficiency with no
false counts.

% WPM-BLOCK:B0638
For the generic perturbative expansion, consider a one-sided quench from
\(\bm{\lambda}\) to \(\bm{\lambda}+\delta y\).  Expanding the pre-quench
ground state in the post-quench eigenbasis gives, for a nondegenerate spectral
channel,
% WPM-BLOCK:B0340
\begin{equation}
  \ket{0(\bm{\lambda})}
  =\ket{0(\bm{\lambda}+\delta y)}
   -\delta\,\partial_y\ket{0}+O(\delta^2).
\end{equation}
% WPM-BLOCK:B0341
Define the transition amplitude into channel $c>0$ by
$a_c(\delta)\equiv
\braket{c(\bm\lambda+\delta y)}{0(\bm\lambda)}$.  With all coefficient
states and derivatives evaluated at the base point, its expansion is
% WPM-BLOCK:B0342
\begin{equation}
  a_c=\delta a_c^{(1)}+\delta^2a_c^{(2)}+O(\delta^3),
  \qquad
  a_c^{(1)}=-\braket{c}{\partial_y0},
\end{equation}
% WPM-BLOCK:B0343
where $a_c^{(2)}$ denotes the second-order coefficient of the same overlap.
Equation~\textup{(\ref*{eq:spectral-channel-weight-app})} in Appendix A of the main text gives
$x_c=\abs{a_c^{(1)}}^2=\abs{\braket{c}{\partial_y0}}^2$, so the probability is
% WPM-BLOCK:B0344
\begin{equation}
  \begin{aligned}
  p_c=\abs{a_c}^2
  &=\delta^2x_c+\delta^3z_c+O(\delta^4),\\
  z_c&=2\Real\!\left((a_c^{(1)})^*a_c^{(2)}\right).
  \end{aligned}
\end{equation}
% WPM-BLOCK:B0345
The cubic correction is allowed by the generic one-sided convention used in
the main text; a protocol symmetrized around the base point can instead remove
this odd-order term.

% WPM-BLOCK:B0346
Define the cubic-response sums
% WPM-BLOCK:B0347
\begin{equation}
  Z_1=\sum_c z_c,
  \qquad
  Z_2=\sum_c x_c z_c.
  \label{eq:quench-cubic-sums}
\end{equation}

% WPM-BLOCK:B0348
For independent elementary Bernoulli channels, each resolved $c$ is a binary
event with $n_c\in\{0,1\}$ and $N=\sum_{c} n_c$.  A coarse projector need not
obey this binary model.  Up to the order needed here,
% WPM-BLOCK:B0349
\begin{align}
  \langle N\rangle&=\sum_{c} p_c
  =\delta^2\Ptwo+\delta^3Z_1+O(\delta^4),\\
  \Var(N)&=\sum_{c} p_c(1-p_c)
  =\langle N\rangle-\sum_{c} p_c^2,\\
  \langle N\rangle-\Var(N)
  &=\sum_{c} p_c^2
  =\delta^4\Pfour+2\delta^5Z_2+O(\delta^6).
\end{align}
% WPM-BLOCK:B0350
Expanding the numerator and denominator of the normalized variance deficit
through first order in $\delta$ gives
% WPM-BLOCK:B0351
\begin{equation}
  \begin{aligned}
  R(\delta)
  &\equiv \frac{\langle N\rangle-\Var(N)}{\langle N\rangle^2}\\
  &=\KF\,\left[
   1+2\delta\left(\frac{Z_2}{\Pfour}-\frac{Z_1}{\Ptwo}\right)
   \right]+O(\delta^2).
  \end{aligned}
  \label{eq:quench-KF-correction}
\end{equation}
% WPM-BLOCK:B0352
This expansion refines the leading counting-statistics relation in
Eq.~\eqref{supp:eq:quench-KF} and identifies the exact cancellation criterion: a common cubic factor
\(z_c=\eta x_c\) drops out of the normalized statistic even though it remains
present in every excitation probability.

% WPM-BLOCK:B0514
\section{Finite-amplitude and fixed-\texorpdfstring{$u$}{u} weak-quench feasibility}
% WPM-BLOCK:B0515
\label{app:weak-quench-feasibility}

% WPM-BLOCK:B0516
\subsection{Finite-\texorpdfstring{$\delta$}{delta} error budget for the TFIM figure}
% WPM-BLOCK:B0517
\label{app:weak-quench-budget}

This component supplies the finite-amplitude error budget for the conditional
readout identity in main-text Eq.~\textup{(\ref*{eq:quench-KF})}.

% WPM-BLOCK:B0518
For the free-fermion TFIM data in Fig.~\ref{fig:quench}, set
$h_i=\hc=1$ and $h_f=h_i+\delta$.  The exact mode probability for this
one-sided field quench is
% WPM-BLOCK:B0519
\begin{equation}
  p_k(\delta)
  =\sin^2\left[\frac{\theta_k(h_f)-\theta_k(h_i)}{2}\right].
  \label{eq:exact-quench-pk}
\end{equation}
% WPM-BLOCK:B0520
At \(h=\hc\), differentiation of
\(\tan\theta_k=\sin k/(h-\cos k)\) gives
% WPM-BLOCK:B0521
\begin{equation}
  \begin{aligned}
  \left.\partial_h\theta_k\right|_{h=\hc}
  &=-\frac12\cot\frac{k}{2},\\
  \left.\partial_h^2\theta_k\right|_{h=\hc}
  &=+\frac12\cot\frac{k}{2}
  =-\left.\partial_h\theta_k\right|_{h=\hc},\\
  \left.\partial_h^3\theta_k\right|_{h=\hc}
  &=\frac14\left(
    \cot^3\frac{k}{2}-3\cot\frac{k}{2}
  \right).
  \end{aligned}
  \label{eq:tfim-angle-derivative-cancellation}
\end{equation}
% WPM-BLOCK:B0522
To obtain $p_k$ through $O(\delta^4)$, the angle increment is needed through
third order and the sine-squared expansion through fourth order.  The required
increment is
% WPM-BLOCK:B0523
\begin{equation}
  \begin{aligned}
  \Delta\theta_k
  &\equiv\theta_k(h_f)-\theta_k(h_i)\\
  &=-\frac12\cot\frac{k}{2}\,\delta
    +\frac14\cot\frac{k}{2}\,\delta^2\\
  &\quad
    +\frac1{24}\left(
      \cot^3\frac{k}{2}-3\cot\frac{k}{2}
    \right)\delta^3
    +O(\delta^4).
  \end{aligned}
  \label{eq:tfim-angle-increment-expansion}
\end{equation}
% WPM-BLOCK:B0524
Using
\(\sin^2(\Delta\theta_k/2)
=\Delta\theta_k^2/4-\Delta\theta_k^4/48+O(\Delta\theta_k^6)\)
and \(x_k=\tfrac1{16}\cot^2(k/2)\),
Eq.~\eqref{eq:exact-quench-pk} yields, mode by mode,
% WPM-BLOCK:B0525
\begin{equation}
  p_k(\delta)
  =\delta^2x_k(1-\delta+\tfrac34\delta^2)
   -3\delta^4x_k^2+O(\delta^5).
  \label{eq:tfim-quench-pk-expansion}
\end{equation}
% WPM-BLOCK:B0526
Thus $z_k/x_k=-1$ for every mode.  Hence \(Z_1=-\Ptwo\) and
\(Z_2=-\Pfour\), proving that
the nominal linear correction in Eq.~\eqref{eq:quench-KF-correction} cancels exactly for
this benchmark.

% WPM-BLOCK:B0527
The first nonzero drift involves \(\Pn{6}\equiv\sum_kx_k^3\).  The same NS
half-grid construction used for main-text Eq.~\textup{(\ref*{eq:tfim-critical-moments-exact})},
now through the sixth-power identity
Eq.~\textup{(\ref*{eq:tfim-critical-cot6-sum})} in Appendix D of the main text, gives
% WPM-BLOCK:B0528
\begin{equation}
  \Pn{6}(\hc,L)
  =\frac{L(L-1)(2L^4+2L^3-8L^2-8L+15)}{122880}.
  \label{eq:tfim-critical-P6-exact}
\end{equation}
% WPM-BLOCK:B0529
Indeed, main-text Eq.~\textup{(\ref*{eq:xk-crit-exact})} implies that \(\Pn{6}\) is
\(16^{-3}\) times the cotangent sixth-power sum.  Summing
Eq.~\eqref{eq:tfim-quench-pk-expansion} and its square gives
% WPM-BLOCK:B0530
\begin{align}
  \sum_kp_k
  &=\delta^2\Ptwo\left[1-\delta+\tfrac34\delta^2\right]
    -3\delta^4\Pfour+O(\delta^5),
  \nonumber\\
  \sum_kp_k^2
  &=\delta^4\Pfour\left[1-2\delta+\tfrac52\delta^2\right]
    -6\delta^6\Pn{6}+O(\delta^7).
  \label{eq:tfim-quench-moment-expansions}
\end{align}
% WPM-BLOCK:B0531
Taking their normalized ratio gives
% WPM-BLOCK:B0532
\begin{equation}
  \begin{aligned}
  R(\delta)&=\KF+c_2(L)\delta^2+O(\delta^3),\\
  c_2(L)&=6\KF\left(\frac{\Pfour}{\Ptwo}
                         -\frac{\Pn{6}}{\Pfour}\right).
  \end{aligned}
  \label{eq:tfim-quench-quadratic-drift}
\end{equation}
% WPM-BLOCK:B0533
At the critical TFIM point \(h=\hc=1\) and \(L=256\), the exact geometric value is
% WPM-BLOCK:B0534
\begin{equation}
  \KF(\hc,256)=0.672,
\end{equation}
% WPM-BLOCK:B0535
which lies \(5.20\times10^{-3}\) above \(2/3\), consistent with the leading
finite-size correction \(4/(3L)\).  Table~\ref{tab:delta-budget} gives the systematic
drift of the directly extracted ratio \(R(\delta)\):
% WPM-BLOCK:B0536
\begin{table*}[t]
  \caption{Finite-amplitude systematic error for the weak-quench extraction at
  the critical TFIM point, using the exact two-level formula
  \eqref{eq:exact-quench-pk} at \(L=256\).  These are exact finite-size
  evaluations of the stated protocol, not thermodynamic extrapolations.
  Here \(\langle N\rangle\) is the mean counted pair occupation,
  \(R(\delta)=[\langle N\rangle-\Var(N)]/\langle N\rangle^2\), and
  \(\KF=\lim_{\delta\to0}R(\delta)\) is the finite-size geometric concentration.}
  \label{tab:delta-budget}
  \begin{ruledtabular}
  \begin{tabular}{lccc}
  \(\delta\) & \(\langle N\rangle\) & \(R(\delta)\) & \(\abs{R-\KF}\) \\
  \hline
  \(10^{-4}\) & \(2.04\times10^{-5}\) & 0.672 & \(1.08\times10^{-5}\) \\
  \(2\times10^{-4}\) & \(8.16\times10^{-5}\) & 0.672 & \(4.32\times10^{-5}\) \\
  \(5\times10^{-4}\) & \(5.09\times10^{-4}\) & 0.672 & \(2.70\times10^{-4}\) \\
  \(10^{-3}\) & \(2.03\times10^{-3}\) & 0.671 & \(1.08\times10^{-3}\) \\
  \(2\times10^{-3}\) & \(8.01\times10^{-3}\) & 0.668 & \(4.27\times10^{-3}\) \\
  \end{tabular}
  \end{ruledtabular}
\end{table*}
% WPM-BLOCK:B0537
A two-parameter least-squares fit
\(R(\delta)-\KF=a\delta+b\delta^2\) to the entries in
Table~\ref{tab:delta-budget} gives
\(a=-1.22\times10^{-2}\) and \(b=-1061.37\), whereas the analytic result is
\(a=0\) and \(c_2(256)=-1.08\times10^3\).  The fitted discrepancies are
higher-order finite-window leakage, not a physical linear derivative at
\(\delta=0\).  At \(\delta=10^{-3}\), the finite-amplitude bias is already of
order \(10^{-3}\), roughly one fifth of the finite-size offset of
\(\KF(\hc,256)\) from \(2/3\).  The generic one-sided expansion allows an
\(O(\delta)\) term, whereas this critical benchmark has a proved
\(O(\delta^2)\) leading correction.

% WPM-BLOCK:B0539
\subsection{Fixed-\texorpdfstring{$u$}{u} weak-quench feasibility}
% WPM-BLOCK:B0540
\label{app:weak-feasibility}

This component supplies the fixed-scaling-amplitude bias and shot-cost analysis for the conditional
readout identity in main-text Eq.~\textup{(\ref*{eq:quench-KF})}.

% WPM-BLOCK:B0541
The fixed-\(u\) analysis below quantifies the shot-count implications of the
full-counting-statistics (FCS) extraction formula in
Sec.~\ref{supp:sec:operational-access}.  Critical perturbation theory is
controlled by the scaling amplitude
% WPM-BLOCK:B0542
\begin{equation}
  u=\delta L,
  \label{eq:weak-quench-scaling-u}
\end{equation}
% WPM-BLOCK:B0543
rather than by \(\delta\) alone.  Indeed, the softest geometric weight grows as
\(x_0\sim L^2\), so fixed nonzero \(\delta\) eventually leaves the weak-channel
window as \(L\) increases.

% WPM-BLOCK:B0544
In the joint limit \(L\to\infty\), \(\delta=u/L\), let
\(q_n=\pi(2n+1)\) for nonnegative integer \(n\), so that \(k_n=q_n/L\).
Define the initial and final fields by $h_i=\hc=1$ and
$h_f=\hc+u/L$.  The exact lattice angles obey
% WPM-BLOCK:B0545
\begin{equation}
  \begin{aligned}
  \tan\theta_{k_n}(h_i)
  &=\frac{\sin(q_n/L)}{1-\cos(q_n/L)}
    \longrightarrow\infty,\\
  \tan\theta_{k_n}(h_f)
  &\longrightarrow\frac{q_n}{u}.
  \end{aligned}
  \label{eq:weak-quench-scaling-angle-limit}
\end{equation}
% WPM-BLOCK:B0546
For \(u>0\), the continuous branch has
$\theta_{k_n}(h_i)\to\pi/2$ and
$\theta_{k_n}(h_f)\to\arctan(q_n/u)$, so their difference is
$-\arctan(u/q_n)$.  The result at \(u=0\) follows continuously.  Applying
the half-angle identity to Eq.~\eqref{eq:exact-quench-pk} gives the exact
fixed-\(u\) scaling-limit probability
% WPM-BLOCK:B0547
\begin{equation}
  p_n^{\mathrm{sc}}(u)
  =\frac12\left[
   1-\frac{1}{\sqrt{1+(u/q_n)^2}}
   \right].
  \label{eq:weak-quench-scaling-probability}
\end{equation}
% WPM-BLOCK:B0548
Its small-\(u\) expansion is
% WPM-BLOCK:B0549
\begin{equation}
  p_n^{\mathrm{sc}}(u)
  =\frac{u^2}{4q_n^2}
   -\frac{3u^4}{16q_n^4}
   +\frac{5u^6}{32q_n^6}
   +O(u^8).
  \label{eq:weak-quench-scaling-mode-expansion}
\end{equation}
% WPM-BLOCK:B0550
Assume independent binary mode occupations $n_k\in\{0,1\}$ and set
$N=\sum_k n_k$.  The second-factorial-moment signal and the normalized
variance deficit are
$F_2=\langle N(N-1)\rangle=(\sum_np_n)^2-\sum_np_n^2$ and
$R_{\mathrm{sc}}(u)\equiv\sum_np_n^2/(\sum_np_n)^2$, respectively.
Substituting $q_n=\pi(2n+1)$ into
Eq.~\eqref{eq:weak-quench-scaling-mode-expansion} uses the
Dirichlet-lambda definition in main-text Eq.~\textup{(\ref*{eq:dirichlet-lambda-def})}:
% WPM-BLOCK:B0551
\begin{equation}
  \sum_{n\ge0}q_n^{-2}=\frac18,
  \qquad
  \sum_{n\ge0}q_n^{-4}=\frac1{96},
  \qquad
  \sum_{n\ge0}q_n^{-6}=\frac1{960}.
\end{equation}
% WPM-BLOCK:B0552
Combining these sums in
$F_2=(\sum_np_n)^2-\sum_np_n^2$ gives
% WPM-BLOCK:B0553
\begin{align}
  \langle N\rangle_{\mathrm{sc}}
  &=\frac{u^2}{32}-\frac{u^4}{512}+O(u^6),
  \nonumber\\
  F_{2,\mathrm{sc}}
  &=\frac{u^4}{3072}-\frac{u^6}{40960}+O(u^8),
  \nonumber\\
  R_{\mathrm{sc}}(u)
  &=\frac23-\frac{u^2}{60}+O(u^4).
  \label{eq:weak-quench-scaling-expansions}
\end{align}
% WPM-BLOCK:B0554
The last line agrees with the finite-size coefficient
Eq.~\eqref{eq:tfim-quench-quadratic-drift}, since
\(c_2(L)/L^2\to-1/60\).

% WPM-BLOCK:B0555
Table~\ref{tab:weak-feasibility} reports exact finite-\(L\) values and their
convergence to the fixed-\(u\) scaling limit.  The direct-estimator shot column
used there follows from the factorial-moment variance derived next.
% WPM-BLOCK:B0676
% WPM-BLOCK:B0556
\begin{table*}[t]
  \caption{Critical TFIM feasibility at fixed \(u=\delta L\), evaluated from
  the exact one-sided two-level probabilities.  The factorial signal is
  \(F_2=\langle N(N-1)\rangle\), and the direct-estimator requirement is
  \(M_{10\%}=100\Var(Y)/F_2^2\), with \(Y=N(N-1)\).
  All tabulated quantities are dimensionless; \(R\) is the normalized variance
  deficit and \(\KF=\lim_{\delta\to0}R(\delta)\) at the same finite \(L\).
  The rows approach
  size-independent values at fixed \(u\), as predicted by
  Eqs.~\eqref{eq:weak-quench-scaling-expansions} and
  \eqref{eq:weak-quench-fixed-u-shots}; they are scale estimates rather than
  apparatus-specific requirements.}
  \label{tab:weak-feasibility}
  \begin{ruledtabular}
  \begin{tabular}{c c c c c c c c}
  \(L\) & \(u\) & \(\delta\) & \(\langle N\rangle\) & \(F_2\) & \(\Var(Y)/F_2\) & \(M_{10\%}\) & \(\abs{R-\KF}\) \\
  \hline
   64 & 0.128 & \(2.0\!\times\!10^{-3}\) & \(5.02\!\times\!10^{-4}\) & \(7.90\!\times\!10^{-8}\) & 2.00 & \(2.53\!\times\!10^{9}\) & \(2.59\!\times\!10^{-4}\) \\
  128 & 0.128 & \(1.0\!\times\!10^{-3}\) & \(5.07\!\times\!10^{-4}\) & \(8.31\!\times\!10^{-8}\) & 2.00 & \(2.41\!\times\!10^{9}\) & \(2.66\!\times\!10^{-4}\) \\
  256 & 0.128 & \(5.0\!\times\!10^{-4}\) & \(5.09\!\times\!10^{-4}\) & \(8.52\!\times\!10^{-8}\) & 2.00 & \(2.35\!\times\!10^{9}\) & \(2.70\!\times\!10^{-4}\) \\
   64 & 0.256 & \(4.0\!\times\!10^{-3}\) & \(2.00\!\times\!10^{-3}\) & \(1.25\!\times\!10^{-6}\) & 2.00 & \(1.60\!\times\!10^{8}\) & \(1.03\!\times\!10^{-3}\) \\
  128 & 0.256 & \(2.0\!\times\!10^{-3}\) & \(2.02\!\times\!10^{-3}\) & \(1.32\!\times\!10^{-6}\) & 2.00 & \(1.51\!\times\!10^{8}\) & \(1.06\!\times\!10^{-3}\) \\
  256 & 0.256 & \(1.0\!\times\!10^{-3}\) & \(2.03\!\times\!10^{-3}\) & \(1.36\!\times\!10^{-6}\) & 2.00 & \(1.48\!\times\!10^{8}\) & \(1.08\!\times\!10^{-3}\) \\
  \end{tabular}
  \end{ruledtabular}
\end{table*}

% WPM-BLOCK:B0557
For one ideal independent-mode shot, use the single-shot estimator
\(Y=N(N-1)\), whose expectation is \(F_2\).  More generally, writing
\(F_m=\langle(N)_m\rangle\) for factorial moments, the exact identity
\(Y^2=(N)_4+4(N)_3+2(N)_2\) gives
% WPM-BLOCK:B0558
\begin{equation}
  \Var(Y)
  =F_4+4F_3+2F_2-F_2^2.
  \label{eq:weak-quench-factorial-variance}
\end{equation}
% WPM-BLOCK:B0559
For the mean of \(M\) independent shots, imposing the 10\% relative-error
condition $\sqrt{\Var(Y)/(M F_2^2)}=0.1$ gives
% WPM-BLOCK:B0560
\begin{equation}
  M_{10\%}
  =\frac{100\Var(Y)}{F_2^2}
  \simeq\frac{200}{F_2}
  \simeq\frac{6.144\times10^5}{u^4},
  \label{eq:weak-quench-fixed-u-shots}
\end{equation}
% WPM-BLOCK:B0561
In the rare-pair limit, $F_3$, $F_4$, and $F_2^2$ are all $o(F_2)$, so
Eq.~\eqref{eq:weak-quench-factorial-variance} gives
$\Var(Y)/F_2\to2$.  The resulting factor of two relative to $100/F_2$ is a
property of the direct estimator $Y$.
Thus neither the signal nor the leading shot requirement improves with \(L\)
at fixed controlled \(u\).  Apparent \(L^4\) growth at fixed \(\delta\) is
simultaneously growth of \(u^4\), not a free large-system statistical gain.

% WPM-BLOCK:B0562
The controlled tradeoff is therefore governed by \(u\): increasing \(u\)
improves the factorial signal as \(u^4\), but increases the systematic bias as
\(u^2\).  Increasing \(L\) at fixed \(u\) reduces the bare displacement
\(\delta=u/L\) without changing this leading bias--variance balance.

% WPM-BLOCK:B0563
\FloatBarrier
% END INLINED WEAK-QUENCH MODULE

\section{Conformal-moment cross-check}
\label{supp:cft-cross-check}

This check independently evaluates the conformal moments supporting the Potts comparator in main-text Eq.~\textup{(\ref*{eq:potts-cft-target})}.

For the complete zero-momentum conformal-level weights of the main text, write
\begin{equation}
  W_r(\Delta)=\sum_{n=0}^{\infty}
  \left\{\frac{[(\Delta)_n/n!]^2}{(\Delta+2n)^2}\right\}^{r},
  \qquad K_F(\Delta)=\frac{W_2(\Delta)}{W_1(\Delta)^2}.
  \label{eq:cft-moment-cross-check-supp}
\end{equation}
The frozen evaluation uses two independent routes.  The first iterates
\(A_{n+1}/A_n=[(n+\Delta)/(n+1)]^2\) and encloses the remainder with a
gamma-ratio asymptotic expansion and Hurwitz-zeta tails.  The second evaluates
the beta/Laplace representations as unit-argument generalized hypergeometric
functions: \({}_4F_3(1)\) for \(W_1\) and \({}_8F_7(1)\) for \(W_2\).

At \(\Delta=4/5\), 80-digit arithmetic gives
\begin{align}
  W_1&=1.6958145904545225831157952249\ldots,\notag\\
  W_2&=2.4487201414979997506944888453\ldots,\notag\\
  K_F&=0.8514956201179281748625366353\ldots .
  \label{eq:cft-potts-moments-supp}
\end{align}
The maximum cross-route difference is \(6.4\times10^{-80}\), well below the
registered \(10^{-20}\) tolerance.  At \(\Delta=1\), the same calculation
returns \(W_1=\pi^2/8\), \(W_2=\pi^4/96\), and \(K_F=2/3\), providing an exact
oracle.  This cross-check evaluates the stated conformal CC only; it
neither fits the Potts lattice data nor identifies numerical lattice
projectors with conformal levels.

% WPM-BLOCK:B0004
\section{Models, sector choices, and finite-size data}
% WPM-BLOCK:B0012
\label{app:interacting-table}

This section supplies the scalar inputs, correction diagnostics, and
retained-spectrum checks for main-text
Figs.~\ref*{fig:interacting-benchmarks} and
\ref*{fig:potts-weight-comparators}.

\subsection{Model, probe, and retained-spectrum contracts}
\label{supp:model-contracts}

These contracts specify the models and probes used in main-text Sec.~\ref*{sec:interacting-benchmarks}.

% WPM-BLOCK:B0013
The self-dual three-state Potts chain is written in the
clock/shift convention \(\sigma\ket{m}=\omega^m\ket{m}\),
\(\tau\ket{m}=\ket{m+1\bmod 3}\), and
\(\sigma\tau=\omega\tau\sigma\), with
% WPM-BLOCK:B0014
\begin{equation}
  H(g)=-\sum_j(\tau_j+\tau_j^\dagger)
  -g\sum_j(\sigma_j^\dagger\sigma_{j+1}+\sigma_{j+1}^\dagger\sigma_j),
  \label{eq:potts-hamiltonian}
\end{equation}
% WPM-BLOCK:B0015
at \(g=1\).  The thermal perturbation is
\(V_g=-\sum_j(\sigma_j^\dagger\sigma_{j+1}
+\sigma_{j+1}^\dagger\sigma_j)\).  The production calculation diagonalizes
the full translation-invariant \(k=0\) block without applying a charge projector
to the quoted block dimensions.  The Hamiltonian, ground state, and thermal
response preserve the global \(\mathbb Z_3\) color shift; explicit charge checks
below confirm \(q=0\) response support.  The next-nearest-neighbor
transverse-field Ising (NNN-TFIM) control is
% WPM-BLOCK:B0016
\begin{equation}
  H(h,J_2)
  =-\sum_j\sigma_j^z\sigma_{j+1}^z
   -J_2\sum_j\sigma_j^z\sigma_{j+2}^z
   -h\sum_j\sigma_j^x,
  \label{eq:nnn-tfim-hamiltonian}
\end{equation}
% WPM-BLOCK:B0017
with \(V_h=-\sum_j\sigma_j^x\).  For \(J_2>0\), the gap-crossing
pseudo-critical field \(h_\times(L;J_2)\) satisfies
\((L-2)\Delta_{L-2}(h,J_2)=L\Delta_L(h,J_2)\), where
\(\Delta_L=E_{0,-}^{(L)}-E_{0,+}^{(L)}\) is the lowest \(k=0\) spin-flip
odd--even gap.  The response calculation uses the \(k=0\), spin-flip-even
sector.  The following \(1/L\) form is an empirical sensitivity basis, not a
theory-derived correction exponent:
% WPM-BLOCK:B0018
\begin{equation}
  \KF(L;J_2)=K_{F,\infty}^{\mathrm{lin}}(J_2)+\frac{a(J_2)}{L}.
  \label{eq:nnn-linear-consistency-fit}
\end{equation}
% WPM-BLOCK:B0019
For Potts, the single-power fit diagnostic is fixed separately.  The
fixed-eigenpair-count calculation evaluates each \(L=6,\ldots,14\) once, using
the full available \(k=0\) sector at \(L=6\) and calibrated
\(N_{\mathrm{eig}}^{\mathrm{req}}=256\) results for \(L=7,\ldots,14\).
Here \(k\) denotes lattice momentum throughout.  The legacy configuration field
\texttt{k\_target=256} denotes the requested retained eigenpair cutoff, not a
momentum.  At \(L=14\),
\(N_{\mathrm{eig}}^{\mathrm{eff}}=256\); one guard eigenpair gives 257 solver
eigenpairs, which form \(N_{\mathrm{ch}}=109\) tolerance-stable numerical
degeneracy projectors.  The channel rank \(r\) is assigned only after ordering these grouped response
weights.  For the thermal scaling dimension \(\Delta_\varepsilon=4/5\), the
singular response scales as \(\Ptwo\sim L^{12/5}\).  The center-of-mass
decomposition derived in the main-text appendix gives a lattice-regulated
short-distance contribution \(\Ptwo^{\mathrm{UV}}\sim L\), and dividing it by
the leading response produces the relative correction
\(L^{1-12/5}=L^{-7/5}\).
For an accessible-size diagnostic, the deterministic midpoints of the archived
Potts \(K_F\) entries enter the unweighted, free-intercept fit
\begin{equation}
  \KF(L)=K_{F,\mathrm{fit}}^{(7/5)}+bL^{-7/5}.
  \label{eq:potts-theory-guided-fit-supp}
\end{equation}
The spectral-tail calibration supplies
outward-enclosed endpoint intervals for \(L=12,\ldots,14\), which are propagated separately
from fit-window and correction-form sensitivity.  The CFT value
supplies a no-fit theory comparator.  The lattice calculation resolves
numerically distinct levels rather than complete conformal levels.  Thus the
fit intercept is not identified with the CFT value; the grouped matching and
moment-expansion assumptions of the main text have not been established for
this observable.  The chosen correction powers test candidate mechanisms and
do not prove its asymptotic correction hierarchy.

% WPM-BLOCK:B0020
The exact \(J_2=0\) sequence uses NS momentum blocks, whereas the
exact-diagonalization (ED) control uses
many-body level projectors.  At \(h=1\), the matched \(k=0\), spin-flip-even
calculation groups numerical degeneracies and compares \(\Ptwo\), \(\Pfour\),
and \(\KF\) at the tested sizes with the exact NS momentum-block formulas
% WPM-BLOCK:B0021
\begin{equation}
  \Ptwo=\frac{L(L-1)}{32},\qquad
  \Pfour=\frac{L(L-1)(L^2+L-3)}{1536}.
\end{equation}
% WPM-BLOCK:B0022
Table~\ref{tab:j2-zero-resolution-match} reports the primary
\(10^{-7}\) numerical-degeneracy tolerance.  Repeating the grouping at
\(10^{-10},10^{-8},10^{-7},10^{-6}\) produces zero \(\KF\) spread at the
displayed precision.  The complete \(\Ptwo\), \(\Pfour\), and tolerance-scan rows
are retained in the accompanying numerical data.  This scan checks \(\KF\)
stability at the displayed precision across the stated grouping tolerances.
Basis invariance within a complete degenerate projector follows from its
projector definition; the scan does not merge finite-size lattice splittings
into conformal levels.

\subsection{Distribution construction and exact calibration}
\label{supp:distribution-construction}

This construction supplies the rank-sorted comparisons in main-text Fig.~\ref*{fig:potts-weight-comparators} and their normalization and calibration.

For the distribution-level NNN-TFIM comparison, no additional eigensolve is
performed.  A deterministic postprocessor reads the frozen checkpoints and,
for every \((J_2,L)\), selects the largest retained-level cutoff whose scalar
\(\KF\) agrees with the released scalar reference within \(5\times10^{-10}\).  It
divides retained historical mean-gap weights $x_a$ by the independently
solved $\Ptwo$ and sorts them in descending order.  The plotted heights
are $x_a/\Ptwo$, not self-normalized $x_a/\sum_R x_a$.  An aggregate
tail bin $1-\sum_R x_a/\Ptwo$ closes the archived ranked-distance
distribution by convention; without a global mean-gap conversion bound it
is not a certified omitted exact-response probability.  NNN retained-count
and scalar-agreement checks are diagnostics, not a same-response enclosure.
Figure~\ref*{fig:potts-weight-comparators}(b) of the main text
displays the leading ten weights for the most strongly perturbed
\(J_2=0.20\) control at \(L=6,10,14,20\).  The black reference is the exact
\(L=20\) TFIM sequence
\(\pi_r^{\rm TFIM}\propto\cot^2[(2r-1)\pi/(2L)]\).  The displayed ranks contain
more than \(99.97\%\) of the projected-solve normalization for each selected size.
The accompanying machine-readable data and selection receipt record the
weights, selected cutoffs, checkpoint hashes, and historical check outcomes.  The
rank ordering compares weight spectra and does not identify individual
channels between the interacting and exact models.

% WPM-BLOCK:B0023
\begin{table}[H]
  \caption{Matched \(J_2=0\) resolution calibration.  The ED result uses
  numerical-degeneracy-grouped level projectors in the \(k=0\), spin-flip-even sector with grouping
  tolerance \(10^{-7}\); the exact column uses the NS momentum-block formula.
  The final column is the absolute \(\KF\) difference.}
  \label{tab:j2-zero-resolution-match}
  \begin{ruledtabular}
  \begin{tabular}{c r c c c}
  \(L\) & dim & \(\KF^{\mathrm{projector}}\) & \(\KF^{\mathrm{exact}}\) & abs. error \\
  \hline
  6  & 8   & 0.866666666667 & 0.866666666667 & \(3.3\times10^{-16}\) \\
  8  & 20  & 0.821428571429 & 0.821428571429 & \(2.1\times10^{-15}\) \\
  10 & 56  & 0.792592592593 & 0.792592592593 & \(4.9\times10^{-15}\) \\
  12 & 180 & 0.772727272727 & 0.772727272727 & \(4.6\times10^{-15}\)
  \end{tabular}
  \end{ruledtabular}
\end{table}

\subsection{Primary finite-size sequences}
\label{supp:primary-sequences}

The sequences below are the numerical inputs displayed in main-text Fig.~\ref*{fig:interacting-benchmarks}.

% WPM-BLOCK:B0024
Table~\ref{tab:interacting-data} lists the exact TFIM reference and three
NNN-TFIM sequences plotted in
Fig.~\ref*{fig:interacting-benchmarks}(b).  Their pseudocritical fields,
sector dimensions, retained-level counts, and response moments are preserved
in the accompanying machine-readable data.  These short exact-diagonalization
sequences define finite-size extrapolation windows, not precision
thermodynamic estimates.  Because \(\KF=\sum_a\pi_a^2\) is dominated by the
largest normalized weights, retained-level convergence controls \(\Pfour\),
as tested in Tables~\ref{tab:retained-level-checks} and
\ref{tab:nnn-l20-stability}.  The Potts inputs for
Fig.~\ref*{fig:interacting-benchmarks}(a) and their fit diagnostics appear
in Tables~\ref{tab:potts-outcome-aware-data} and
\ref{tab:potts-fit-diagnostics}; the full-precision sequence accompanies the
reproducibility package.

% WPM-BLOCK:B0025
The global color-shift generator \(C=\prod_j\tau_j\) was constructed as a
permutation in the full \(k=0\) orbit basis.  For \(L=4,\ldots,8\), both
\(\norm{(1-P_{q=0})\ket{0}}\) and
\(\norm{(1-P_{q=0})QV_g\ket{0}}/\norm{QV_g\ket{0}}\) are below
\(2\times10^{-14}\), compared with the stated tolerance
\(10^{-10}\).  Table~\ref{tab:potts-charge-sector} gives the required
three-representation comparison.  These checks establish neutral response
support without relabeling the diagonalized full-\(k=0\) dimension as a
charge-neutral block.

% WPM-BLOCK:B0026
\begin{table}[H]
  \footnotesize
  \caption{Canonical scalar sequences for the main-text NNN-TFIM comparison.
    The exact column is the critical NS TFIM result at \(h=1\).  The three
    interacting columns are evaluated at the corresponding gap-crossing
    pseudocritical fields \(h_\times(L;J_2)\) in the \(k=0\), spin-flip-even
    response sector.  These are the values plotted in
    Fig.~\ref*{fig:interacting-benchmarks}(b).
    The interacting columns retain the historical low-level mean-gap
diagnostic with projected-solve normalization, as distinguished from
the exact projector response in Sec.~\ref{supp:conditional-tail}.}
  \label{tab:interacting-data}
  \begin{ruledtabular}
  \begin{tabular}{c c c c c}
    \(L\) & \(\KF^{\rm TFIM}\) & \(J_2=0.05\) & \(J_2=0.10\) & \(J_2=0.20\) \\
    \hline
    6  & 0.866666667 & 0.876010982 & 0.885561146 & 0.904106457 \\
    8  & 0.821428571 & 0.827386912 & 0.834071756 & 0.848277879 \\
    10 & 0.792592593 & 0.795924831 & 0.800186461 & 0.810171957 \\
    12 & 0.772727273 & 0.774276969 & 0.776816878 & 0.783619373 \\
    14 & 0.758241758 & 0.758594523 & 0.759936611 & 0.764426353 \\
    16 & 0.747222222 & 0.746762541 & 0.747264400 & 0.750075123 \\
    18 & 0.738562092 & 0.737543871 & 0.737448809 & 0.739025914 \\
    20 & 0.731578947 & 0.730172967 & 0.729644737 & 0.730308153 \\
  \end{tabular}
  \end{ruledtabular}
\end{table}

\begin{table}[H]
  \footnotesize
  \caption{Archived fixed-retained-count Potts inputs for the finite-size
comparison. The printed endpoints are rounded representations of the
historical calculations, not directed-rounding bounds. The $L=6$ point
self-normalizes complete-sector mean-gap weights; its repeated endpoint
is an archived scalar, not a zero-width physical enclosure. Sparse rows
use retained mean-gap numerators and the projected-solve denominator,
with $N_{\mathrm{eig}}^{\mathrm{req}}=256$ for $L=7,\ldots,14$.
For $L=12,13,14$, ``separately enclosed'' refers to the exact-response
enclosures described in Sec.~\ref{supp:conditional-tail}, conditional
on spectral identification, complete groups, and valid error bounds.
Those separate enclosures are narrower than $10^{-8}$ and below the
graphical resolution of Fig.~\ref*{fig:interacting-benchmarks}(a),
which therefore shows no visible Potts error bars. Other sizes retain
their full-sector or retained-count diagnostic status.}
  \label{tab:potts-outcome-aware-data}
  \begin{ruledtabular}
  \begin{tabular}{c c c c}
  \(L\) & rounded lower & rounded upper & evidence status \\
  \hline
  6  & 0.919949336245 & 0.919949336245 & full-sector pass \\
  7  & 0.906664272578 & 0.906664272831 & retained-count pass \\
  8  & 0.896828291148 & 0.896828291712 & retained-count pass \\
  9  & 0.889343715085 & 0.889343716904 & retained-count pass \\
  10 & 0.883512065042 & 0.883512066437 & retained-count pass \\
  11 & 0.878875756624 & 0.878875758932 & retained-count pass \\
  12 & 0.875125782742 & 0.875125785831 & separately enclosed \\
  13 & 0.872047513204 & 0.872047516382 & separately enclosed \\
  14 & 0.869488064668 & 0.869488068118 & separately enclosed \\
  \end{tabular}
  \end{ruledtabular}
\end{table}

\begin{table}[H]
  \footnotesize
  \caption{Potts response-scale inputs underlying the correction
  diagnostics.  The last column divides the response by the thermal singular
  scale \(L^{12/5}\).  Values at \(L=12,13,14\) are deterministic midpoints of
  the conditional numerical enclosures; the enclosure widths are immaterial at the shown
  precision.}
  \label{tab:potts-scaled-p2-inputs}
  \begin{ruledtabular}
  \begin{tabular}{c c c}
  \(L\) & \(\Ptwo\) & \(\Ptwo/L^{12/5}\) \\
  \hline
  6  & 2.5724018473  & 0.0348960131 \\
  7  & 3.7853777137  & 0.0354710402 \\
  8  & 5.2748255438  & 0.0358750183 \\
  9  & 7.0558980136  & 0.0361718126 \\
  10 & 9.1426531727  & 0.0363975579 \\
  11 & 11.5482427508 & 0.0365740842 \\
  12 & 14.2850529032 & 0.0367152815 \\
  13 & 17.3648125633 & 0.0368303677 \\
  14 & 20.7986788921 & 0.0369256768 \\
  \end{tabular}
  \end{ruledtabular}
\end{table}

\subsection{Finite-size correction diagnostics}
\label{supp:correction-diagnostics}

These diagnostics assess the conditional correction forms discussed after main-text Eq.~\textup{(\ref*{eq:potts-KF-correction-hierarchy})} and the finite-size sequences in Fig.~\ref*{fig:interacting-benchmarks}.

% WPM-BLOCK:B0027
\begin{table}[H]
  \caption{Fit diagnostics for the NNN-TFIM Ising-line controls through \(L=20\).  The
  unconstrained fit uses Eq.~\eqref{eq:nnn-linear-consistency-fit}; the
  constrained fit fixes $K_{F,\infty}=2/3$ and refits only the slope.  The purpose
  is diagnostic: the all-size linear intercepts remain close to the Ising value,
  but the comparison with Table~\ref{tab:nnn-alt-fit-diagnostics} shows that
  precision extrapolation is correction-form dependent.  RMSE denotes the
  root-mean-square error, and \(\mathrm{RMSE}_{2/3}\) is evaluated with
  $K_{F,\infty}=2/3$.}
  \label{tab:nnn-fit-diagnostics}
  \footnotesize
  \begin{ruledtabular}
  \begin{tabular}{c c c c c}
  $J_2$ & $K_{F,\infty}^{\mathrm{lin}}$ & $a$ & RMSE & \(\mathrm{RMSE}_{2/3}\) \\
  \hline
  $0$ & 0.674875 & 1.16170 & $1.20\times10^{-3}$ & $3.38\times10^{-3}$ \\
  $0.05$ & 0.668597 & 1.25671 & $1.29\times10^{-3}$ & $1.49\times10^{-3}$ \\
  $0.10$ & 0.663652 & 1.34603 & $1.51\times10^{-3}$ & $1.91\times10^{-3}$ \\
  $0.20$ & 0.656813 & 1.50529 & $2.22\times10^{-3}$ & $4.40\times10^{-3}$
  \end{tabular}
  \end{ruledtabular}
\end{table}

% WPM-BLOCK:B0028
\begin{table}[H]
  \footnotesize
  \caption{Alternative fit-window diagnostics for the NNN-TFIM controls through
  \(L=20\).  The columns compare the all-size leading linear fit, a large-window
  linear fit over \(L\ge12\), and the all-size quadratic correction form
  \(K_{F,\infty}+a/L+b/L^2\).  The \(1/L\) and \(1/L^2\) terms are empirical
  sensitivity bases, not theory-derived correction exponents.  Their spread is
  a fit-form sensitivity check, not a statistical confidence interval.}
  \label{tab:nnn-alt-fit-diagnostics}
  \begin{ruledtabular}
  \begin{tabular}{c c c c c}
  \(J_2\) & all linear & \(L\ge12\) linear & all quadratic & \(\mathrm{RMSE}_{2/3}\) \\
  \hline
  0 & 0.674875 & 0.669977 & 0.665770 & \(3.38\!\times\!10^{-3}\) \\
  0.05 & 0.668597 & 0.664028 & 0.658920 & \(1.49\!\times\!10^{-3}\) \\
  0.10 & 0.663652 & 0.658814 & 0.652442 & \(1.91\!\times\!10^{-3}\) \\
  0.20 & 0.656813 & 0.650147 & 0.640507 & \(4.40\!\times\!10^{-3}\)
  \end{tabular}
  \end{ruledtabular}
\end{table}

% WPM-BLOCK:B0029
\begin{table}[H]
  \footnotesize
  \caption{Theory-guided Potts extrapolation and correction-form diagnostics
  from Table~\ref{tab:potts-outcome-aware-data}.  The first row is the
  free-intercept diagnostic for the finite-size data in
  Fig.~\ref*{fig:interacting-benchmarks}(a), which displays no fitted curve;
  its exhaustive archived-endpoint
  envelope is \(0.8475891035\)--\(0.8475891084\).  This precision records the
  deterministic computation, not a correction-model-independent physical
  limit.  The next two rows vary the lower fit cutoff.  Fixed-CFT rows compare
  correction mechanisms rather than estimate an intercept.  RMSE is evaluated
  on the stated window.}
  \label{tab:potts-fit-diagnostics}
  \begin{ruledtabular}
  \begin{tabular}{l c c c}
  fit form & window & intercept or target & RMSE \\
  \hline
  \(K_{F,\mathrm{fit}}^{(7/5)}+bL^{-7/5}\) & \(L=6\)--\(14\) & \(0.847589\) & \(3.49\times10^{-4}\) \\
  \(K_{F,\mathrm{fit}}^{(7/5)}+bL^{-7/5}\) & \(L=8\)--\(14\) & \(0.846548\) & \(7.28\times10^{-5}\) \\
  \(K_{F,\mathrm{fit}}^{(7/5)}+bL^{-7/5}\) & \(L=10\)--\(14\) & \(0.846187\) & \(1.13\times10^{-5}\) \\
  \(K_{F,\infty}^{\mathrm{lin}}+a/L\) & \(L=6\)--\(14\) & \(0.830914\) & \(4.28\times10^{-4}\) \\
  \(K_{F,\infty}+a/L+b/L^2\) & \(L=6\)--\(14\) & \(0.837328\) & \(9.44\times10^{-5}\) \\
  \(K_F^{\mathrm{CFT}}+C_{\mathrm{UV}}L^{-7/5}\) & \(L=6\)--\(14\) & \(0.851496\) fixed & \(1.48\times10^{-3}\) \\
  \(K_F^{\mathrm{CFT}}+C_{\mathrm{irr}}L^{-4/5}+C_{\mathrm{UV}}L^{-7/5}\) & \(L=6\)--\(14\) & \(0.851496\) fixed & \(4.86\times10^{-4}\) \\
  \(K_F^{\mathrm{CFT}}+C_{2\mathrm{irr}}L^{-8/5}\) & \(L=6\)--\(14\) & \(0.851496\) fixed & \(8.98\times10^{-4}\) \\
  \end{tabular}
  \end{ruledtabular}
\end{table}

The known leading \(L^{12/5}\) growth of \(\Ptwo\) makes the response scale a
more direct test of candidate correction powers.  An unweighted nonlinear least-squares
fit of the deterministic midpoint values in
Table~\ref{tab:potts-scaled-p2-inputs} to
\(\Ptwo/L^{12/5}=\mathcal A_{P_2}^{\rm sing}+C_\omega L^{-\omega}\) gives \(\omega=1.4511\) on
\(L=6\)--\(14\) and \(1.4477\) on
\(L=10\)--\(14\).  Table~\ref{tab:potts-scale-correction-diagnostics}
compares the fixed powers used to organize the main-text hierarchy.
The \(L^{-8/5}\) rows in the two diagnostic tables are second-order
irrelevant-power comparators, \(2\omega_{\mathrm{irr}}=8/5\); they represent a
quadratic correction in the same scaling field without asserting a nonzero
amplitude for that mechanism.

\begin{table}[H]
  \footnotesize
  \caption{Fixed-power diagnostics for the scaled Potts response
  \(\Ptwo/L^{12/5}\) over \(L=6\)--\(14\).  The \(L^{-7/5}\) row is the
  best-supported fixed single-power description over this window.  LOOCV
  denotes leave-one-out cross-validation.  The two-term row
  resolves a
  smaller \(L^{-4/5}\) amplitude but has a substantially more correlated
  design matrix; it is retained as a mechanism diagnostic rather than the
  main single-power scale diagnostic.}
  \label{tab:potts-scale-correction-diagnostics}
  \begin{ruledtabular}
  \begin{tabular}{l c c c}
  fit form & \(\mathcal A_{P_2}^{\rm sing}\) & RMSE & LOOCV RMSE \\
  \hline
  \(\mathcal A_{P_2}^{\rm sing}+C_{\mathrm{irr}}L^{-4/5}\) & \(0.0390692\) & \(4.97\times10^{-5}\) & \(7.77\times10^{-5}\) \\
  \(\mathcal A_{P_2}^{\rm sing}+C_{\mathrm{UV}}L^{-7/5}\) & \(0.0378228\) & \(3.85\times10^{-6}\) & \(6.27\times10^{-6}\) \\
  \(\mathcal A_{P_2}^{\rm sing}+C_{2\mathrm{irr}}L^{-8/5}\) & \(0.0376156\) & \(1.12\times10^{-5}\) & \(1.83\times10^{-5}\) \\
  \(\mathcal A_{P_2}^{\rm sing}+C_2L^{-2}\) & \(0.0373263\) & \(4.05\times10^{-5}\) & \(6.85\times10^{-5}\) \\
  \(\mathcal A_{P_2}^{\rm sing}+C_{\mathrm{irr}}L^{-4/5}+C_{\mathrm{UV}}L^{-7/5}\) & \(0.0377172\) & \(2.35\times10^{-7}\) & \(6.09\times10^{-7}\) \\
  \end{tabular}
  \end{ruledtabular}
\end{table}

\subsection{Retained-spectrum and symmetry-sector checks}
\label{supp:retained-checks}

These checks support the finite-size evidence of main-text Sec.~\ref*{sec:interacting-benchmarks} and Fig.~\ref*{fig:interacting-benchmarks}; they track truncation, symmetry, and field sensitivity separately.

% WPM-BLOCK:B0030
\begin{table}[H]
  \footnotesize
  \caption{Retained-level and sector-resolution checks.  For Potts the
  half-retained column recomputes \(\Pfour\) after discarding the upper half of
  the retained low-energy levels while keeping the residual-controlled projected-solve
  \(\Ptwo\).  For NNN-TFIM at \(L=20\), Table~\ref{tab:nnn-l20-stability}
  gives the more relevant retained-level scan.}
  \label{tab:retained-level-checks}
  \begin{ruledtabular}
  \begin{tabular}{l r c c c}
  check & dim & $n$ & $\KF$ & $\KF^{(n/2)}$ \\
  \hline
  Potts $6,k=0$ & 130 & 45 & 0.919949 & -- \\
  Potts $6$, full & 729 & 269 & 0.919949 & -- \\
  Potts $12,k=0$ & 44368 & 47 & 0.875126 & 0.875110
  \end{tabular}
  \end{ruledtabular}
\end{table}

% WPM-BLOCK:B0031
\begin{table}[H]
  \footnotesize
  \caption{Potts \(L=6\) charge-sector representation check using complete
  dense spectra and the same \(10^{-7}\) distinct-level grouping rule.  The
  full space, full translation-invariant \(k=0\) block, and explicit
  \((k=0,q=0)\) block agree in all response moments within the stated
  \(10^{-10}\) tolerance.  The distinct dimensions show why the production
  \(k=0\) dimension must not be labeled as charge neutral.}
  \label{tab:potts-charge-sector}
  \begin{ruledtabular}
  \begin{tabular}{l r c c c}
  representation & dim & \(\Ptwo\) & \(\Pfour\) & \(\KF\) \\
  \hline
  full space & 729 & 2.5724018473 & 6.0875359081 & 0.9199493362 \\
  full \(k=0\) & 130 & 2.5724018473 & 6.0875359081 & 0.9199493362 \\
  explicit \((k=0,q=0)\) & 46 & 2.5724018473 & 6.0875359081 & 0.9199493362
  \end{tabular}
  \end{ruledtabular}
\end{table}

% WPM-BLOCK:B0032
\begin{table}[H]
  \footnotesize
  \caption{\(L=20\) retained-level and field-sensitivity diagnostics for the
  NNN-TFIM sector calculation in the released numerical baseline.
  The last column gives the maximum projected-solve residual over each
  retained-level scan.  The range \(n_{\mathrm{low}}\) lists the retained
  low-energy levels used in the convergence scan; \(\Delta\KF\) is the
  difference between the largest and smallest \(\KF\) in that scan.  For
  \(J_2=0.20\), shifting the field confirmed by the gap crossing by
  \(\pm10^{-4}\) changes
  \(\KF\) by only \(1.0\times10^{-6}\).}
  \label{tab:nnn-l20-stability}
  \begin{ruledtabular}
  \begin{tabular}{c c c c c c}
  $J_2$ & $h_\times(20;J_2)$ & $\KF(20;J_2)$ & $n_{\mathrm{low}}$ & $\Delta\KF$ & solve res. \\
  \hline
  0.05 & 1.08399831 & 0.730172967 & 16--64 & $1.4\times10^{-5}$ & $8.8\times10^{-8}$ \\
  0.10 & 1.16606077 & 0.729644737 & 16--32 & $5.7\times10^{-6}$ & $8.2\times10^{-8}$ \\
  0.20 & 1.32555518 & 0.730308153 & 16--96 & $3.4\times10^{-6}$ & $6.5\times10^{-8}$
  \end{tabular}
  \end{ruledtabular}
\end{table}

% WPM-BLOCK:B0033
\section{Operator-dependence finite-size diagnostics}
% WPM-BLOCK:B0034
\label{app:operator-dependence}

This small-size exact-diagonalization comparison supports the operator-dependent
RG discussion in main-text Sec.~\ref*{sec:tfim:envelope}.  The RG
eigenvalue \(y_{\mathcal O}=d+z-\Delta_{\mathcal O}\) organizes the leading
scaling of \(\Ptwo\); fitted exponents compare perturbations, while the archived
low-level ratio is an approximate channel-shape diagnostic with a different
evidence limit.

% WPM-BLOCK:B0035

% WPM-BLOCK:B0036
For the TFIM we use the real-space convention of the main text,
\(H=-\sum_j\sigma_j^z\sigma_{j+1}^z-\sum_j\sigma_j^x\), and compare
\(V_h=-\sum_j\sigma_j^x\), \(V_J=-\sum_j\sigma_j^z\sigma_{j+1}^z\), and the uniform spin-field
perturbation \(V_\sigma=-\sum_j\sigma_j^z\).  The first two are the Ising energy field
in dual representatives, whereas the last couples to the Ising spin primary.
Here \(V_\sigma\) is a diagnostic perturbing operator, not a term present in the
base TFIM Hamiltonian.  Operationally it means evaluating the channel weights of
the infinitesimal source family \(H(\lambda)=H_c+\lambda V_\sigma\) at
\(\lambda=0\), using the eigenstates and gaps of the unperturbed critical TFIM.
For the Potts chain we use the clock/shift convention of
Eq.~\eqref{eq:potts-hamiltonian} and compare the thermal interaction
representative with the uniform order-field perturbation
\(V_\sigma=-\sum_j(\sigma_j+\sigma_j^\dagger)\).  The Potts thermal calculation
diagonalizes the full translation-invariant \(k=0\) block and has independently
confirmed neutral response support.  The order-field
perturbation preserves momentum but changes \(\mathbb Z_3\) charge, so it is
used only as a small-size operator-selectivity diagnostic rather than as the main
thermal finite-size comparison.

% WPM-BLOCK:B0037
% WPM-BLOCK:B0038
\begin{table}[H]
  \caption{Operator-dependence diagnostic from projected exact diagonalization.
  The fitted exponent is obtained from a log--log fit of \(\Ptwo(L)\) over the
  listed small sizes.  The expected exponent is
  \(2(d+z-\Delta_{\mathcal O})\).  The final column reports the largest-size low-level
  \(P_{4,\mathrm{low}}/P_{2,\mathrm{solve}}^2\) diagnostic from the archived
  calculation.  Its retained-eigenpair count is not recorded, so this column
  carries no certified truncation error and is not a thermodynamic extrapolation.}
  \label{tab:operator-dependence-check}
  \begin{ruledtabular}
  \begin{tabular}{l l c c c c c}
  Model & perturbing operator / field & sizes & \(\Delta_{\mathcal O}\) & expected \(\Ptwo\) exponent & fitted exponent & \shortstack{largest-size\\low-level ratio} \\
  \hline
  TFIM & energy, \(-\sum_j\sigma_j^x\) & 6,8,10,12 & 1 & 2 & 2.138 & 0.77198 \\
  TFIM & energy, \(-\sum_j\sigma_j^z\sigma_{j+1}^z\) & 6,8,10,12 & 1 & 2 & 2.138 & 0.77198 \\
  TFIM & spin, \(-\sum_j\sigma_j^z\) & 6,8,10,12 & 1/8 & 15/4 & 3.763 & 0.99988 \\
  Potts & thermal, \(-\sum_j(\sigma_j^\dagger\sigma_{j+1}+\mathrm{H.c.})\) & 4--8 & 4/5 & 12/5 & 2.538 & 0.89677 \\
  Potts & order, \(-\sum_j(\sigma_j+\sigma_j^\dagger)\) & 4--8 & 2/15 & 56/15 & 3.782 & 0.99982 \\
  \end{tabular}
  \end{ruledtabular}
\end{table}

The archived largest-size energy rows have relative solve residuals
\(4.55\times10^{-9}\) (TFIM) and \(8.69\times10^{-9}\) (Potts).
Their retained-group counts, 10 and 12, do not determine the missing
eigenpair-cutoff metadata.  A separate controlled check retains 96 nominal
eigenpairs plus one guard and discards incomplete boundary groups.  It gives
\(K_{F,\mathrm{low}}=0.772640860386\) for TFIM at \(L=12\), versus the
complete NS value \(0.772727272727\), with solve residual
\(1.42\times10^{-10}\).  For the Potts thermal response at \(L=8\), it gives
\(0.896828291430\), agreeing within \(3\times10^{-14}\) with full
diagonalization of the 834-dimensional translation block; the solve residual
is \(4.52\times10^{-9}\).  These separately recorded checks clarify the role
of truncation without replacing the archived diagnostics or certifying them.

% WPM-BLOCK:B0039
The TFIM energy representatives give identical \(\Ptwo\) values at criticality,
as expected from self-duality.  Over \(L=6,8,10,12\), the spin-field
perturbation instead yields a fitted exponent \(3.763\), \(0.013\) above the
expected \(15/4\).  The Potts thermal and order perturbations also have distinct
\(\Ptwo\) growth over \(L=4,\ldots,8\); their fitted exponents differ from the
expected values by about \(0.14\) and \(0.049\), respectively.  These stated
size sets and deviations make the comparison a small-size operator-selectivity
diagnostic rather than a thermodynamic extrapolation.  The displayed
low-level ratios are archived finite-size channel-shape diagnostics under the
stated level resolution, not universal functions of
\(\Delta_{\mathcal O}\).  The exact projector concentration depends on the
full normalized form-factor distribution beyond the total-response exponent.

% WPM-BLOCK:B0040
\section{Numerical implementation and convergence controls}
% WPM-BLOCK:B0041
\label{app:numerical-reproducibility}

% WPM-BLOCK:B0042
This deterministic checklist records the basis reductions, solvers,
tolerances, and validation checks for the reported ED diagnostics.

OpenAI ChatGPT (GPT-5.6 Sol) was used to assist with code organization,
review, and verification. The authors specified the algorithms, equations,
numerical tolerances, and validation criteria governing these tasks. All
scientific judgments concerning the physical models, computational methods,
validation standards, interpretation of numerical results, and reported
conclusions were made exclusively by the authors. All AI-assisted code changes
and checks were manually reviewed by the authors and verified against the
stated analytical formulas, numerical controls, and reproduced outputs. The
authors retain full responsibility for the code and reported results.

% WPM-BLOCK:B0043
% WPM-BLOCK:B0044
\begin{enumerate}
\item \textbf{Potts Hilbert space.}
Standard three-state clock/shift basis, projected only to the full
translation-invariant \(k=0\) block for the production thermal calculation.
The global color-shift projector is applied independently for the
charge-leakage and explicit-\(q=0\) checks, not to the production block
dimension.

\item \textbf{Level grouping.}
Eigenstates with energies differing by at most \(10^{-7}\) are grouped before
forming \(\Pi_a\), so \(\Pfour\) is invariant under rotations inside numerical
multiplets.  This tolerance is a numerical degeneracy rule, not a
scale-dependent conformal-level identification.

\item \textbf{Potts diagonalization.}
Sparse Hermitian diagonalization of retained low-energy levels, with
retained-level checks in Table~\ref{tab:retained-level-checks} and
full/full-\(k=0\)/explicit-\(q=0\) checks, including the \(10^{-10}\) leakage
threshold, in Table~\ref{tab:potts-charge-sector}.

\item \textbf{NNN-TFIM Hilbert space.}
For \(J_2>0\), the response uses the \(k=0\), spin-flip-even sector selected by
\(V_h=-\sum_j\sigma_j^x\).  The \(J_2=0\) reference is the exact critical TFIM
sequence; matched sector ED reproduces \(\Ptwo\), \(\Pfour\), and \(\KF\) in
Table~\ref{tab:j2-zero-resolution-match}.

\item \textbf{NNN pseudo-critical field.}
The phenomenological gap-crossing field \(h_\times(L;J_2)\) is defined by the
condition
\[
(L-2)\Delta_{L-2}(h,J_2)=L\Delta_L(h,J_2),
\]
where
\(\Delta_L=E_{0,-}^{(L)}-E_{0,+}^{(L)}\) is the lowest \(k=0\) spin-flip
odd--even gap defined above.  The same construction is used through \(L=20\);
shifting the \(L=20\), \(J_2=0.20\) field by \(\pm10^{-4}\) changes \(\KF\) by
about \(10^{-6}\).

\item \textbf{\(\Ptwo\) evaluation.}
A projected linear solve for \((H-E_0)\ket{\chi}=Q V\ket{0}\) lifts the
ground-state null vector by a rank-one term. The normalized linear-system
residual is recomputed independently of the stopping criterion; its maximum
in the \(L=20\) retained-eigenpair scans is \(8.8\times10^{-8}\)
(Table~\ref{tab:nnn-l20-stability}).

\item \textbf{Historical \(\Pfour\) numerator.}
The archived numerator sums the squared retained mean-gap weights
\(x_a\), and the concentration diagnostic divides it by the
independently solved \(\Ptwo^2\). Section~\ref{supp:conditional-tail}
distinguishes this diagnostic from the exact projector response.
The half-retained-level check recomputes \(\Pfour\) after discarding the upper half
of retained levels while keeping the residual-controlled projected-solve \(\Ptwo\).

\item \textbf{Precision and validation.}
Calculations use double precision; parity-gap eigensolver tolerance
\(10^{-10}\), scaled-gap-crossing root absolute and relative tolerances
\(10^{-10}\), response
eigensolver tolerance \(10^{-9}\), MINRES stopping parameter
\(10^{-9}\), and primary level-grouping tolerance \(10^{-7}\).  Even/odd
eigenpair residuals and scaled crossing residuals are checked explicitly.
The matched \(J_2=0\) resolution check documents agreement between the
level-projector and momentum-block representations at the tested sizes in
Table~\ref{tab:j2-zero-resolution-match}. Independent full-space
parity tests at \(L=4,6,8\) were also reported in the historical validation;
their separate test records are unavailable in the released numerical baseline.
\end{enumerate}
% WPM-BLOCK:B0045

\subsection{Conditional spectral-tail enclosure}
\label{supp:conditional-tail}

This derivation supplies the conditional numerical enclosure used in main-text
Sec.~\ref*{sec:interacting-benchmarks} and Fig.~\ref*{fig:interacting-benchmarks},
and the retained-rank and omitted-mass bounds used in
Fig.~\ref*{fig:potts-weight-comparators}.
Let \(P_-\le\Ptwo\le P_+\), with \(P_->0\), enclose the total
squared norm of the reduced-resolvent response, and let
\(0\le r_a^-\le r_a\le r_a^+\) enclose each retained, complete
spectral-group mass. For the retained set \(R\), define
\begin{align}
 Q_r^-&=\sum_{a\in R}(r_a^-)^2,
 &Q_r^+&=\sum_{a\in R}(r_a^+)^2,\notag\\
 T_r^+&=P_+-\sum_{a\in R}r_a^-.
 \label{eq:supp-tail-mass}
\end{align}
Consistency requires \(T_r^+\ge0\). Positivity of omitted response
masses gives
\[
 \sum_{a\notin R}r_a^2
 \le\left(\sum_{a\notin R}r_a\right)^2
 \le(T_r^+)^2.
\]
The normalized projector concentration therefore obeys
\begin{equation}
 \begin{split}
 \frac{Q_r^-}{P_+^2}
 &\le K_F^{\mathrm{proj}}\\
 &\le\min\!\left\{1,
   \frac{Q_r^++(T_r^+)^2}{P_-^2}\right\}.
 \end{split}
 \label{eq:supp-tail-enclosure}
\end{equation}
All endpoint operations in the machine-readable enclosure calculation use outward rounding. Inconsistent operands,
including retained lower mass above \(P_+\), are rejected rather than
clipped. The bound uses resolvent masses in its numerator, normalization
and tail; no omitted mean-gap conversion is needed.

The historical mean-gap weights define a different self-normalized
concentration, \(K_F^{(x)}=\sum_a x_a^2/S_x^2\), where
\(S_x=\sum_a x_a\). A numerator normalized by the resolvent total is
instead \(\widetilde K_F=\sum_a x_a^2/\Ptwo^2\), so that
\(\widetilde K_F=K_F^{(x)}(S_x/\Ptwo)^2\).
Neither is identified with \(K_F^{\mathrm{proj}}\) without controlling
the mean-gap approximation. A retained-only numerator carries the
additional superscript \(\mathrm{low}\).

For positive individual-gap and mean-gap intervals
\(\Delta_n\in[d_a^-,d_a^+]\) and
\(\bar\Delta_a\in[m_a^-,m_a^+]\), positivity gives
\begin{equation}
 \left(\frac{d_a^-}{m_a^+}\right)^2r_a
 \le x_a\le
 \left(\frac{d_a^+}{m_a^-}\right)^2r_a.
\end{equation}
This follows by writing
\(x_a=\sum_{n\in a}(\Delta_n/\bar\Delta_a)^2
|\langle n|V|0\rangle|^2/\Delta_n^2\); it assumes no independence
between the two weights. Retained-cluster bounds do not constrain the
unresolved mean-gap tail.

For the total response, a positive lower bound \(\beta\) on the augmented
projected operator and an effective residual bound \(\epsilon\), including
right-hand-side and ground-projector error, imply
\(\|y-y_*\|\le\epsilon/\beta\).  Thus its norm squared lies between
\(\max(0,\|y\|-\epsilon/\beta)^2\) and
\((\|y\|+\epsilon/\beta)^2\).  The stored calculation forms \(\beta\) from
the rank-one lift and the estimated ground gap, and forms retained-projector
error bounds from eigenpair residuals and intercluster separation.  One guard
eigenpair excludes an incomplete retained boundary cluster; stability is
checked at the recorded grouping tolerances.

These steps are conditional on correct ground-state and low-spectrum
identification, complete retained spectral clusters, and valid separation and
operator-error bounds.  Small Ritz residuals locate nearby eigenvalues but do
not exclude an uncomputed lower eigenvalue or prove spectral completeness.
The archived receipts therefore supply numerical premises for
Eq.~\eqref{eq:supp-tail-enclosure}, rather than a standalone proof of all
spectral premises.  Their reported interval widths quantify propagated
finite-size numerical uncertainty under those premises; they do not quantify
finite-size extrapolation or CFT matching error.

Without a new eigensolve, a replay uses all nine stored resolvent-mass rungs at
\(L=12,13,14\).
Each same-response enclosure is contained in the corresponding archived
interval.  At the finest retained count their widths are at most
\(3.09\times10^{-9}\), \(3.18\times10^{-9}\) and
\(3.46\times10^{-9}\), respectively.  Their endpoints round to the same
seven decimal places, but not eight; additional printed digits specify
interval endpoints rather than an equally precise point estimate.

The archived scientific tables and all plotted central values are retained.
The accompanying reproducibility package supplies the machine-readable
Potts projector-response enclosure dataset (R3). Records are indexed
by system size and nominal retained-eigenpair count and distinguish
the outward exact-response endpoints from the historical intervals.
The printed endpoints in Table~\ref{tab:potts-outcome-aware-data} are
rounded historical displays and do not inherit the directed-rounding
guarantee.

For the \(L=14\) distribution the first fifteen retained cluster identities
have strictly separated normalized response intervals in the same order.
The omitted probability is at most \(1.051\times10^{-6}\), which protects
only the first ten ranks against any omitted channel.  Ranks eleven through
fifteen therefore denote retained ordering, not certified full-spectrum
ordering.  The ratio of the first to second exact-response masses is enclosed
by \([18.83196928898,18.83196929648]\), consistent with the plotted
second-to-leading ratio \(0.0531\).

% WPM-BLOCK:B0046
% ======================================================================
% END NUMERICAL SUPPLEMENT

\end{document}